\documentclass[12pt]{amsart}
\usepackage[T1]{fontenc}
\usepackage{amssymb}
\usepackage{a4wide}
\usepackage{graphicx}
\usepackage{verbatim}
\usepackage{amsmath}
\usepackage{version}

\newcommand{\be}{\begin{equation}}
\newcommand{\ee}{\end{equation}}

 \theoremstyle{plain}    
 \newtheorem{thm}{Theorem}[section]
 \numberwithin{equation}{section} 
 \numberwithin{figure}{section} 
 \theoremstyle{plain}
 \newtheorem{prop}[thm]{Proposition} 
 \theoremstyle{plain}    
 \newtheorem{lem}[thm]{Lemma} 
 \theoremstyle{plain}    
 \newtheorem{cor}[thm]{Corollary} 
 \theoremstyle{plain}    

 \theoremstyle{definition}
 \newtheorem{rem}[thm]{Remark} 
 \theoremstyle{definition}

\begin{document}
\newcommand{\nwc}{\newcommand}
\nwc{\nwt}{\newtheorem}
\nwt{coro}{Corollary}
\nwt{ex}{Example}


\nwc{\mf}{\mathbf} 
\nwc{\blds}{\boldsymbol} 
\nwc{\ml}{\mathcal} 


\nwc{\lam}{\lambda}
\nwc{\del}{\delta}
\nwc{\Del}{\Delta}
\nwc{\Lam}{\Lambda}
\nwc{\elll}{\ell}

\nwc{\IA}{\mathbb{A}} 
\nwc{\IB}{\mathbb{B}} 
\nwc{\IC}{\mathbb{C}} 
\nwc{\ID}{\mathbb{D}} 
\nwc{\IE}{\mathbb{E}} 
\nwc{\IF}{\mathbb{F}} 
\nwc{\IG}{\mathbb{G}} 
\nwc{\IH}{\mathbb{H}} 
\nwc{\IN}{\mathbb{N}} 
\nwc{\IP}{\mathbb{P}} 
\nwc{\IQ}{\mathbb{Q}} 
\nwc{\IR}{\mathbb{R}} 
\nwc{\IS}{\mathbb{S}} 
\nwc{\IT}{\mathbb{T}} 
\nwc{\IZ}{\mathbb{Z}} 
\def\bbbone{{\mathchoice {1\mskip-4mu {\rm{l}}} {1\mskip-4mu {\rm{l}}}
{ 1\mskip-4.5mu {\rm{l}}} { 1\mskip-5mu {\rm{l}}}}}



\nwc{\va}{{\bf a}}
\nwc{\vb}{{\bf b}}
\nwc{\vc}{{\bf c}}
\nwc{\vd}{{\bf d}}
\nwc{\ve}{{\bf e}}
\nwc{\vf}{{\bf f}}
\nwc{\vg}{{\bf g}}
\nwc{\vh}{{\bf h}}
\nwc{\vi}{{\bf i}}
\nwc{\vI}{{\bf I}}
\nwc{\vj}{{\bf j}}
\nwc{\vk}{{\bf k}}
\nwc{\vl}{{\bf l}}
\nwc{\vm}{{\bf m}}
\nwc{\vM}{{\bf M}}
\nwc{\vn}{{\bf n}}
\nwc{\vo}{{\it o}}
\nwc{\vp}{{\bf p}}
\nwc{\vq}{{\bf q}}
\nwc{\vr}{{\bf r}}
\nwc{\vs}{{\bf s}}
\nwc{\vt}{{\bf t}}
\nwc{\vu}{{\bf u}}
\nwc{\vv}{{\bf v}}
\nwc{\vw}{{\bf w}}
\nwc{\vx}{{\bf x}}
\nwc{\vy}{{\bf y}}
\nwc{\vz}{{\bf z}}
\nwc{\bal}{\blds{\alpha}}
\nwc{\bep}{\blds{\epsilon}}
\nwc{\barbep}{\overline{\blds{\epsilon}}}
\nwc{\bnu}{\blds{\nu}}
\nwc{\bmu}{\blds{\mu}}



\nwc{\bk}{\blds{k}}
\nwc{\bm}{\blds{m}}
\nwc{\bM}{\blds{M}}
\nwc{\bp}{\blds{p}}
\nwc{\bq}{\blds{q}}
\nwc{\bn}{\blds{n}}
\nwc{\bv}{\blds{v}}
\nwc{\bw}{\blds{w}}
\nwc{\bx}{\blds{x}}
\nwc{\bxi}{\blds{\xi}}
\nwc{\by}{\blds{y}}
\nwc{\bz}{\blds{z}}


\nwc{\cA}{\ml{A}}
\nwc{\cB}{\ml{B}}
\nwc{\cC}{\ml{C}}
\nwc{\cD}{\ml{D}}
\nwc{\cE}{\ml{E}}
\nwc{\cF}{\ml{F}}
\nwc{\cG}{\ml{G}}
\nwc{\cH}{\ml{H}}
\nwc{\cI}{\ml{I}}
\nwc{\cJ}{\ml{J}}
\nwc{\cK}{\ml{K}}
\nwc{\cL}{\ml{L}}
\nwc{\cM}{\ml{M}}
\nwc{\cN}{\ml{N}}
\nwc{\cO}{\ml{O}}
\nwc{\cP}{\ml{P}}
\nwc{\cQ}{\ml{Q}}
\nwc{\cR}{\ml{R}}
\nwc{\cS}{\ml{S}}
\nwc{\cT}{\ml{T}}
\nwc{\cU}{\ml{U}}
\nwc{\cV}{\ml{V}}
\nwc{\cW}{\ml{W}}
\nwc{\cX}{\ml{X}}
\nwc{\cY}{\ml{Y}}
\nwc{\cZ}{\ml{Z}}


\nwc{\tA}{\widetilde{A}}
\nwc{\tB}{\widetilde{B}}
\nwc{\tk}{\tilde k}
\nwc{\tN}{\tilde N}
\nwc{\tP}{\widetilde{P}}
\nwc{\tQ}{\widetilde{Q}}
\nwc{\tR}{\widetilde{R}}
\nwc{\tV}{\widetilde{V}}
\nwc{\tW}{\widetilde{W}}
\nwc{\ty}{\tilde y}
\nwc{\teta}{\tilde \eta}
\nwc{\tdelta}{\tilde \delta}
\nwc{\tlambda}{\tilde \lambda}
\nwc{\ttheta}{\tilde \theta}
\nwc{\tvartheta}{\tilde \vartheta}
\nwc{\tPhi}{\widetilde \Phi}
\nwc{\tpsi}{\tilde \psi}

\nwc{\To}{\longrightarrow} 

\nwc{\ad}{\rm ad}
\nwc{\eps}{\epsilon}
\nwc{\ep}{\epsilon}
\nwc{\vareps}{\varepsilon}

\def\ep{\epsilon}
\def\tr{{\rm tr}}
\def\Tr{{\rm Tr}}
\def\i{{\rm i}}
\def\mi{{\rm i}}
\def\e{{\rm e}}
\def\sq2{\sqrt{2}}
\def\sqn{\sqrt{N}}
\def\vol{\mathrm{vol}}
\def\defi{\stackrel{\rm def}{=}}
\def\t2{{\mathbb T}^2}
\def\s2{{\mathbb S}^2}
\def\hn{\mathcal{H}_{N}}
\def\shbar{\sqrt{\hbar}}
\def\A{\mathcal{A}}
\def\N{\mathbb{N}}
\def\T{\mathbb{T}}
\def\R{\mathbb{R}}
\def\Z{\mathbb{Z}}
\def\C{\mathbb{C}}
\def\O{\mathcal{O}}
\def\Sp{\mathcal{S}_+}
\def\Lap{\triangle}
\nwc{\lap}{\bigtriangleup}
\nwc{\rest}{\restriction}
\nwc{\Diff}{\operatorname{Diff}}
\nwc{\diam}{\operatorname{diam}}
\nwc{\Res}{\operatorname{Res}}
\nwc{\Spec}{\operatorname{Spec}}
\nwc{\Vol}{\operatorname{Vol}}
\nwc{\Op}{\operatorname{Op}}
\nwc{\supp}{\operatorname{supp}}
\nwc{\Span}{\operatorname{span}}

\nwc{\dia}{\varepsilon}
\nwc{\cut}{f}
\nwc{\qm}{u_\hbar}

\def\hto0{\xrightarrow{\hbar\to 0}}
\def\htoo{\stackrel{h\to 0}{\longrightarrow}}
\def\rto0{\xrightarrow{r\to 0}}
\def\rtoo{\stackrel{r\to 0}{\longrightarrow}}
\def\ntoinf{\xrightarrow{n\to +\infty}}

\providecommand{\abs}[1]{\lvert#1\rvert}
\providecommand{\norm}[1]{\lVert#1\rVert}
\providecommand{\set}[1]{\left\{#1\right\}}

\nwc{\la}{\langle}
\nwc{\ra}{\rangle}
\nwc{\lp}{\left(}
\nwc{\rp}{\right)}

\nwc{\bequ}{\begin{equation}}
\nwc{\ben}{\begin{equation*}}
\nwc{\bea}{\begin{eqnarray}}
\nwc{\bean}{\begin{eqnarray*}}
\nwc{\bit}{\begin{itemize}}
\nwc{\bver}{\begin{verbatim}}

%\nwc{\eal}{\end{align}}
\nwc{\eequ}{\end{equation}}
\nwc{\een}{\end{equation*}}
\nwc{\eea}{\end{eqnarray}}
\nwc{\eean}{\end{eqnarray*}}
\nwc{\eit}{\end{itemize}}
\nwc{\ever}{\end{verbatim}}

\newcommand{\defeq}{\stackrel{\rm{def}}{=}}

\title[Half-delocalization]
{Half-delocalization of eigenfunctions for the Laplacian on an Anosov manifold}

\author[N. Anantharaman]{Nalini Anantharaman}
\author[S. Nonnenmacher]{St\'ephane Nonnenmacher}
\address{Unit\'e de Math\'ematiques Pures et Appliqu\'ees, \'Ecole Normale Sup\'erieure,
6, all\'ee d'Italie, 69364 LYON Cedex 07, France}
\email{nanantha@umpa.ens-lyon.fr}
\address{Service de Physique Th\'eorique, 
CEA/DSM/PhT, Unit\'e de recherche associ\'e CNRS,
CEA/Saclay,
91191 Gif-sur-Yvette, France}
\email{snonnenmacher@cea.fr}

\begin{abstract}
We study the high-energy eigenfunctions of 
the Laplacian on a compact Riemannian manifold with Anosov geodesic flow.
The localization of a semiclassical 
measure associated with a sequence of eigenfunctions is characterized by
the Kolmogorov-Sinai entropy of this measure. 
We show that this entropy is necessarily bounded
from below by a constant which, in the case of constant negative curvature,
equals {\em half} the maximal entropy. In this sense, high-energy eigenfunctions are
at least half-delocalized.
\end{abstract}
\maketitle

The theory of quantum chaos tries to understand how the chaotic behaviour of a classical
Hamiltonian system is reflected in its quantum version. 
For instance, let $M$ be a compact Riemannian $C^\infty$ manifold, such that the geodesic flow has the Anosov
property --- the ideal chaotic behaviour. The corresponding quantum dynamics 
is the unitary flow generated by the Laplace-Beltrami operator on $L^2(M)$. One expects that the 
chaotic properties of the geodesic flow influence the spectral theory
of the Laplacian. The Random Matrix conjecture \cite{bohigas} asserts that the 
high-lying eigenvalues should, after proper renormalization, statistically resemble those of a 
large random matrix, at least for a generic Anosov metric. 
The Quantum
Unique Ergodicity conjecture \cite{RudSar94} (see also \cite{berry77, voros77}) deals with 
the corresponding eigenfunctions $\psi$: it claims that the probability density
$|\psi(x)|^2 dx$ should approach (in a weak sense) the Riemannian volume,
when the eigenvalue corresponding to $\psi$ tends to infinity. In fact a stronger property should hold for the 
{\em Wigner transform} $W_{\psi}$, a distribution on the cotangent bundle $T^* M$ 
which describes the distribution of the wave function $\psi$ on the phase space
$T^* M$. We will adopt a semiclassical point of view, that is consider the
eigenstates of eigenvalue unity of the semiclassical Laplacian $-\hbar^2\lap$, in
the semiclassical limit $\hbar\to 0$.
Weak limits of the distributions $W_\psi$
are called {\em semiclassical measures}: they are invariant measures of the geodesic flow 
on the unit energy layer $\cE$.
The Quantum Unique Ergodicity conjecture asserts that on an Anosov manifold 
there exists a unique semiclassical measure, 
namely the Liouville measure on $\cE$; in other words, in the semiclassical r\'egime all
eigenfunctions become uniformly distributed over $\cE$. 

For manifolds with an ergodic geodesic flow (with respect to the Liouville measure), 
it has been shown by Schnirelman, Zelditch and Colin de Verdi\`ere 
that {\em almost all} eigenfunctions 
become uniformly distributed over $\cE$, in the semiclassical limit: this property 
is dubbed as Quantum Ergodicity \cite{Shni74,Zel87,CdV85}.
The possibility of {\em exceptional sequences} of eigenstates with
different semiclassical limits remains open in general. The Quantum Unique Ergodicity conjecture
states that such sequences do not exist for an Anosov manifold \cite{RudSar94}.

So far the most precise results
on this question were obtained for Anosov manifolds $M$ with {\em arithmetic} properties:
see Rudnick--Sarnak \cite{RudSar94},  Wolpert \cite{Wol01}. Recently, 
Lindenstrauss \cite{Linden06} proved the asymptotic equidistribution of all
``arithmetic'' eigenstates (these are believed to exhaust
the full family of eigenstates). The proof, unfortunately, cannot be extended to general Anosov manifolds.

To motivate the conjecture, 
one may instead invoke the following dynamical explanation.
By the Heisenberg uncertainty principle, 
an eigenfunction
cannot be strictly localized on a submanifold in phase space. Its
microlocal support must contain a symplectic cube of volume $\hbar^{d}$, 
where $d$ is the dimension of $M$. Since $\psi$
is invariant under the quantum dynamics, which is semiclassically approximated by the geodesic flow, 
the {\em fast mixing} property of the latter will spread this cube throughout
the energy layer, showing that the support of the eigenfunction must also spread throughout $\cE$.

This argument is however too simplistic. First,
Colin de Verdière and Parisse 
showed that, on a surface of revolution of negative curvature,
eigenfunctions can
concentrate on a single periodic orbit in the semiclassical limit, despite the
exponential unstability of that orbit \cite{CdVP94}. Their construction shows that one cannot
use purely local features, such as instability, to rule out
localization of eigenfunctions on closed geodesics.
Second,  the argument above is based on the classical dynamics, and does not take into
account the {\em interferences} of the wavefunction with itself, after a long
time.
Faure, Nonnenmacher and De Bi\`evre exhibited in \cite{FNdB03} a simple example of a symplectic Anosov
dynamical system, namely the action of a linear hyperbolic automorphism on the 2-torus (also called
``Arnold's cat map''), 
the quantization of which does not
satisfy the Quantum Unique Ergodicity conjecture. Precisely, they construct 
a family of eigenstates for which the semiclassical measure consists in two ergodic components:
half of it is the Liouville measure, while the other half is a Dirac peak on 
a single unstable periodic orbit. It was also shown
that --- in the case of the ``cat map'' --- this half-localization 
on a periodic orbit is {\em maximal} \cite{FN04}.
Another type of semiclassical measures was recently exhibited by Kelmer for quantized automorphisms
on higher-dimensional tori and some of their perturbations \cite{Kelmer05,Kelmer06}: 
it consists in the Lebesgue measure on some invariant co-isotropic 
subspace of the torus. In those cases, the existence of exceptional 
eigenstates is due to some nongeneric algebraic properties of the 
classical and quantized systems.

In a previous paper \cite{AN06}, we discovered how to use an information-theoretic variant 
of the uncertainty principle \cite{Kraus87, MaaUff}, called the {\em Entropic Uncertainty Principle}, 
to constrain the localization
properties of eigenfunctions in the case of another toy model, the Walsh-quantized baker's map. 
For any dynamical system, the complexity of an invariant measure
can be described through its {\em Kolmogorov--Sinai entropy}. In the case of the Walsh-baker's map, 
we showed
that the entropy of semiclassical measures must be at least half the entropy of the Lebesgue measure. 
Thus, our result can be
interpreted as a ``half-delocalization'' of eigenstates. The Walsh-baker model being very
special, 
it was not clear whether the strategy could be generalized to more realistic systems, like geodesic flows 
or more general symplectic systems quantized {\em \`a la} Weyl. 

In this paper we show that it is the case: the strategy used in \cite{AN06} is rather general, and its
implementation to the case of Anosov geodesic flows
only requires more technical
suffering. 

\section{Main result.}
Let $M$ be a compact Riemannian manifold. We will denote by
$\abs{\cdot}_x$ the norm on $T^*_x M$ given by the metric.
The geodesic
flow $(g^t)_{t\in \R}$ is the Hamiltonian flow on $T^* M$ generated by the
Hamiltonian 
$$
H(x, \xi)=\frac{\abs{\xi}^2_x}2\,.
$$ 
In the semiclassical setting, the corresponding quantum operator is $-\frac{\hbar^2\lap}2$, which
generates the unitary flow $(U^t)= (\exp(it\hbar\frac\lap{2}))$ acting on 
$L^2(M)$.

We denote by $(\psi_k)_{k\in \N}$ an orthonormal basis
of $L^2(M)$ made of eigenfunctions of the Laplacian, and by $(\frac1{\hbar_k^{2}})_{k\in \N}$
the corresponding eigenvalues: 
$$
-\hbar_k^2\,\Lap \psi_k=\psi_k,\quad \text{with} \quad \hbar_{k+1}\leq \hbar_{k}\,.
$$ 
We are interested in the high-energy eigenfunctions of $-\lap$, in other words the
semiclassical limit $\hbar_k\to 0$.

The Wigner distribution associated to an eigenfunction $\psi_k$ is defined by
$$
W_k(a)=\langle \Op_{\hbar_k}(a)\psi_k, \psi_k\rangle_{L^2(M)},\qquad a\in C_c^\infty(T^* M)\,.
$$
Here $\Op_{\hbar_k}$ is a quantization procedure, set at the scale $\hbar_k$, 
which associates a bounded
operator on $L^2(M)$ to any smooth phase space function $a$ with nice behaviour at infinity 
(see for instance \cite{DS99}). If $a$ is a function
on the manifold $M$, we have $W_k(a)= \int_M a(x) |\psi_k(x)|^2 dx$: the distribution $W_k$ is
a {\em microlocal lift} of the probability measure $|\psi_k(x)|^2 dx$ into a phase space distribution. 
Although the definition
of $W_k$ depends on a certain number of choices, like the choice of local coordinates,
or of the quantization procedure (Weyl, anti-Wick, ``right'' or ``left'' quantization...),
its asymptotic
behaviour when $\hbar_k\To 0$ does not. Accordingly, we call 
{\em semiclassical measures} the limit points
of the sequence $(W_k)_{k\in\IN}$, in the distribution topology.

Using standard semiclassical arguments, one easily shows
the following \cite{CdV85}:
\begin{prop}\label{e:semiclass-measure} 
Any semiclassical measure is a probability measure carried on the energy
layer $\cE=H^{-1}(\frac12)$ (which coincides with the unit cotangent bundle $\cE=S^* M$). 
This measure is invariant under the geodesic flow.
\end{prop}
If the geodesic flow has the Anosov property --- for instance if $M$ has negative sectional curvature ---
then there exist many invariant probability measures on $\cE$, in addition to the Liouville measure.
The geodesic flow has countably many periodic orbits, each of them
carrying an invariant probability measure. There are still many others, like the equilibrium states
obtained by variational principles \cite{KatHas95}.
The Kolmogorov--Sinai entropy, also called
metric entropy, of a $(g^t)$-invariant probability measure $\mu$ is a nonnegative number $h_{KS}(\mu)$
that describes, in some sense, the complexity of a
$\mu$-typical orbit of the flow. For instance, a measure carried on a closed geodesic
has zero entropy.  An upper bound on the entropy
is given by the Ruelle inequality: since the geodesic flow has the Anosov
property, the energy layer $\cE$ is foliated into unstable manifolds of the flow,
and for any invariant probability measure $\mu$ one has
\begin{equation}\label{Ruelle}
h_{KS}(\mu)\leq\left| \int_{\cE} \log J^u(\rho)d\mu(\rho)\right|\,.
\end{equation}
In this inequality, $J^u(\rho)$ is the {\em unstable Jacobian} of the flow at the point $\rho\in \cE$, 
defined as the Jacobian of the map $g^{-1}$ restricted to the unstable manifold at the
point $g^1 \rho$ (the average of $\log J^u$ over any invariant measure is negative).
If $M$ has dimension $d$ and has constant sectional curvature $-1$, this inequality just reads
$h_{KS}(\mu)\leq d-1$.
The equality holds in \eqref{Ruelle} if and only if $\mu$ is the Liouville measure on $\cE$ \cite{LY85}.
Our central result is the following
\begin{thm}\label{thethm} 
Let $\mu$ be a semiclassical measure
associated to the eigenfunctions of the Laplacian on $M$. Then its metric entropy satisfies
\begin{equation}\label{e:main1}
h_{KS}(\mu)\geq \frac32\left|\int_{\cE} \log J^u(\rho)d\mu(\rho) \right|- (d-1)\lambda_{\max}\,,
\end{equation}
where $d=\dim M$ and $\lambda_{\max}=\lim_{t\to\pm \infty}\frac{1}t
\log \sup_{\rho\in \cE} |dg^t_\rho|$ is the maximal expansion rate of the geodesic flow on $\cE$.

In particular, if $M$ has constant sectional curvature $-1$, this means that
\begin{equation}\label{e:main2}
h_{KS}(\mu)\geq \frac{d-1}2.
\end{equation}
\end{thm}
The first author proved in \cite{An} that the entropy of such a semiclassical measure is
bounded from below by a positive (hardly explicit) constant. 
The bound \eqref{e:main2} in the above theorem is much sharper
in the case of constant curvature. On the other hand, 
if the curvature varies a lot (still being negative everywhere),
the right hand side of \eqref{e:main1} may actually be negative, 
in which case the above bound is trivial.
We believe this to be but a technical shortcoming of our 
method\footnote{Herbert Koch has recently managed to improve the above lower bound to 
$\left|\int_{\cE} \log J^u(\rho)d\mu(\rho) \right|- \frac{(d-1)\lambda_{\max}}{2}$.}, 
and would actually expect the following bound
to hold:
\begin{equation}\label{e:main3}
h_{KS}(\mu)\geq \frac12\left|\int_{\cE} \log J^u(\rho)d\mu(\rho) \right|\,.
\end{equation}
\begin{rem}
Proposition~\ref{e:semiclass-measure} and Theorem~\ref{thethm} still apply if $\mu$ is
not associated to a subsequence of eigenstates,
but rather a sequence $(\qm)_{\hbar\to 0}$ of
{\em quasimodes} of the Laplacian, of the following order:
$$
\norm{(-\hbar^2\lap-1)\qm}=o(\hbar|\log\hbar|^{-1})\norm{\qm}\,,\qquad\hbar\to 0\,.
$$
This extension of the theorem requires little modifications, which we leave to the
reader.
It is also possible to prove lower bounds on the entropy
in the case of quasimodes of the type
$$
\norm{(-\hbar^2\lap-1)\qm}\leq c\,\hbar|\log\hbar|^{-1}\norm{\qm}\,,\qquad\hbar\to 0\,,
$$
as long as $c>0$ is sufficiently small. However, this extension is not as straightforward as 
in \cite{An}, so we defer it to a future work. 
\end{rem}
\begin{rem}
In this article we only treat the case of the geodesic flow, but our methods can obviously be adapted
to the case of a more general Hamiltonian flow, assumed to be Anosov on some compact energy layer. 
The quantum operator can then be any standard $\hbar$-quantization of
the Hamiltonian function.
\end{rem}
Although this paper is overall in the same spirit as \cite{An}, certain
aspects of the proof are quite different. We recall that the proof given in \cite{An}
required to study the quantum dynamics far beyond the Ehrenfest time --- i.e. 
the time needed by the classical flow to transform wavelengths $\sim 1$ into wavelengths $\sim \hbar$.
In this paper we will study the dynamics until twice the Ehrenfest time, but not beyond.
In variable curvature, the fact that the Ehrenfest time depends on the initial
position seems to be the reason why the bound \eqref{e:main1} is not optimal.

Quantum Unique Ergodicity would mean that
$h_{KS}(\mu)= \left|\int_{\cE} \log J^u(\rho)\,d\mu(\rho) \right|$.
We believe however that \eqref{e:main3}
is the optimal result that can be obtained without using more precise information,
like for instance upper bounds on the multiplicities of eigenvalues. 
Indeed, in the above mentioned examples of Anosov systems where
Quantum Unique Ergodicity fails, the
bound \eqref{e:main3} is actually {\em sharp} \cite{FNdB03, Kelmer05, AN06}. 
In those examples, the spectrum has
high degeneracies in the semiclassical limit, 
which allows for a lot of freedom to select the eigenstates.
Such high degeneracies are not expected to happen in the case of the Laplacian 
on a negatively curved manifold. 
Yet, for the moment we have no clear understanding
of the relationship between spectral degeneracies and failure of Quantum Unique Ergodicity.

\subsection*{ Acknowledgements} 
Both authors were partially supported by the Agence Nationale de la Recherche, under
the grant ANR-05-JCJC-0107-01. They are grateful to Yves Colin de Verdi\`ere for his
encouragement and his comments.
S.~Nonnenmacher also thanks Maciej Zworski and Didier Robert for interesting discussions, and
Herbert Koch for his enlightening remarks on the Riesz-Thorin theorem.

\section{Outline of the proof}
\subsection{Weighted entropic uncertainty principle}\label{p:WEIP}

Our main tool is an adaptation of the entropic uncertainty principle conjectured
by Kraus in \cite{Kraus87} and proven by Maassen and Uffink \cite{MaaUff}. 
This principle states that if a unitary matrix has ``small'' entries, then
any of its eigenvectors must have a ``large'' Shannon entropy.
For our purposes, we need an elaborate version of this uncertainty principle,
which we shall prove in Section \ref{s:proofWEUP}. 

Let $(\cH, \la.,.\ra)$ be a complex Hilbert space, and denote $\norm{\psi}=\sqrt{\la\psi,\psi\ra}$ 
the associated norm. Let $\pi=(\pi_k)_{k=1,\ldots,\cN}$ be an quantum partition of unity,
that is, a family of operators on $\cH$ such that 
\begin{equation}\label{e:unity}
\sum_{k=1}^{\cN}\pi_{k}\pi_{k}^{*}=Id.
\end{equation}
In other words, for all $\psi\in \cH$ we have
$$
\norm{\psi}^2=\sum_{k=1}^{\cN} \norm{\psi_k}^2\,\qquad\text{where we denote}\ \psi_k=\pi_k^*\psi\quad
\text{for all }k=1,\ldots,\cN\,.
$$
If $\norm{\psi}=1$, we define the entropy of $\psi$
with respect to the partition $\pi$ as
$$
h_\pi(\psi)=-\sum_{k=1}^{\cN} \norm{\psi_k}^2\log\norm{\psi_k}^2\,.
$$
We extend this definition by introducing the notion of pressure, associated to a family 
$(\alpha_k)_{k=1,\ldots,\cN}$ of positive real numbers: it is defined by
$$
p_{\pi,\alpha}(\psi)=-\sum_{k=1}^\cN\norm{\psi_k}^2\log\norm{\psi_k}^2-\sum_{k=1}^\cN 
\norm{\psi_k}^2\log\alpha_k^2.
$$
In Theorem \ref{t:WEUP} below, we use two families of weights $(\alpha_k)_{k=1,\ldots,\cN}$,
$(\beta_j)_{j=1,\ldots,\cN}$, and consider the corresponding pressures $p_{\pi,\alpha}$,
$p_{\pi,\beta}$.

Besides the appearance of the weights
$\alpha, \beta$, we also modify the statement in \cite{MaaUff}
by introducing an auxiliary operator $O$ --- for reasons that should become clear later.
\begin{thm}
\label{t:WEUP}
Let $O$ be a bounded operator and $\cU$ an isometry on $\cH$.
Define  $A=\max_k \alpha_k$, $B=\max_j \beta_j$ and 
$$
c_O^{(\alpha, \beta)}(\cU)\defi\sup_{j,k}\alpha_k\beta_j \norm{\pi_j^*\,\cU\,\pi_{k}\, O}_{\cL(\cH)}\,.
$$
 Then, for any $\eps\geq 0$, for any normalized $\psi\in\cH$ satisfying
$$
\forall k=1,\ldots,\cN,\qquad \norm {(Id-O)\pi^*_k\psi}\leq \eps\,,
$$
the pressures $p_{\pi,\beta}\big(\cU\psi \big)$, $p_{\pi,\alpha}\big(\psi\big)$ satisfy
$$
p_{\pi,\beta}\big(\cU\psi \big) + p_{\pi,\alpha}\big(\psi\big)
\geq - 2 \log \big(c_O^{(\alpha, \beta)}(\cU)+\cN\,A\,B\,\eps\big)\,.
$$
\end{thm}
\begin{rem}
The result of \cite{MaaUff} corresponds to the case where $\cH$ is an $\cN$-dimensional Hilbert space,
$O=Id$, $\eps=0$, $\alpha_k=\beta_j=1$,
and the operators $\pi_k$ are orthogonal projectors on an orthonormal basis of $\cH$. 
In this case, the theorem reads
$$
h_\pi(\cU\psi)+h_\pi(\psi) \geq -2\log c(\cU)\,,
$$
where $c(\cU)$ is the supremum of all matrix elements of $\cU$ in the orthonormal 
basis defined by $\pi$.
\end{rem}

\subsection{Applying the entropic uncertainty principle to the Laplacian eigenstates}\label{s:appl0}
In the whole article, we consider a certain 
subsequence of eigenstates $(\psi_{k_j})_{j\in\IN}$ of the Laplacian, such
that the corresponding sequence of Wigner functions $(W_{k_j})$ converges
to a certain semiclassical measure $\mu$ (see the discussion preceding 
Proposition~\ref{e:semiclass-measure}). The subsequence $(\psi_{k_j})$
will simply be denoted by $(\psi_\hbar)_{\hbar\to 0}$, using the slightly abusive 
notation $\psi_\hbar=\psi_{\hbar_{k_j}}$ for the eigenstate $\psi_{k_j}$. Each state $\psi_\hbar$ satisfies
\begin{equation}\label{e:eigenstate}
(-\hbar^2\lap-1)\psi_\hbar=0\,.
\end{equation}
In this section we define the data to input in Theorem~\ref{t:WEUP},
in order to obtain informations on the eigenstates $\psi_\hbar$ and the measure $\mu$. 
Only the Hilbert space is fixed, $\cH\defeq L^2(M)$. All other data depend on the semiclassical 
parameter $\hbar$:
the quantum partition $\pi$, the operator $O$, the positive real number $\eps$, the
weights $(\alpha_j)$, $(\beta_k)$ and the unitary operator $\mathcal{U}$.
 
\subsubsection{Smooth partition of unity}\label{s:smooth-partition}
As usual when computing the Kolmogorov--Sinai entropy, we start by decomposing the manifold $M$
into small cells of diameter $\dia>0$. More precisely, let $(\Omega_k)_{k=1,\ldots,K}$ be
an open cover of $M$ such that all $\Omega_k$ have diameters $\leq\dia$, and 
let $(P_k)_{k=1,\ldots,K}$ be a family of smooth real functions on $M$, 
with $\supp P_k\Subset \Omega_k$, such that 
\begin{equation}\label{e:partition}
\forall x\in M,\qquad \sum_{k=1}^K P_k^2(x)= 1\,.
\end{equation}
Most of the time, the notation $P_k$ will actually denote
the operator of multiplication by $P_k(x)$ on the Hilbert space $L^2(M)$: 
the above equation
shows that they form a quantum partition of unity \eqref{e:unity}, which we will call $\cP^{(0)}$.

\subsubsection{Refinement of the partition under the Schr\"odinger flow}\label{s:refine}
We denote the quantum propagator by $U^t=\exp(it\hbar\lap~/~2)$. With no loss of generality, we 
will assume that
the injectivity radius of $M$ is greater than $2$, and work with the
propagator at time unity, $U=U^1$. This propagator quantizes the
flow at time one, $g^1$. The $\hbar$-dependence of $U$ will be implicit in our notations.

As one does to compute the Kolmogorov--Sinai entropy of an invariant measure, 
we define a new quantum partition of unity by evolving and refining
the initial partition $\cP^{(0)}$ under
the quantum evolution. 
For each time $n\in\IN$ and any 
sequence of symbols $\bep=(\ep_0 \cdots \ep_n)$, 
$\ep_i\in [1,K]$ (we say that the sequence $\bep$ is of {\em length} $|\bep|=n$),
we define the operators
\begin{equation}\label{e:P_bep}
\begin{split}
P_{\bep}&=P_{\ep_n} U P_{\ep_{n-1}}\ldots U P_{\ep_0}\\
\tP_{\bep}&= U^{-n}P_{\bep}= P_{\ep_n}(n)P_{\ep_{n-1}}(n-1)\ldots P_{\ep_0}\,.
\end{split}
\end{equation}
Throughout the paper we will use the notation $A(t)=U^{-t} A U^t$ for the quantum evolution
of an operator $A$.
From \eqref{e:partition} and the unitarity of $U$, 
the family of operators $\set{P_{\bep}}_{|\bep|=n}$ obviously satisfies the resolution of identity
$\sum_{|\bep|=n} P_{\bep}\,P_{\bep}^*=Id_{L^2}$, and therefore forms a quantum partition
which we call $\cP^{(n)}$. The operators $\tP_{\bep}$ also have this property, 
they will be used in the proof of the
subadditivity, see sections~\ref{s:subadd} and \ref{s:subadditivity}.

\subsubsection{Energy localization}\label{s:energy}
In the semiclassically setting, the eigenstate $\psi_\hbar$ of \eqref{e:eigenstate} is associated with the
energy layer $\cE=\cE(1/2)=\set{\rho\in T^* M,\ H(\rho)=1/2}$. 
Starting from the cotangent bundle $T^* M$, we restrict ourselves 
to a compact phase space by introducing an energy cutoff (actually, several cutoffs) near $\cE$. 
To optimize our estimates, we will need this cutoff to depend on $\hbar$ in a sharp way.
For some fixed $\delta\in (0, 1)$, we consider a smooth function $\chi_\delta\in C^\infty(\IR;[0,1])$,
with $\chi_\delta(t)= 1$ for $|t|\leq \e^{-\delta/2}$ and $\chi_\delta(t)=0$ for $|t|\geq 1$. 
Then, we rescale that function to obtain a family of
$\hbar$-dependent cutoffs near $\cE$:
\begin{equation}\label{e:chi_n}
\forall \hbar\in(0,1),\ \forall n\in\IN,\ 
\forall \rho\in T^*M,\qquad\chi^{(n)}(\rho;\hbar)\defeq 
\chi_\delta\big(\e^{-n\delta}\,\hbar^{-1+\delta}(H(\rho)-1/2)\big)\,.
\end{equation}
The cutoff $\chi^{(0)}$ is localized in an energy interval of length $ 2\hbar^{1-\delta}$.
Choosing $0<C_\del<\delta^{-1}-1$, we will only consider indices 
$n\leq C_\del|\log\hbar|$, such that the
``widest'' cutoff will be supported in an interval of {\em microscopic} length 
$ 2\hbar^{1-(1+C_\del)\delta}<<1$. In our applications, we will take $\delta$ small 
enough, so that we may assume $C_{\delta}>4/\lambda_{\max}$.

These cutoffs can be quantized into pseudodifferential operators
$\Op(\chi^{(n)})=\Op_{\cE,\hbar}(\chi^{(n)})$ described in Section \ref{s:PDO} (the quantization
uses a nonstandard pseudodifferential calculus drawn from \cite{SZ99}). 
It is shown there (see Proposition~\ref{p:division})
that the eigenstate $\psi_\hbar$ satisfies
\bequ\label{e:local}
\norm{\big(\Op(\chi^{(0)})-1\big)\psi_\hbar}=\cO(\hbar^\infty)\,\norm{\psi_\hbar}.
\end{equation}
Here and below, the norm $\norm{\cdot}$ will either denote
the Hilbert norm on $\cH=L^2(M)$, or the corresponding operator norm.
\begin{rem}\label{r:cutoffs}
We will constantly use the fact that sharp energy localization is almost preserved by 
the operators $P_{\bep}$. 
Indeed, using results of section \ref{s:cutoff-props}, 
namely the first statement of Corollary~\ref{c:highlytechnical} and the norm estimate \eqref{e:norm},
we obtain that for $\hbar$ small enough and any $m,\,m'\leq C_\del |\log\hbar|/2$,
\bequ\label{e:en-conserv}
\forall |\bep|= m,\qquad
\norm{\Op(\chi^{(m'+m)})\,P^*_{\bep}\,\Op(\chi^{(m')}) - P^*_{\bep}\Op(\chi^{(m')})} 
=\cO(\hbar^\infty)\,.
\end{equation}
Here the implied constants are uniform with respect to $m,\,m'$ --- and of course
the same estimates hold if we replace
$P^*_{\bep}$ by $P_{\bep}$.
Similarly, from \S\ref{s:psiDO} one can easily show that
$$ 
\forall |\bep|=m,\ \qquad\norm{P_{\bep}\Op(\chi^{(m')})-P^\cut_{\bep}\Op(\chi^{(m')})}=\cO(\hbar^\infty)\,,
$$
where $P^\cut_{\ep_j}\defeq \Op_\hbar(P_{\ep_j}\,\cut)$,
$\cut$ is a smooth, compactly supported function in $T^*M$ which takes
the value $1$ in a neighbourhood of $\cE$ --- and 
$P^\cut_{\bep}=P^\cut_{\ep_m} U P^\cut_{\ep_{m-1}}\ldots U P^\cut_{\ep_0}$.
\end{rem}

In the whole paper, we will fix a small $\delta'>0$, and call
``Ehrenfest time'' the $\hbar$-dependent integer
\begin{equation}\label{e:Ehrenf}
n_E(\hbar)\defeq\big\lfloor\frac{(1-\delta')|\log \hbar|}{\lambda_{\max}}\big\rfloor\,.
\end{equation}
Unless indicated otherwise, the integer $n$ will always be taken equal to $n_E$.
For us, the significance of the Ehrenfest time is that it is the largest time interval
on which the (non--commutative) dynamical system formed by $(U^t)$ acting on
pseudodifferential operators can be treated as being, approximately, commutative (see
\eqref{e:ref-egorov}). 

Using the estimates \eqref{e:en-conserv} with $m=n$, $m'=0$ together with \eqref{e:local},
one easily checks the following
\begin{prop}
For any fixed $L>0$, there exists $\hbar_L$ such that, for any $\hbar\leq \hbar_L$, any $n\leq n_E(\hbar)$
and any sequence $\bep$ of length $n$, the Laplacian 
eigenstate $\psi_\hbar$ satisfies
\bequ\label{e:psi-cone}
\norm{\big(\Op(\chi^{(n)})-Id \big) P^*_{\bep}\psi_\hbar}\leq \hbar^L\norm{\psi_\hbar}\,. 
\eequ
\end{prop}
\subsubsection{Applying the entropic uncertainty principle}\label{s:mainapp}
We now precise some of 
the data we will use in the entropic uncertainty principle, Theorem~\ref{t:WEUP}:
\begin{itemize}
\item the quantum partition $\pi$ is given by the family of operators
$\set{P_{\bep},\ \abs{\bep}=n=n_E}$.
In the semiclassical limit, this partition 
has cardinality $\mathcal{N}=K^{n}\asymp \hbar^{-K_0}$ for some fixed $K_0>0$.
\item the operator $O$ is $O=\Op(\chi^{(n)})$, and $\eps=\hbar^L$, where $L$ will be 
chosen very large (see \S \ref{s:weights}).
\item the isometry will be $\cU=U^{n}=U^{n_E}$. 
\item the weights $\alpha_{\bep},\ \beta_{\bep}$ will be selected in \S\ref{s:weights}. They will be 
semiclassically tempered, meaning that there exists $K_1>0$ such that, for $\hbar$ small enough, 
all $\alpha_{\bep}$, $\beta_{\bep}$ are contained in the interval $[1,\hbar^{-K_1}]$.
\end{itemize}
As in Theorem~\ref{t:WEUP}, the entropy and pressures associated with a normalized 
state $\phi\in\cH$ are given by
\begin{align}\label{e:entropy}
h_n(\phi)&=h_{\cP^{(n)}}(\phi)=-\sum_{|\bep|=n}\norm{P^*_{\bep}\phi}^2\log\big(\norm{P^*_{\bep}\phi}^2\big),\\
p_{n, \alpha}(\phi)&= h_n(\phi)
-2\sum_{|\bep|=n}\norm{P^*_{\bep}\phi}^2\log \alpha_{\bep}.\label{e:Jpressures}
\end{align}
We may apply Theorem~\ref{t:WEUP} to any sequence of states satisfying
\eqref{e:psi-cone}, in particular the eigenstates $\psi_\hbar$.
\begin{cor}\label{c:WEUP} 
Define
\bequ\label{e:c_C}
c_{\Op(\chi^{(n)})}^{\alpha, \beta}(U^n)\defeq
\max_{|\bep|=|\bep'|=n}\Big(\alpha_{\bep}\, \beta_{\bep'}
\norm{P^*_{\bep'}\,U^n\,P_{\bep}\Op(\chi^{(n)})}\Big)\,,
\end{equation}
Then for $\hbar$ small enough and
for any normalized state $\phi$ satisfying \eqref{e:psi-cone},
$$
p_{n,\beta}(U^n\,\phi) + p_{n,\alpha}(\phi) \geq 
-2\log\Big(c_{\Op(\chi^{(n)})}^{\alpha, \beta}(U^n)+h^{L-K_0-2K_1}\Big)\,.
$$
\end{cor}
Most of Section~\ref{s:mainest} 
will be devoted to obtaining a good upper bound for the norms 
$\norm{P_{\bep'}^*\, U^n\, P_{\bep} \Op(\chi^{(n)})}$ involved in the above quantity.
The bound is given in
Theorem~\ref{t:main} below. Our choice for 
the weights $\alpha_{\bep}$, $\beta_{\bep}$ will then be guided by
these upper bounds.

\subsubsection{Unstable Jacobian for the geodesic flow}\label{s:unstable-Jac}
We need to recall a few definitions pertaining to
Anosov flows. For any $\lambda>0$, the geodesic flow $g^t$ is Anosov on
the energy layer $\cE(\lambda)=H^{-1}(\lambda)\subset T^*M$. 
This implies that for each 
$\rho\in \cE(\lambda)$, the tangent space $T_\rho\cE(\lambda)$ splits into
$$
T_\rho\cE(\lambda)=E^u(\rho)\oplus E^s(\rho) \oplus \IR\,X_H(\rho)\,
$$
where $E^u$ is the unstable subspace and $E^s$ the stable subspace.
The unstable Jacobian $J^u(\rho)$ at
the point $\rho$ is defined as the Jacobian of the map $g^{-1}$, restricted
to the unstable subspace at the point $g^1\rho$: $J^u(\rho)=\det\big(dg^{-1}_{|E^u(g^1\rho)}\big)$
(the unstable spaces at $\rho$ and $g^1\rho$ are equipped with the induced Riemannian metric). 
This Jacobian can be ``coarse-grained'' as follows in a neighbourhood 
$\cE^{\dia}\defeq\cE([1/2-\dia, 1/2+\dia])$
of $\cE$. 
For any pair $(\ep_0,\ep_1)\in [1,K]^2$, we define
\begin{equation}\label{e:coarse-Jac}
J^u_1(\ep_0,\ep_1)\defi \sup\set{J^u(\rho)\ :\ \rho\in T^*\Omega_{\ep_0}\cap
\cE^{\dia},\ g^1 \rho\in T^*\Omega_{\ep_1}}
\end{equation}
if the set on the right hand side is not empty, and $J^u_1(\ep_0,\ep_1)=e^{-\Lambda}$ otherwise,
where $\Lambda>0$ is a fixed large number.
For any sequence of symbols $\bep$ of length $n$, we define the coarse-grained Jacobian
\begin{equation}\label{e:multi}
J^u_n(\bep)\defi J^u_1(\ep_0,\ep_1)\ldots J^u_1(\ep_{n-1}, \ep_n)\,.
\end{equation}
Although $J^u$ and $J^u_1(\ep_0,\ep_1)$ are not necessarily everywhere smaller than unity,
there exists $C,\lambda_+,\ \lambda_->0$ such that, 
for any $n>0$, all the coarse-grained Jacobians of length $n$ satisfy
\begin{equation}\label{e:decay}
C^{-1}\,\e^{-n(d-1)\,\lambda_{+}}\leq J^u_n(\bep)\leq C\,\e^{-n(d-1)\,\lambda_{-}}\,.
\end{equation}
One can take $\lambda_+=\lambda_{\max}(1+\dia)$.
We can now give our central estimate, proven in Section~\ref{s:mainest}.
\begin{thm} \label{t:main} 
Given a partition $\cP^{(0)}$ and $\delta, \delta'>0$ small enough, there exists 
$\hbar_{\cP^{(0)},\del,\del'}$ such that, for any $\hbar\leq \hbar_{\cP^{(0)},\del,\del'}$, 
for any positive integer
$n\leq n_E(\hbar)$, and
any pair of sequences $\bep$, $\bep'$ of length $n$,
\begin{equation}\label{e:main}
\norm{P_{\bep'}^*\, U^n\, P_{\bep} \Op(\chi^{(n)})}
 \leq C\,\hbar^{-(d-1+c\delta)}\, J^u_n(\bep)^{1/2}\,J^u_n(\bep')\,.
\end{equation}
The constants $c$, $C$ only depend on the Riemannian manifold $(M,g)$. 
\end{thm}

\subsubsection{Choice of the weights}\label{s:weights}
There remains to choose
the weights $(\alpha_{\bep},\beta_{\bep})$ to use in Theorem~\ref{t:WEUP}.
Our choice is guided by the following idea: 
in the quantity \eqref{e:c_C}, the weights should balance the variations (with respect to $\bep,\bep'$)
in the norms, such as to make all terms in \eqref{e:c_C} of the same order. 
Using the upper bounds \eqref{e:main}, we end up with the following choice for all
$\bep$ of length $n$:
\bequ\label{e:weights1}
\alpha_{\bep}\defeq J^u_n(\bep)^{-1/2}\quad\text{and}\quad\beta_{\bep}\defeq J^u_n(\bep)^{-1}\,.
\end{equation}
All these quantities are defined using the Ehrenfest time $n=n_E(\hbar)$. From \eqref{e:decay},
there exists $K_1>0$ such that, for $\hbar$ small enough, all the weights are bounded by
\bequ\label{e:weights-bounds}
1\leq |\alpha_{\bep}|\leq \hbar^{-K_1},\qquad 1\leq |\beta_{\bep}|\leq \hbar^{-K_1}\,,
\end{equation}
as announced in \S\ref{s:mainapp}. The estimate \eqref{e:main} 
can then be rewritten as
$$
c_{\Op(\chi^{(n)})}^{\alpha, \beta}(U^n)\leq C\,\hbar^{-(d-1+c\delta)}\,.
$$
We now apply Corollary~\ref{c:WEUP} to the particular case of 
the eigenstates $\psi_\hbar$. We choose $L$ large enough such that $\hbar^{L-K_0-2K_1}$
is negligible in comparison with $\hbar^{-(d-1+c\delta)}$.
\begin{prop}\label{p:WEUP} 
Let $(\psi_\hbar)_{\hbar\to 0}$ be our sequence of eigenstates \eqref{e:eigenstate}.
Then, in the semiclassical limit, the pressures of $\psi_\hbar$ at time $n=n_E(\hbar)$
satisfy
\begin{equation}\label{e:ineg}
p_{n,\alpha}(\psi_\hbar)+p_{n,\beta}(\psi_\hbar)\geq 2(d-1+c\delta)\log \hbar+\cO(1)
\geq -2\frac{(d-1+c\delta)\lambda_{\max}}{(1-\delta')}\; n +\cO(1)\,. 
\end{equation}
\end{prop}

\subsubsection{Subadditivity until the Ehrenfest time}\label{s:subadd}
Before taking the limit $\hbar\to 0$, we prove
that a similar lower bound holds if we replace $n \asymp |\log\hbar|$ by some fixed $n_o$,
and $\cP^{(n)}$ by the corresponding partition $\cP^{(n_o)}$. This is due to
the following {\em subadditivity property}, which is the semiclassical analogue 
of the classical subadditivity of pressures for invariant measures.
\begin{prop} [Subadditivity]\label{p:subadd}  
Let $\del'>0$ and define the Ehrenfest time $n_E(\hbar)$ as in \eqref{e:Ehrenf}. 
There exists a real number $R>0$ independent of $\delta'$ and a function $R(\bullet,\bullet)$
on $\IN\times (0,1]$ such that
$$
\forall n_o\in\IN,\qquad\limsup_{\hbar\to 0}\abs{R(n_o, \hbar)}\leq R
$$
and with the following properties.
For any $\hbar\in(0,1]$, any
$n_o,m\in\IN$ with $n_o+m\leq n_E(\hbar)$,  
for $\psi_\hbar$ any normalized 
eigenstate satisfying \eqref{e:eigenstate}, we have
$$
p_{n_o+m,\alpha}(\psi_\hbar)\leq p_{n_o, \alpha}(\psi_\hbar)+p_{m-1, \alpha}(\psi_\hbar) 
+R(n_o, \hbar)\,.
$$
The same inequality holds for $p_{n_o+m,\beta}(\psi_\hbar)$.
 \end{prop}
The proof is given in \S\ref{s:subadditivity}. 
The time $n_o+m$ needs to be smaller than the Ehrenfest time because, 
in order to show the subadditivity, the various operators $P_{\eps_i}(i)$
composing $\tP_{\bep}$ have to approximately commute with each other. Indeed,
for $m\geq n_E(\hbar)$ the commutator  $[P_{\eps_m}(m),P_{\eps_0}]$ may have a norm of
order unity.

Equipped with this subadditivity, we may finish the proof of Theorem \ref{thethm}.
Let $n_o\in\IN$ be fixed and $n=n_E(\hbar)$. 
Using the Euclidean division $n=q(n_o+1)+r $, with $r\leq n_o$, Proposition~\ref{p:subadd} implies that
for $\hbar$ small enough,
$$
\frac{p_{n, \alpha}(\psi_\hbar)}{n}\leq \frac{p_{n_o, \alpha}(\psi_\hbar)}{n_o}
+\frac{p_{r, \alpha}(\psi_\hbar)}{n}+\frac{R(n_o,\hbar)}{n_o}\,.
$$
Using \eqref{e:ineg} and the fact that $p_{r, \alpha}(\psi_\hbar)+p_{r, \beta}(\psi_\hbar)$ 
stays uniformly bounded (by a quantity depending on $n_o$) when $\hbar\to 0$, we find
\begin{equation}\label{e:bientotfini}
\frac{p_{n_o, \alpha}(\psi_\hbar)}{n_o}+
\frac{p_{n_o, \beta}(\psi_\hbar)}{n_o}\geq -2\frac{(d-1+c\delta)\lambda_{\max}}{(1-\delta')} -2\frac{R(n_o, \hbar)}{n_o}+\cO_{n_o}(1/n)\,.
\end{equation}
We are now dealing with the partition $\cP^{(n_o)}$, $n_0$ being independent of $\hbar$. 

\subsubsection{End of the proof}\label{e:endofproof}
As explained at the beginning of \S\ref{s:appl0}, the subsequence $(\psi_{\hbar})_{\hbar\to 0}$ 
has the property that the Wigner measures $W_{\psi_{\hbar}}$
converge to the semiclassical measure $\mu$ on $\cE$.  
Because $\psi_{\hbar}$ are eigenstates 
of $U$, the norms appearing in the definition of $h_{n_o}(\psi_{\hbar})$ can be alternatively
written as
\begin{equation}\label{e:weights}
\norm{P_{\bep}^*\,\psi_{\hbar}}=\norm{\tP_{\bep}^*\,\psi_{\hbar}}=
\norm{P_{\ep_0}P_{\ep_1}(1)\cdots P_{\ep_{n_o}}(n_o)\,\psi_{\hbar}}\,.
\end{equation}
We may take the limit 
$\hbar\to 0$ (so that $n\to \infty$) in \eqref{e:bientotfini}.
For any sequence $\bep$ of length $n_o$, 
the above convergence property 
implies that each $\norm{\tP^*_{\bep}\,\psi_{\hbar}}^2$ converges to
$\mu(\{\bep\})$, where $\{\bep\}$ is
the function $P^2_{\ep_0}\,(P^2_{\ep_1}\circ g^1)\ldots (P^2_{\ep_{n_o}}\circ g^{n_o})$
on $T^*M$.
Thus $h_{n_o}(\psi_{\hbar})$ semiclassically converges to the classical entropy
$$
h_{n_o}(\mu)=h_{n_o}(\mu,(P^2_k))=-\sum_{|\bep|=n_o}\mu(\{\bep\})\log \mu(\{\bep\})\,.
$$
As a result,
the left hand side of \eqref{e:bientotfini} converges to
\begin{equation}\label{e:pressure0}
\frac2 n_o h_{n_o}(\mu)+\frac3{n_o}
\sum_{|\bep|=n_o}\mu(\{\bep\})\;\log J^u_{n_o}(\bep)\,.
\end{equation}
Since the semiclassical measure $\mu$ is $g^t$-invariant and $J^u_{n_o}$ has the multiplicative structure
\eqref{e:multi}, the second term in \eqref{e:pressure0} can be simplified:
$$
\sum_{|\bep|=n_o} \mu(\{\bep\})\,\log J^u_{n_o}(\bep)=
n_o\sum_{\ep_0, \ep_1}\mu(\{\ep_0 \ep_1\})\,\log J^u_1(\ep_0,\ep_1)\,.
$$
We have thus obtained the lower bound
\begin{equation}\label{e:classicentropy}
\frac{ h_{n_o}(\mu)}{n_o} 
\geq -\frac32 \sum_{\ep_0, \ep_1}\mu(\{\ep_0 \ep_1\})\,\log J^u_1(\ep_0,\ep_1)
-\frac{(d-1+c\delta)\lambda_{\max}}{(1-\delta')}-2\frac{R}{n_o}\,.
\end{equation}
$\delta$ and $ \delta'$ could be taken arbitrarily small, and at this stage they
can be let vanish.

The Kolmogorov--Sinai entropy of $\mu$ is by definition the limit
of the first term $\frac{ h_{n_o}(\mu)}{n_o} $ when $n_o$ goes to infinity, 
with the notable difference that
the smooth functions $P_k$ should be replaced by characteristic functions associated
with some partition of $M$, $M=\bigsqcup_k O_k$.
Thus, let us consider such a partition of diameter $\leq \dia/2$, 
such that $\mu$ does not charge the boundaries of the $O_k$.
By convolution we can smooth the characteristic functions $(\bbbone_{O_k})$
into a smooth partition of unity $(P_k)$ satisfying the conditions of 
section~\ref{s:smooth-partition} (in particular, each $P_k$ is supported on a set $\Omega_k$
of diameter $\leq \dia$). The lower bound \eqref{e:classicentropy} holds with respect to the smooth
partition $(P^2_k)$, and does not depend on the derivatives
of the $P_k$: as a result, the same bound carries over to the
characteristic functions $(\bbbone_{O_k})$.

We can finally let $n_o$ tend to $+\infty$, then let the diameter $\dia/2$ of the partition tend to $0$. 
From the definition \eqref{e:coarse-Jac} of the coarse-grained Jacobian,
the first term in the right hand side of \eqref{e:classicentropy} converges to the integral
$-\frac32\int_{\cE} \log J^u(\rho)d\mu(\rho)$ as $\dia\to 0$, which proves \eqref{e:main1}.

$\hfill\square$

\medskip

The next sections are devoted to proving, successively,
Theorem~\ref{t:main},
Proposition~\ref{p:subadd} and Theorem~\ref{t:WEUP}.

\section{The main estimate: proof of Theorem~\ref{t:main}}\label{s:mainest}

\subsection{Strategy of the proof}

We want to bound from above the norm of the operator $P^*_{\bep'}\,U^n\,P_{\bep}\Op(\chi^{(n)})$. 
This norm can be obtained
as follows:
$$
\norm{P_{\bep'}^*\, U^n\, P_{\bep}\Op(\chi^{(n)})  }=
\sup\set{|\la P_{\bep'}\Phi,\, U^n\, P_{\bep}\Op(\chi^{(n)}) \Psi \ra|\;:\;
\Psi,\,\Phi\in \cH,\  \norm{\Psi}=\norm{\Phi}=1}\,.
$$
Using 
Remark \ref{r:cutoffs}, we may insert $\Op(\chi^{(4n)})$ on the right of
$P_{\bep'}$, up to an error $\cO_{L^2}(\hbar^\infty)$.
In this section we will prove the following
\begin{prop}\label{p:main}
For $\hbar$ small enough, for any time $n\leq n_E(\hbar)$,
for any sequences $\bep$, $\bep'$ of length $n$ and any normalized states $\Psi$, $\Phi\in L^2(M)$,
one has
\begin{equation}\label{e:ineq}
|\la P_{\bep'}\,\Op(\chi^{(4n)})\,\Phi,\, U^n\, P_{\bep} \Op(\chi^{(n)})\Psi \ra|\leq
C\,\hbar^{-(d-1)-c\delta}\, J^u_n(\bep)^{1/2}J^u_n(\bep')\,.
\end{equation}
Here we have taken $\delta$ small enough such that $C_\del>4/\lambda_{\max}$, and 
$n_E(\hbar)$ is the Ehrenfest time \eqref{e:Ehrenf}.
The constants $C$ and $c=2+5/\lambda_{\max}$ only depend on the Riemannian manifold $M$.
\end{prop}
For such times $n$, the right hand side in the above bound is larger than 
$C\,\hbar^{\frac12(d-1)}$, in comparison to which
the errors $\cO(\hbar^{\infty})$ are negligible. 
Theorem~\ref{t:main} therefore follows from the above proposition.

The idea in Proposition \ref{p:main} is rather simple, although the technical implementation
becomes cumbersome. We first show that any state of the form $\Op(\chi^{(*)})\Psi$, as those appearing
on both sides of the scalar product \eqref{e:ineq}, can be decomposed as a superposition of
essentially
$\hbar^{-\frac{(d-1)}2}$
normalized Lagrangian states, supported on Lagrangian manifolds transverse to the stable leaves
of the flow: see \S \ref{s:decompo-elementary}. In fact the states we start with are
truncated $\delta$--functions (see \eqref{e:delta}), which are microlocally
supported on Lagrangians of the form $\cup_{t} g^t S_z^*M$, where $S_z^*M$ is
the unit sphere at the point $z$.
The action of the operator
$P_{\bep}=P_{\ep_n}UP_{\ep_{n-1}}U\cdots UP_{\ep_0}$ on such Lagrangian
states is intuitively simple to understand: each application of $U$ stretches the Lagrangian
in the unstable direction (the rate of elongation being described by the unstable Jacobian) whereas
each multiplication by $P_{\ep}$ cuts a small piece of Lagrangian. This iteration of stretching and
cutting accounts for the exponential decay, see \S \ref{s:expdec}.  

\subsection{Decomposition of $\Op(\chi)\Psi$ into elementary Lagrangian states}\label{s:decompo-elementary}

In Proposition~\ref{p:main}, we apply the cutoff $\Op(\chi^{(n)})$ on $\Psi$, respectively
$\Op(\chi^{(4n)})$ on $\Phi$. To avoid too cumbersome  notations, we treat both cases at the same time,
denoting both cutoffs by $\chi=\chi^{(*)}$, and their associated quantization by $\Op(\chi)$. 
The original
notations will be restored only when needed.
The energy cutoff
$\chi$ is supported on a microscopic energy interval, where it varies between 
$0$ and $1$. In spite of those fast variations
in the direction transverse to $\cE$, it can be
quantized such as to satisfy some sort of pseudodifferential
calculus.
As explained in Section~\ref{s:cutoff}, the quantization 
$\Op\defeq\Op_{\cE,\hbar}$ (see \eqref{e:Op-Sigma})
uses a {\it finite} family of Fourier Integral Operators $(U_{\kappa_j})$ associated with
local canonical maps $(\kappa_j)$. Each $\kappa_j$ sends an open bounded set 
$\cV_j\subset T^*M$ intersecting $\cE$ 
to $\cW_j\subset \R^{2d}$, endowed with coordinated $(y, \eta)=(y_1,\ldots, y_d, \eta_1, \ldots, \eta_d)$, 
such that $H\circ\kappa_j^{-1}=\eta_1+1/2$. In other words,
each $\kappa_j$ defines a set of local flow-box coordinates $(y,\eta)$, such that $y_1$ is 
the time variable and $\eta_1+1/2$ the energy, 
while $(y',\eta')\in \R^{2(d-1)}$ are symplectic coordinates
in a Poincar\'e section transverse to the flow.

\subsubsection{Integral representation of $U_{\kappa_j}$}\label{s:U_kappa}

Since $\kappa_j$ is defined only on $\cV_j$, one may
assume that $U_{\kappa_j}u=0$ for functions $u\in L^2(M\setminus \pi\cV_j)$ 
(here and below $\pi$ will represent either the projection from $T^*M$ to
$M$ along fibers, or from $\IR^{2d}_{y,\eta}$ to $\IR^d_y$).
If $\cV_j$ is small enough,
the action of $U_{\kappa_j}$ on a function $\Psi\in L^2(M)$
can be represented as follows:
$$
[U_{\kappa_j} \Psi](y)=(2\pi \hbar)^{-\frac{D+d}2}\int_{\pi\cV_j} 
\e^{\frac{i}{\hbar} S(z, y, \theta)}\, a_\hbar(z, y, \theta)\,\Psi(z)\,dz\, d\theta\,,
$$
where\\
-- $\theta$ takes values in an open neighbourhood $\Theta_j\subset\R^D$ for some integer $D\geq 0$,\\
-- the Lagrangian manifold generated by $S$ is the graph of $\kappa_j$,\\
-- $a_\hbar(z,y,\theta)$ has an asymptotic expansion $a_\hbar\sim \sum_{l\geq 0} \hbar^l\, a_l$,
and it is supported on $\pi\cV_j\times \pi\cW_j\times \Theta_j$.

When applying the definition \eqref{e:Op-Sigma} to the cutoff $\chi$, we notice that the product 
$\chi(1-\phi)\equiv 0$, so that $\Op(\chi)$ is given 
by the sum of operators $\Op(\chi)_j=U_{\kappa_j}^*\,\Op_\hbar^w(\chi_j)\,U_{\kappa_j}$, each of
them effectively acting from $L^2(\pi\cV_j)$ to itself.
We denote by $\delta_j(x;z)$ the kernel of the operator $\Op(\chi)_j$: it is given by the integral
\begin{multline}\label{e:delta}
\delta_j(x;z)=
(2\pi \hbar)^{-(D+2d)}\int e^{-\frac{i}{\hbar}S(x,y, \theta)} e^{\frac{i}{\hbar}\langle y-\ty,\eta\rangle}
\e^{\frac{i}{\hbar} S(z,\ty, \ttheta)}\times\\
\bar a_\hbar(x,y,\theta)\,a_\hbar(z, \ty, \ttheta)\,\varphi_j(y,\eta)\,\chi(\eta_1)\, 
dy\, d\theta\,d\ty\, d\ttheta\,d\eta\,.
\end{multline}
For any wavefunction $\Psi\in L^2(M)$, we have therefore
\begin{equation}\label{e:deltabasis}
[\Op(\chi)\Psi](x)=\sum_j \int_{\pi\cV_j}  \Psi(z) \delta_j(x;z)\,dz \,.
\end{equation}
We temporarily restore the dependence of $\delta_j(x;z)$ on the cutoffs, calling
$\delta_j^{(n)}(x;z)$ the kernel of the operator $\Op(\chi^{(n)})_j$.
In order to prove Proposition~\ref{p:main}, we will for each set $(j,j',z,z')$, 
obtain approximate expressions for the
wavefunctions $U^t\, P_{\bep}\,\delta^{(n)}_j(z)$, respectively 
$P_{\bep'}\,\delta^{(4n)}_{j'}(z')$, and use these expressions to bound from above 
their overlaps:
\begin{lem}\label{l:main}
Under the assumptions and notations of Proposition~\ref{p:main}, the upper bound 
$$
|\la U^{-n/2}P_{\bep'}\,\delta^{(4n)}_{j'}(z'),\, U^{n/2}\, P_{\bep}\, \delta^{(n)}_j(z) \ra|\leq
C\,\hbar^{-(d-1)-c\delta}\, J^u_n(\bep)^{1/2}J^u_n(\bep')\,.
$$
holds uniformly for any $j,j'$, any points
$z\in \pi\cV_j$, $z'\in\pi\cV_{j'}$ and any $n$-sequences $\bep,\bep'$.
\end{lem}
Using \eqref{e:deltabasis} and the Cauchy-Schwarz
inequality $\norm{\Psi}_{L^1}\leq \sqrt{\Vol(M)}\,\norm{\Psi}_{L^2}$, 
this Lemma yields Proposition~\ref{p:main}.

\bigskip

In the following sections we study the action of the operator
$P_{\bep}$ on the state $\delta(z)=\delta_j^{(*)}(z)$ of the form \eqref{e:delta}.
By induction on $n$, we propose an Ansatz
for that state, valid for times $n=|\bep|$ of the order of $|\log\hbar|$. 
Apart from the sharp energy cutoff, this Ansatz is similar to the one described 
in \cite{An}.

\subsection{WKB Ansatz for the first step}

The first step of the evolution consists in applying the operator $U P_{\ep_0}$ to 
$\delta(z)$. For this aim, we will use the decomposition \eqref{e:delta} into WKB states
of the form $a(x)\e^{iS(x)/\hbar}$, and
evolve such states individually through the above operator. 
We briefly review how the propagator $U^t=\e^{it\hbar\lap/2}$ 
evolves such states.

\subsubsection{Evolution of a WKB state}\label{s:evolWKB}

Consider an initial state $u(0)$ of 
the form $u(0,x)=a_\hbar(0,x)\,\e^{\frac{i}{\hbar}S(0,x)}$, where $S(0,\bullet)$, $a_\hbar(0,\bullet)$
are smooth functions defined on a subset $\Omega\subset M$,
and $a_\hbar$ expands as $a_\hbar\sim \sum_k \hbar^k\,a_k$. 
This represents a WKB (or Lagrangian) state, 
supported on the Lagrangian manifold $\cL(0)=\set{(x,d_x S(0,x)),\ x\in\Omega}$.

Then, for any integer $N$, 
the state $\tilde u(t)\defi U^t u(0)$ can be approximated, to order $\hbar^N$, by a
WKB state $u(t)$ of the following form:
\begin{equation}\label{e:as-dev}
u(t,x)=\e^{\frac{iS(t, x)}{\hbar}}\,a_\hbar(t, x)=
\e^{\frac{iS(t, x)}{\hbar}} \sum_{k=0}^{N-1}\hbar^k a_k(t,x)\,.
\end{equation}
Since we want $u(t)$ to solve
$\frac{\partial u}{\partial t}=i\hbar\frac{\lap_x u}2$
up to a remainder of order $\hbar^N$,
the functions $S$ and $a_k$ must satisfy the following
partial differential equations:
\begin{equation} \label{e:mainBKW}
\begin{cases}\frac{\partial S}{\partial t}+H(x, d_x S)=0 \quad \mbox{ (Hamilton-Jacobi equation)}\\
\\
\frac{\partial a_0}{\partial t}=-\langle d_x a_0,d_x S(t, x)\rangle-a_0\frac{\lap_x S(t, x) }2\quad
\mbox{($0$-th transport equation)}\,,
\\
\\
\frac{\partial a_k}{\partial t}=\frac{i\lap a_{k-1}}2-\langle d_x a_k, d_x S\rangle-a_k\frac{\lap S }2
\quad\mbox{($k$-th transport equation)}\,.
\end{cases}
\end{equation}
Assume that, on a certain time interval --- say $s\in [0,1]$ --- the above equations
have a well defined smooth solution $S(s,x)$, meaning that the transported Lagrangian manifold $\cL(s)$ is
of the form $\cL(s)=\set{(x,d_x S(s,x))}$, where $S(s)$ is a smooth function on the open
set $\pi \cL(s)$. Under these conditions, we denote as follows the 
induced flow on $M$:
\begin{equation}
g_{S(s)}^{t}: x\in \pi \cL(s) \mapsto \pi g^{t}\big(x, d_x S(s, x)\big)\in \pi\cL(s+t)\,,
\end{equation}
This flow satisfies the property
$g_{S(s+\tau)}^{t}\circ g_{S(s)}^{\tau}=g_{S(s)}^{t+\tau}$.
We then introduce the following (unitary) operator $T_{S(s)}^{t}$,
which transports functions on $\pi\cL(s)$ into functions on $\cL(s+t)$:
\begin{equation}\label{e:transport}
T_{S(s)}^{t}(a)(x)= a\circ g_{S(s+t)}^{-t}(x)\; \big(J_{S(s+t)}^{-t}(x)\big)^{1/2}\,.
\end{equation}
Here $J_{S(s)}^{t}(x)$ is the Jacobian
of the map $ g_{S(s)}^{t}$ at the point $x$ (measured with respect to the Riemannian volume on $M$). 
It is given by
\begin{equation}\label{e:Jacobian}
J_{S(s)}^{t}(x)=
\exp\Big\{\int_{0}^{t} \lap S\big(s+\tau, g_{S(s)}^{\tau}(x))\big)\,d\tau\Big\}\,.
\end{equation}
The $0$-th transport equation in \eqref{e:mainBKW} is explicitly solved by
\begin{equation}\label{e:transport-class}
a_0(t)=T_{S(0)}^{t}\,a_0\,,\quad t\in[0,1]\,,
\end{equation}
and the higher-order terms $k\geq 1$ are given by
\begin{equation}\label{e:solution-transport}
a_k(t)=T_{S(0)}^{t}a_k+ \int_0^t T_{S(s)}^{t-s}\left(\frac{i\lap a_{k-1}}2(s)\right)ds\,.
\end{equation}
The function $u(t,x)$ defined by \eqref{e:as-dev} satisfies the approximate equation
$$
\frac{\partial u}{\partial t}=i\hbar\frac{\lap u}{2} -i\hbar^N \, \e^{\frac{i}{\hbar}S(t, x)}
\frac{\lap a_{N-1}}2(t,x)\,.
$$
From Duhamel's principle and the unitarity of $U^t$, 
the difference between $u(t)$ and the exact solution 
$\tilde u(t)$ is bounded, for $t\in [0,1]$, by
\begin{equation}\label{e:Duham}
\norm{u(t)-\tilde u(t)}_{L^2}\leq \frac{\hbar^N}2 \int_0^t \norm{\lap a_{N-1}(s)}_{L^2}\,ds
\leq C\,t\, \hbar^{N}\big(\sum_{k=0}^{N-1}\norm{a_k(0)}_{C^{2(N-k)}}\big)\,.
\end{equation}
The constant $C$ is controlled by the volumes of the sets $\pi \cL(s)$ ($0\leq s\leq t\leq 1$),
and by a certain number of derivatives of the flow $g_{S(s+t)}^{-t}$ ($0\leq s+t\leq 1$).

\subsubsection{The Ansatz for time $n=1$}\label{s:n=1}

We now apply the above analysis to study the evolution of the state $\delta(z)$ given
by the integral \eqref{e:delta}. Until section~\ref{s:decompo}, 
we will consider a single point $z$. 
Selecting in \eqref{e:delta} a pair $(y,\theta)$ in the support of $a_\hbar$, 
we consider the state
$$
u(0,x)=\e^{-\frac{i}{\hbar}S(x,y,\theta)}\,
\bar a^{\ep_0}_\hbar(x, y,\theta),\quad \text{where}\quad 
a^{\ep_0}_\hbar(x, y,\theta)\defeq P_{\ep_0}(x)\,a_\hbar(x, y,\theta)\,.
$$ 
Notice that this state is compactly supported in $\Omega_{\ep_0}$. We will choose a (large)
integer $N>0$ (see the condition at the very end of \S\ref{s:overlaps}), 
truncate the $\hbar$-expansion of 
$\bar a^{\ep_0}_\hbar$ to the order
$\tN=N+D+2d$, and 
apply to that state the WKB evolution described in the previous section,
up to order $\tN$ and for times $0\leq t\leq 1$. 
We then obtain an approximate state 
$\bar a^{\ep_0}_\hbar(t,x;y,\theta)\,\e^{-\frac{i}{\hbar}S(t,x;y,\theta)}$. By the superposition
principle, we get the following representation for the state $U^t P_{\ep_0}\,\delta(z)$:
\begin{equation}\label{e:decompo-eta1}
[U^t P_{\ep_0}\,\delta(z)](x)=(2\pi\hbar)^{-\frac{d+1}2}
\int v(t,x;z,\eta_1) \,\chi(\eta_1)\,d\eta_1+\cO_{L^2}(\hbar^N)\,,
\end{equation}
where for each energy parameter $\eta_1$ we took
\begin{multline}\label{e:firstansatz}
v(t,x;z,\eta_1)=
(2\pi \hbar)^{-D-\frac{3d-1}2}
\int \e^{-\frac{i}{\hbar}S(t,x;y,\theta)} \,
\e^{\frac{i}{\hbar}\langle y-\ty,\eta\rangle}\,
\e^{\frac{i}{\hbar} S(z,\ty, \ttheta)}\times\\
\bar a^{\ep_0}_\hbar(t,x;y,\theta)\,a_\hbar(z; \ty, \ttheta)\,\varphi_j(y,\eta)\, 
dy\, d\theta\,d\ty\, d\ttheta\,d\eta'\,
\end{multline}
(here $\eta'=(\eta_2,\ldots, \eta_d)$). 
The reason why we integrate over all variables but $\eta_1$
lies in the sharp cutoff $\chi$: due to this cutoff 
one cannot apply a stationary phase analysis in the variable $\eta_1$. 

At time $t=0$, the state $v(0,\bullet;z,\eta_1)$ is a WKB state, supported on the 
Lagrangian manifold
$$
\cL^0_{\eta_1}(0)=\set{\rho\in\cE(1/2+\eta_1),\ \pi(\rho)\subset\Omega_{\ep_0},\ \exists \tau\in
[-1,1],\ g^{-\tau}\rho\in T^*_z M }\,.
$$
This Lagrangian is obtained by propagating the sphere 
$S^*_{z,\eta_1}M=\set{\rho=(z,\xi),\ |\xi|_z=\sqrt{1+2\eta_1}}$
on the interval $\tau\in[-1,1]$, and keeping only the points situated above $\Omega_{\ep_0}$.
The projection of $\cL^0_{\eta_1}(0)$ on $M$ is not one-to-one: 
the point $z$ has infinitely many preimages, while other points $x\in \Omega_{\ep_0}$ have
in general two preimages $(x,\xi_x)$ and $(x,-\xi_x)$.

\begin{figure}
\begin{center}
\includegraphics[width=13cm]{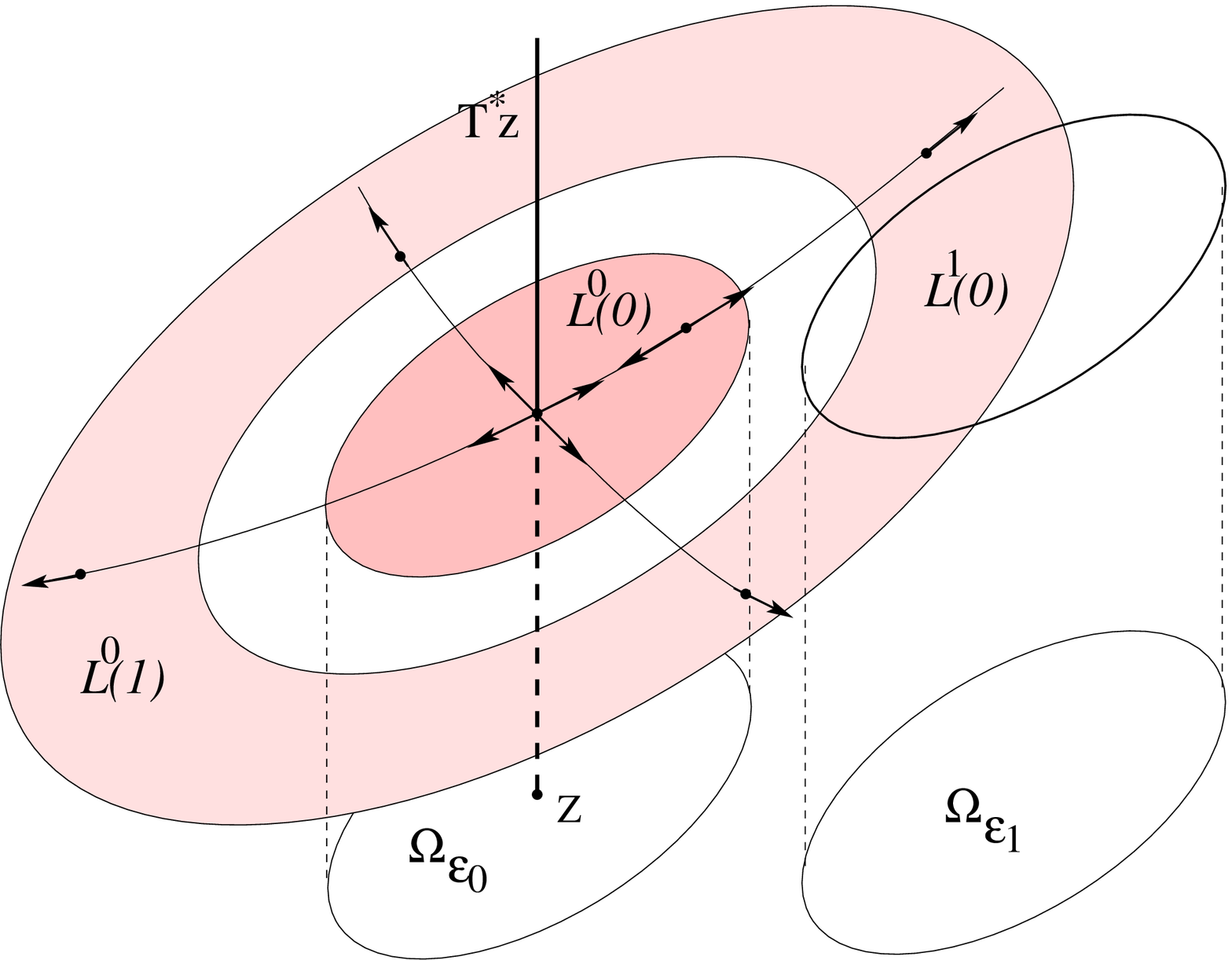}
\caption{\label{f:Lagrangian} Sketch of the Lagrangian manifold $\cL^0_{\eta_1}(0)$ situated above
$\Omega_{\eps_0}$ and centered at $z$ (center ellipse, dark pink), its image
$\cL^0_{\eta_1}(1)$ through the flow (external annulus, light pink) 
and the intersection $\cL^1_{\eta_1}(0)$ of
the latter with $T^*\Omega_{\eps_1}$. The thick arrows show the possible momenta at points 
$x\in M$ (black dots)}
\end{center}
\end{figure}
Let us assume that the diameter of the partition $\dia$ is less than $1/6$.
For $0<t\leq 1$, $v(t;z,\eta_1)$ is a WKB state supported on
$\cL^0_{\eta_1}(t)=g^t\,\cL^0_{\eta_1}(0)$. If the time is small, $\cL^0_{\eta_1}(t)$
still intersects $T^*_zM$. On the other hand,
all points in $\cE(1/2+\eta_1)$ move at a speed
$\sqrt{1+2\eta_1}\in [1-2\dia,1+\dia]$, so for times $t\in [3\dia, 1]$ any point
$x\in\pi\cL^0_{\eta_1}(t)$ is at distance greater than $ \dia$ from $\Omega_{\ep_0}$. Since the
injectivity radius of $M$ is $\geq 2$, such a point $x$ is connected to $z$ by a single
short geodesic arc. Furthermore, since $x$ is outside $\Omega_{\ep_0}$,
there is no ambiguity about the sign of the
momentum at $x$: in conclusion, there is a unique $\rho\in\cL^0_{\eta_1}(t)$ sitting above $x$ 
(Fig.~\ref{f:Lagrangian}).

For times $3\dia\leq t\leq 1$, the Lagrangian $\cL^0_{\eta_1}(t)$ 
can therefore be generated by a single function $S^0(t,\bullet;z,\eta_1)$.
Equivalently, for any $x$ in the support of $v(t,\bullet;z,\eta_1)$, 
the integral \eqref{e:firstansatz} is stationary at a unique set of parameters 
$\bullet_c=(y_c, \theta_c, \ty_c, \ttheta_c, \eta'_c)$, and leads to an expansion 
(up to order $\hbar^N$):
\begin{equation}\label{e:v(t,eta1)}
v(t;z,\eta_1)=v^0(t;z,\eta_1)+\cO(\hbar^N)\,,\quad
\text{where}\quad
v^0(t,x;z,\eta_1)=\e^{\frac{i}{\hbar}S^0(t, x; z, \eta_1)}\,b_\hbar^0(t, x; z, \eta_1)\,.
\end{equation}
The above discussion shows that 
$\cL^0_{\eta_1}\defeq\cup_{3\dia\leq t\leq 1}\cL^0_{\eta_1}(t)$ is a Lagrangian 
manifold which can be generated
by a single function $S^0(\bullet;z,\eta_1)$ defined on 
$\pi\cL^0_{\eta_1}$.
The phase functions $S^0(t,\bullet;z,\eta_1)$ 
obtained through the stationary phase
analysis depend very simply on time:
$$
S^0(t,x;z,\eta_1)=S^0(x;z,\eta_1)-(1/2+\eta_1)\,t\,.
$$
The symbol $b^0_\hbar$ is given by a truncated expansion
$b_\hbar^0=\sum_{k=0}^{N-1}\hbar^k\, b^0_k$. The principal symbol reads
$$
b^0_0(t, x; z,\eta_1)=\bar a^{\ep_0}_0(t, x; y_c, \theta_c)\, a_0(z; \ty_c, \ttheta_c)\,,
$$
while higher order terms $b^0_k$ are given by
linear combination of derivatives of $\bar a^{\ep_0}_\hbar(t, x;\bullet)\,a_\hbar(z; \bullet)$ 
at the critical point $\bullet=\bullet_c$.
Since $\bar a^{\ep_0}_\hbar(0, \bullet;y_c, \theta_c)$ is supported inside $\Omega_{\ep_0}$,
the transport equation \eqref{e:solution-transport} shows that 
$b^0_\hbar(t,\bullet;z,\eta_1)$ is supported inside
$\pi\cL^0_{\eta_1}(t)$.

If we take in particular $t=1$, the state 
\begin{equation}
v^0(1;z)=(2\pi \hbar)^{-\frac{d+1}2}\int v^0(1;z,\eta_1)\,\chi(\eta_1)\,d\eta_1
\end{equation}
provides an approximate expression for $UP_{\ep_0}\delta(z)$, up to a remainder
$\cO_{L^2}(|\supp\chi|\,\hbar^{N-\frac{d+1}2})$.

\subsection{Iteration of the WKB Ans\"atze}\label{p:ansatz-n}

In this section we will obtain an approximate Ansatz
for $P_{\ep_{n}}\ldots U P_{\ep_1}U P_{\ep_0} \delta(z)$. 
Above we have already performed the first step, obtaining an approximation $v^0(1;z)$ of 
$UP_{\ep_0}\delta(z)$, which was decomposed into fixed-energy WKB states
$v^0(1;z,\eta_1)$.
The next steps will be performed by evolving
each component $v^0(1;z,\eta_1)$ individually, and integrating over $\eta_1$ only at the end.
Until Lemma~\ref{l:transfer} we will fix the variables
$(z,\eta_1)$, and omit them in our notations when no confusion may arise.

Applying the multiplication operator $P_{\ep_1}$ to the state $v^0(1)=v^0(1;z,\eta_1)$, 
we obtain another WKB state which we denote as follows:
$$
v^1(0,x)=b^1_\hbar(0,x)\,\e^{\frac{i}{\hbar}S^1(0,x)}\,,\qquad\text{with}\qquad
\begin{cases} S^1(0,x)=S^0(1,x;z,\eta_1)\,,\\ 
b^1_\hbar(0,x)=P_{\ep_1}(x)\,b^0_\hbar(1,x;z,\eta_1)\,.
\end{cases}
$$ 
This state is associated with the manifold
$$
\cL^1(0)= \cL^0_{\eta_1}(1)\cap T^*\Omega_{\ep_1}\,.
$$
If this intersection is empty, then $v^1(0)=0$, which means that
$P_{\ep_1} U\, v(0;z,\eta_1)=\cO(\hbar^{N})$.
In the opposite case, we can evolve $v^1(0)$ 
following the procedure described in \S\ref{s:evolWKB}.
For $t\in [0,1]$, and up to an error $\cO_{L^2}(\hbar^N)$,
the evolved state $U^t v^1(0)$ is given by the WKB Ansatz
$$
v^1(t,x)=b^1_\hbar(t,x)\,\e^{\frac{i}{\hbar}S^1(t,x)}\,,\qquad b^1_\hbar(t)=\sum_{k=0}^{N-1}b^1_k(t)\,.
$$
The state $v^1(t)$ is associated with the Lagrangian $\cL^1(t)=g^t\,\cL^1(0)$, 
and the function $b^1_\hbar(t)$ is supported inside $\pi\cL^1(t)$.
The Lagrangian $\cL^1\defeq\cup_{0\leq t\leq 1}\cL^1(t)$ is
generated by the function $S^1(0,x)$, and
for any $t\in [0,1]$ we have $S^1(t,x)=S^1(0,x)-(1/2+\eta_1)\,t$.

\subsubsection{Evolved Lagrangians}

We can iterate this procedure,
obtaining a sequence of approximations 
\begin{equation}\label{e:v^j}
v^j(t)=U^t\,P_{\ep_j}v^{j-1}(1)+\cO(\hbar^N)\,,\quad\text{where}\quad
v^j(t,x)=b^j_\hbar(t,x)\,\e^{\frac{i}{\hbar}S^j(t,x)}\,.
\end{equation}
To show that this procedure is consistent, we must check that the Lagrangian manifold $\cL^j(t)$
supporting $v^j(t)$ does not develop caustics through the evolution ($t\in [0,1]$), and that 
it can be generated by a single function $S^j(t)$. We now show that these properties hold, due
to the assumptions on the classical flow.

The manifolds $\cL^j(t)$ are obtained by the following procedure. Knowing $\cL^{j-1}(1)$, which
is generated by the phase function $S^{j-1}(1)$, we take
for $\cL^{j}(0)$ the intersection
$$
\cL^{j}(0)=\cL^{j-1}(1)\cap T^*\Omega_{\ep_j}\,.
$$ 
If this set is empty, we then stop the construction. Otherwise, this Lagrangian
is evolved into $\cL^{j}(t)=g^t\cL^{j}(0)$ for $t\in [0,1]$. 
Notice that the Lagrangian $\cL^j(t)$ corresponds to the
evolution at time $j+t$ of a piece of $\cL^0(0)$; the latter is contained in the
union $\cup_{|\tau|\leq 1}g^{\tau} S^*_{z,\eta_1} M$, where $S^*_{z,\eta_1} M$ is the sphere of 
energy $1/2+\eta_1$ above $z$. 
If the geodesic flow is Anosov, the 
geodesic flow has no conjugate points \cite{Kl74}. 
This implies that $g^t\cL^0(0)$ will
not develop caustics: in other words, the phase functions $S^j(t)$ will never become singular.

On the other hand, when $j\to\infty$ the Lagrangian $g^{j+t}\cL^0(0)$ will 
spread out over $M$, and cover all points $x\in M$ many times, so that many phase functions are
needed to describe the different sheets (see \S\ref{s:disjoint}). However, 
the small piece $\cL^j(t)\subset g^{j+t}\cL^0(0)$ is generated by only one of them. 
Indeed, because the injectivity radius is $\geq 2$,
any point $x\in \Omega_{\ep_j}$ can be connected to another point $x'\in M$ by
at most one geodesic of length $\sqrt{1+2\eta_1}\leq 1+\dia$.
This ensures that, for any $j\geq 1$, the manifold $\cL^j=\cup_{t\in[0,1]}\cL^j(t)$ is generated by a
single function $S^j(0)$ defined on $\pi\cL^j$, or equivalently by 
$S^j(t)=S^j(0)-(1/2+\eta_1)\,t$ (this $S^j$ is a stationary solution of the Hamilton--Jacobi
equation, and we will often omit to show its time dependence in the notations).

Finally, since the flow on $\cE(1/2+\eta_1)$ is Anosov, 
the sphere bundle $\set{S^*_{z,\eta_1}M, z\in M}$ is uniformly transverse to the strong
stable foliation \cite{Kl74}.
As a result, under the flow a piece of sphere
becomes exponentially close to an unstable leaf when $t\to +\infty$. 
The Lagrangians $\cL^j$ thus become exponentially close to the weak unstable
foliation as $j\to\infty$. This transversality argument is crucial in our choice to
decompose the state $\Psi$ into components $\delta_j(z)$.

\subsubsection{Exponential decay of the symbols\label{s:expdec}}

We now analyze the behaviour of the symbols $b^j_\hbar(t,x)$ appearing in \eqref{e:v^j},
when $j\to\infty$. These symbols are constructed iteratively:
starting from the function $b^{j-1}_\hbar(1)=\sum_{k=0}^{N-1}b^{j-1}_k(1)$ supported inside
$\pi\cL^{j-1}(1)$, we define
\begin{equation}\label{e:recur}
b^j_\hbar(0,x)=P_{\ep_j}(x)\,b^{j-1}_\hbar(1,x)\,,\qquad x\in\pi\cL^{j}(0)\,.
\end{equation}
The WKB procedure of \S\ref{s:evolWKB} shows that for any $t\in [0,1]$,
\begin{equation}\label{e:R^j}
U^t\,v^{j}(0)=v^j(t)+R^j_N(t)\,,
\end{equation}
where the transported symbol
$b^{j-1}_\hbar(t)=\sum_{k=0}^{N-1}\hbar^k\,b^{j-1}_k(t)$ is supported inside $\pi\cL^{j}(t)$. 
The remainder satisfies
\begin{equation}\label{e:R^j-deriv}
\norm{R^j_N(t)}_{L^2}\leq C\,t\,\hbar^N\big(\sum_{k=0}^{N-1}\Vert b^j_k(0)\Vert_{C^{2(N-k)}}\big)\,.
\end{equation}
To control this remainder when $j\to\infty$, we need to bound from above the derivatives of 
$b^j_\hbar$.
Lemma~\ref{l:transfer} below shows that all terms $b^j_k(t)$ and their derivatives 
decay exponentially when $j\to\infty$,
due to the Jacobian appearing in \eqref{e:transport}.

To understand the reasons of the decay, we first consider
the principal symbols $b^j_0(1,x)$. They satisfy the following recurrence:
\begin{equation}\label{e:principal-evol}
b^j_0(1,x)=T_{S^j}^{1}(P_{\ep_j}\times b^{j-1}_0(1))(x)=
P_{\ep_j}( g_{S^j}^{-1}(x))\, b^{j-1}_0(1, g_{S^j}^{-1}(x))\, \sqrt{J_{S^j}^{-1}(x)}\,.
\end{equation}
Iterating this expression, and using the fact that $0\leq P_{\ep_j}\leq 1$,
we get at time $n$ and for any $x\in\pi\cL^n(0)$:
\begin{equation}\label{e:prod-Jac}
|b^n_0(0,x)|\leq |b^0_0(1, g_{S^{n}}^{-n+1}(x))|\times
\Big(J_{S^{n-1}}^{-1}(x)\,J_{S^{n-2}}^{-1}( g_{S^{n}}^{-1}(x))\cdots \,
J_{S^1}^{-1}( g_{S^{n}}^{-n+2}(x))\Big)^{1/2}\,.
\end{equation}
Since the Lagrangians $\cL^j$ converge exponentially fast to the weak unstable foliation, 
the associated Jacobians satisfy for some $C>0$:
$$
\forall j\geq 2,\ \forall \rho=(x,\xi)\in\cL^j(0),\qquad
\left|\frac{J_{S^j}^{-1}(x)}{J^{-1}_{S^u(\rho)}(x)}-1\right|\leq C\,\e^{-j/C}\,.
$$
Here $S^u(\rho)$ generates the local weak unstable manifold
at the point $\rho$ (which is a Lagrangian submanifold of $\cE(1/2+\eta_1)$). 
The product of Jacobians in \eqref{e:prod-Jac} therefore satisfies,
uniformly with respect to $n$ and $\rho\in \cL^n(0)$:
$$
\prod_{j=1}^{n-1} J_{S^{n-j}}^{-1}( g_{S^{n}}^{-j+1}(x))
=\e^{\cO(1)}\, \prod_{j=1}^{n-1} J^{-1}_{S^u(g^{-j+1}\rho)}( g_{S^{n}}^{-j+1}(x))
=\e^{\cO(1)}\, J^{1-n}_{S^u(\rho)}(x)\,,\quad n\to\infty\,.
$$
The Jacobian $J^{-1}_{S^u(\rho)}$ measures
the contraction of $g^{-1}$ along $E^u(\rho)$: so does the Jacobian $J^u(\rho)$ defined in
\S\ref{s:unstable-Jac}, but with respect to different coordinates. When
iterating the contraction $n$ times, the ratio of these Jacobians remains bounded:
$$
J^{1-n}_{S^u(\rho)}(x) = \e^{\cO(1)}\,\prod_{j=1}^{n-1} J^u(g^{-j+1}\rho)\,,\quad n\to\infty\,.
$$
We finally express the upper bound in terms of the ``coarse-grained'' Jacobians 
(\ref{e:coarse-Jac},\ref{e:multi}). Since $\rho\in \cL^n(0)\subset T^*\Omega_{\ep_n}$ and 
$g^{-j}\rho \in T^*\Omega_{\ep_{n-j}}$ for all $j=1,\ldots, n-1 $, 
we obtain the following estimate on the principal symbol $b^n_0(0)$:
\begin{equation}
\forall n\geq 1\,\quad 
\norm{b^n_0(0)}_{L^\infty}\leq 
C\,\norm{b^0_0(1;z,\eta_1)}_{L^\infty}\,J^u_{n-1}(\ep_1 \cdots \ep_{n})^{1/2} \,.
\end{equation}
The constant $C$ only depends on the Riemannian manifold $M$.
Finally, by construction the symbol $b^0_0(1;z,\eta_1)$ is bounded uniformly with respect
to the variables $(z,\eta_1)$ (assuming $|\eta_1|<\dia$).

The following lemma shows that the above bound extends to the full symbol $b_\hbar^n(0,x)$
and its derivatives (which are supported on $\pi\cL^n(0)$).
\begin{lem}\label{l:transfer} 
Take any index $0\leq k\leq N$ and $m\leq 2(N-k)$. Then there exists a constant
$C(k, m)$ such that 
$$
\forall n\geq 1,\ \ \forall x\in \pi\cL^n(0),\qquad
|d^m b^n_k(0, x)|\leq C(k, m)\, n^{m+3k}\,
J^u_n(\ep_0\cdots\ep_{n})^{1/2}\,.
$$
This bound is uniform with respect to the parameters $(z,\eta_1)$. 
For $(k,m)\neq (0,0)$, the constant $C(k,m)$
depends on the partition $\cP^{(0)}$, while $C(0,0)$ does not.
\end{lem}
Before giving the proof of this lemma, we draw somes consequences.
Taking into account the fact that the remainders $R^j_N(1)$ are dominated by the derivatives
of the $b^j_k$ (see \eqref{e:R^j-deriv}), the above statement translates into
$$
\forall j\geq 1,\qquad
\norm{R^j_N(1)}_{L^2}\leq C(N)\,j^{3N}\,J^u_j(\ep_0 \cdots \ep_{j})^{1/2}\,\hbar^N\,.
$$
A crucial fact for us is that the above bound also holds for the propagated remainder 
$P_{\ep_n}U\cdots U P_{\ep_{j+1}}R^j_N(1)$, due to the fact that the operators $P_{\ep_j}U$ are contracting.
As a result, the total error at time $n$ is bounded from above by the sum of the 
errors $\norm{R^j_N(1)}_{L^2}$.
We obtain the following estimate for any $n>0$:
\begin{equation}\label{e:WKB-n}
\norm{P_{\ep_n}UP_{\ep_{n-1}}\cdots P_{\ep_1} U\, v(0;z,\eta_1) - 
v^n(0;z,\eta_1)}_{L^2}\leq C(N)\,\hbar^N \, \sum_{j=0}^n j^{3N}\,J^u_j(\ep_0 \cdots \ep_{j})^{1/2}\,.
\end{equation}
From the fact that the Jacobians $J^u_j$ decay exponentially with $j$, the last term is bounded
by $C(N) \hbar^N$.
This bound is uniform with respect
to the data $(z,\eta_1)$. 

By the superposition principle, we obtain the following
\begin{cor}\label{c:approx-n}
For small enough $\hbar>0$, any point $z\in \pi\cV_j$,
and any sequence $\bep$ of arbitrary length $n\geq 0$, we have
$$
P_{\bep}\,\delta_j(z)= 
(2\pi\hbar)^{-\frac{d+1}2}\int v^n(0;z,\eta_1) \,\chi(\eta_1)\,d\eta_1+
\cO_{L^2}(|\supp\chi|\,\hbar^{N-\frac{d+1}2})\,.
$$
Here we may take $\chi=\chi^{(n')}$  with 
an arbitrary $0\leq n'\leq C_\del|\log\hbar|$ (see \eqref{e:chi_n} and the following discussion).
\end{cor}

\subsubsection*{Proof of Lemma \ref{l:transfer}}
The transport equation (\ref{e:transport-class},\ref{e:solution-transport}) linking $b^j$ to $b^{j-1}$,
\begin{equation}\label{e:rec1}
\begin{split}
b^j_k(t)&=T_{S^j}^{t}\,b^j_k(0)+ (1-\delta_{k,0})\,\int_0^t T_{S^j}^{t-s}\,
\Big(\frac{i\lap b^j_{k-1}(s)}2\Big)ds\,,\qquad k=0,\ldots,N-1\,,\\
b^j_k(0)&=P_{\ep_j}\times b^{j-1}_k(1)\,,
\end{split}
\end{equation}
can be $m$ times differentiated. We can write the recurrence equations for 
the $m$-differential forms $d^m b^j_k(t)$ as follows:
\begin{equation}\label{e:rec2}
d^m b^j_k(t,x)=\sum_{\ell\leq m}T_{S^j}^{t} d^\ell b_k^{j-1}(1,x). \theta^j_{m\ell}(t,x)
+\sum_{\ell\leq m}\int_0^t T_{S^j}^{t-s} d^{\ell+2}b_{k-1}^j(s,x). \alpha^j_{m\ell}(t,s, x)\, ds\,.
\end{equation}
Above we have extended the transport operator $T^{t}_{S}$ defined in
\eqref{e:transport} to multi-differential forms on $M$. Namely, 
$$
(T_{S^j}^t\, d^\ell b)(x)\defeq \sqrt{J_{S^j}^{-t}(x)}\;d^\ell b( g_{S^j}^{-t}(x))
$$
is an $\ell$-form on $(T_{ g_{S}^{-t}(x)}M)^\ell$. 
The linear form $\theta^j_{m\ell}(t,x)$ 
sends  $(T_x M)^m$ to $(T_{ g_{S^j}^{-t}(x)}M)^\ell$ 
(resp. $\alpha^j_{m\ell}(t,s, x)$ sends $(T_x M)^m$ to 
$(T_{ g_{S^j}^{s-t}(x)} M)^{\ell+2}$). 
These forms can be expressed in terms of derivatives of the maps
$ g^{-t}_{S^j}$, $ g^{s-t}_{S^j}$ at the point $x$, and $\theta^j_{m\ell}$ also
depends on $m-\ell$
derivatives of the function $P_{\ep_j}$.
These forms are uniformly bounded with respect to $j$, $x$ and $t\in [0,1]$.
We only need to know the explicit expression for $\theta^j_{mm}$:
\begin{equation}\label{e:theta_mm}
\theta^j_{mm}(t,x)= P_{\ep_j}\big( g^{-t}_{S^j}(x)\big)\times
\Big(d  g^{-t}_{S^j}(x)\Big)^{\otimes m}\,.
\end{equation}
Since the above expressions involve several sets of parameters, to facilitate the
bookkeeping we 
arrange the functions $b^j_k(t,x)$ and the $m$-differential forms 
$d^m b^j_k(t,x)$, $m\leq 2(N-k)$, inside a vector $\vb^j$. 
We will denote the entries by $\vb^j_{(k,m)}=d^m b^j_k$, and
with $0\leq k\leq N-1$, $m\leq 2(N-k)$:
\begin{equation}\label{e:array}
\begin{split}
\vb^j=\vb^j(t, x)\defeq\big(&b^j_0, db^j_0,\ldots\ldots,  d^{2N}b^j_0,\\ 
&b^j_1,db^j_1,\ldots, d^{2(N-1)}b^j_1,\\
&\ldots,\\ &
b^j_{N-1},db^j_{N-1}, d^{2}b^j_{N-1}\big)\,.
\end{split}
\end{equation}
The set of recurrence equations \eqref{e:rec2} may then be cast in
a compact form, using three operator-valued matrices $\vM^j_*$ 
(here the subscript $j$ is not a power,
but refers to the Lagrangian $\cL^j$ on which the transformation is based):
\begin{equation}\label{e:matrix}
(\vI-\vM^j_1)\vb^j = (\vM^j_{0,0}+\vM^j_{0,1})\vb^{j-1}\,.
\end{equation}
The first matrix act as follows on the indices $(k,m)$:
$$
\big(\vM^j_1\,\vb^j\big)_{(k,m)}(t)=\sum_{\ell\leq m}
\int_0^t ds\;T^{t-s}_{S^j}\,\vb^j_{(k-1,\ell+2)}(s)\,.\, \alpha^j_{m\ell}(t,s)\,.
$$ 
Since $\vM^j_1$ relates $b_{k}$ to $b_{k-1}$, it is obviously a nilpotent matrix of order $N$. The matrix
$\vM^j_{0, 1}$:
$$
\big(\vM^j_{0, 1}\vb^{j-1}\big)_{(k,m)}(t)=\sum_{\ell < m}
T_{S^j}^{t}\, \vb^{j-1}_{(k,\ell)}(1)\,.\, \theta^j_{m\ell}(t)\,,
$$
which relates $m$-derivatives to $\ell$-derivatives, $\ell<m$, is also nilpotent. Finally,
the last matrix $\vM^j_{0,0}$ acts diagonally on the indices $(k,m)$:
\begin{equation}\label{e:M_00}
\big(\vM^j_{0, 0}\vb^{j-1}\big)_{(k,m)}(t)=
T_{S^j}^{t}\, \vb^{j-1}_{(k,m)}(1)\,.\, \theta^j_{mm}(t)\,.
\end{equation}
From the nilpotence of $\vM^j_1$, we can invert \eqref{e:matrix} into
$$
\vb^j= \Big( \sum_{k_j=0}^{N-1} [\vM^j_1]^k_j \Big) \big(\vM^j_{0,0}+\vM^j_{0, 1}\big)\vb^{j-1}\,,
$$
where $[\vM]^k$ denotes the $k$-th power of the matrix $\vM$.
The above expression can be iterated:
\begin{equation}\label{e:final}
\vb^n =\sum_{k_1,..., k_n=0}^{N-1}\sum_ {\alpha_1,...,\alpha_n=0}^1 
[\vM^n_1]^{k_n}\,\vM^n_{0, \alpha_n}\,[\vM^{n-1}_1]^{k_{n-1}}\,\vM^{n-1}_{0,\alpha_{n-1}}\ldots 
[\vM^1_1]^{k_1}\,\vM^1_{0, \alpha_1}\,\vb^0\,.
\end{equation}
Notice that the first step
$\vM^1_{0, \alpha_1}\vb^0$ only uses the vector $\vb^0$ at time $t=1$, where it is well-defined.

From the nilpotence of $\vM^j_1$ and $\vM^j_{0,1}$, the terms contributing to $\vb^n_{(k,m)}$
must satisfy $\sum k_j\leq k$ and $\sum \alpha_j\leq m+2(\sum k_j)$.
In particular, $\sum k_j\leq N$, $\sum \alpha_j\leq 2N$, so for $n$ large, all terms in 
\eqref{e:final} are made of few
(long) strings of successive matrices $\vM^j_{0,0}$, separated by a few matrices $\vM^j_{0,1}$ or $\vM^j_{1}$
(the total number of matrices $\vM^j_{0,1}$ or $\vM^j_{1}$ in each term is at most $3N$).
As a result, the total number of terms on the right hand side grows at most like
$\cO(n^{m+3k})$ when $n\to\infty$.

Using the fact that $\theta^j_{m\ell}$ and $\alpha^j_{m\ell}$ are uniformly bounded,
the actions of the nilpotent matrices $\vM^j_1$, $\vM^j_{0,1}$ induce the following bounds on
the sup-norm of $\vb^j_{{k,m}}(t)$:
\begin{equation}\label{e:M1-bound}
\begin{split} 
\sup_{0\leq t\leq 1}\norm{\vM^j_1 \vb^j_{(k, m)}(t)}_{L^\infty} &\leq 
C\, \max_{m'\leq m+2}\sup_{0\leq t\leq 1}\,\norm{\vb^j_{(k-1, m')}(t)}_{L^\infty}\,,\\
\sup_{0\leq t\leq 1}\norm{(\vM^j_{0,1} \vb^{j-1})_{(k, m)}(t)}_{L^\infty}&\leq 
C(m)\, \max_{m'\leq m-1} \norm{\vb^{j-1}_{(k, m')}(1)}_{L^\infty}\,.
\end{split}
\end{equation}
The constant $C(m)$ depends on the partition $\cP^{(0)}$: for a partition of diameter
$\dia$, it is of order $\dia^{-m}$.

On the other hand, for any pair $(k,m)$, the ``diagonal'' action \eqref{e:M_00} 
on $\vb^j_{(k,m)}$ is very similar with its action on $\vb^j_{(0,0)}$, which is the
recurrence relation \eqref{e:principal-evol}. The only difference
comes from the appearance of the $m$-forms $\theta^j_{mm}$ instead of the functions
$\theta^j_{00}$. From the explicit expression \eqref{e:theta_mm}
and the fact that $0\leq P_{\ep_j}\leq 1$,
one easily gets
\begin{equation*}
|(\vM^j_{0,0} \vb^{j-1})_{(k, m)}(t,x)|\leq \sqrt{J^{-t}_{S^j}(x)}\;|d g^{-t}_{S^j}(x)|^m\,
|\vb^{j-1}_{(k, m)}(1, g^{-t}_{S^j}(x))|\,.
\end{equation*}
By contrast with \eqref{e:M1-bound}, in the above bound there is no potentially large
constant prefactor in front of the right hand side. This allows us to 
iterate this inequality, and obtain a bound similar with \eqref{e:prod-Jac}. 
Indeed, using the composition of the maps $ g^{-1}_{S^j}$ and their derivatives,
we get for any $j,j'\in\IN$ and $t\in [0,1]$:
\begin{equation}\label{M0-bound}
|(\vM^{j+j'}_{0,0}\cdots \vM^j_{0,0} \vb^{j-1})_{(k, m)}(t,x)|\leq
\sqrt{J^{-t-j'}_{S^{j'+j}}(x)}\;|d g^{-t-j'}_{S^{j+j'}}(x)|^m\,
|\vb^{j-1}_{(k, m)}(1, g^{-t-j'}_{S^{j'+j}}(x))|\,.
\end{equation}
As we explained 
above, the flow $g^{t}$ acting on $\cL^{j}$ is asymptotically expanding except
in the flow direction, because $g^t\cL^j$ converges to the weak unstable manifold. 
As a result, the inverse flow $g^{-j'}$ acting on $\cL^{j+j'}\subset g^{j'}\cL^j$,
and its projection $ g^{-j'}_{S^{j+j'}}$, have
a tangent map $d g^{-j'}_{S^{j+j'}}$ {\em uniformly} bounded 
with respect to $j,\ j'$.
In each ``string'' of operators $\vM^*_{0,0}$, 
the factor $d g^{-j'}_{S}$ can be replaced by a uniform constant.
For each term in \eqref{e:final}, we can then iteratively
combine the bounds (\ref{e:M1-bound},\ref{M0-bound}), to get
$$
|(\vM^{n}\vM^{n-1}\cdots\vM^{1}\vb^0)_{(k,m)}(t,x)|\leq 
C\,{\sqrt{J^{-t-n+1}_{S^n}(x)}}\; 
\norm{\vb^0(1)}
$$ 
Summing over those terms, we obtain
\begin{equation}\label{e:final2}
|\vb^n_{(k, m)}(t,x)|
\leq \tilde{C}(k,m)\,n^{m+3k}\,{\sqrt{J^{-t-n+1}_{S^n}(x)}} \;\norm{\vb^0(1)}\,.
\end{equation}
The Jacobian on the right hand side is the same as in the bound \eqref{e:prod-Jac}.
We can thus follow the same reasoning and replace $J^{-t-n+1}_{S^n}$ by $J^u_n(\bep)$ 
to obtain the lemma. $\hfill\square$

This ends the proof of Lemma~\ref{l:transfer} and Corollary \ref{c:approx-n}. 
We proceed with the proof of our main Lemma~\ref{l:main},
and now describe the states
$U^{-n/2}P_{\bep'}\,\delta^{(4n)}_{j'}(z')$ and $U^{n/2}\, P_{\bep}\, \delta^{(n)}_j(z)$.

\subsection{Evolution under $U^{-n/2}$ and $U^{n/2}$}

Applying Corollary~\ref{c:approx-n} with $n'=4n$, resp. $n'=n$, we have approximate
expressions for the states appearing in Lemma~\ref{l:main}:
\begin{align}\label{e:decompo1}
P_{\bep}\,\delta^{(n)}_{j}(z)&=(2\pi\hbar)^{-\frac{d+1}2}\int v^n(0;z,\eta_1,\bep) \,\chi^{(n)}(\eta_1)\,d\eta_1+
\cO_{L^2}(\e^{n\delta}\,\hbar^{N-\frac{d-1}2})\,,\\
P_{\bep'}\,\delta^{(4n)}_{j'}(z')&=(2\pi\hbar)^{-\frac{d+1}2}\int v^n(0;z',\eta_1',\bep') \,
\chi^{(4n)}(\eta_1')\,d\eta_1'+
\cO_{L^2}(\e^{4n\delta}\,\hbar^{N-\frac{d-1}2})\,,
\end{align}
we notice that for $n\leq n_E(\hbar)$ the remainders are of the form
$\cO(\hbar^{N-N_1})$ for some fixed $N_1$.

To prove the bound of Lemma~\ref{l:main}, we assume $n$ is an even integer, and
consider the individual overlaps
\begin{equation}\label{e:overlap}
\left\la U^{-n/2}v^{n}(0;z',\eta_1',\bep'), U^{n/2}\,v^{n}(0;z,\eta_1,\bep)\right\ra\,,
\end{equation}
Until the end of the section, we will fix $z,\eta_1,z',\eta'_1$ and omit them in the notations.
On the other hand, we will sometimes make explicit
the dependence on the sequences $\bep'$, $\bep$.
We then need to understand the states $U^{-n/2}\,v^{n}(0;\bep')$ and $U^{n/2}\,v^{n}(0;\bep)$.

\subsubsection{Evolution under $U^{-n/2}$}
We use WKB approximations to describe the backwards-evolved state
$U^{-t}v^n(0;\bep')$. 
Before entering into the details, let us sketch the backwards evolution of the Lagrangian
$\cL^n=\cL^n(0;\bep')$ supporting $v^n(0)=v^n(0;\bep')$ (for a moment we omit to indicate the dependence 
in $\bep'$). Since $\cL^n$ had been obtained by evolving $\cL^0$ and
truncating it at each step, for any $0\leq t\leq n-1$, the Lagrangian
$\cL^n(-t)\defeq g^{-t}\cL^n$ will be contained in $\cL^{n-\lfloor t\rfloor -1}(1-\{t\})$, 
where we decomposed the time $t$
into its integral and fractional part. This Lagrangian projects well onto the base manifold,
and is generated by the function $S^n(-t)=S^{n-\lfloor t\rfloor -1}(1-\{t\})$ (which
satisfies the Hamilton-Jacobi equation for negative times).
This shows that the WKB method of \S\ref{s:evolWKB},
applied to the backwards flow $U^{-t}$ acting on $v^n(0)$, can be {\em formally} 
used for all times $0\leq t\leq n-1$. The evolved state can be written as
\begin{equation}\label{e:hatR_N}
U^{-t}\,v^n(0)=v^n(-t)+\hat R_N(-t)\,,
\end{equation}
and $v^n(-t)$ has the WKB form 
\begin{equation}\label{e:v^n}
v^n(-t)=b^n_\hbar(-t)\,\e^{iS^{n}(-t)/\hbar}\,,\qquad 
b^n_\hbar(-t)=\sum_{k=0}^{N-1}\hbar^{k}\,b^n_k(-t)\,.
\end{equation}
The symbols $b^n_k(-t)$ are obtained
from $b^n_k(0)$ using the backwards transport equations (see Eqs.~(\ref{e:transport-class},
\ref{e:solution-transport})):
\begin{align}
b^n_0(-t)&=T_{S^n(0)}^{-t}\,b^n_0(0)=\big( J_{S^n(-t)}^t\big)^{1/2}\;b^n(0)\circ g^{t}_{S^n(-t)}\,,\\
b^n_k(-t)&=T_{S^n(0)}^{-t}\,b^n_k(0) - 
\int_0^{t} T_{S^n(-t)}^{-t+s}\Big(\frac{i\lap b^n_{k-1}}2(-s)\Big)ds\,.
\end{align}
These symbols are supported on $\pi\cL^n(-t)$.
We need to estimate their $C^m$ norms uniformly in $t$. The inverse of the
Jacobian $J_{S^n(-t)}^t$ approximately measures the volume of the
Lagrangian $\cL^n(-t)$. Since the latter remains close to the
weak unstable manifold as long as $n-t>>1$, the backwards flow has the effect to
shrink it along the unstable directions. Thus, for $n-1\geq t>>1$, $\cL^n(-t)$
consist in a thin, elongated subset of $\cL^{n-\lfloor t \rfloor-1}$ (see figure~\ref{f:decompo}), 
with a volume of order
\begin{equation}\label{e:volume}
\Vol(\cL^n(-t))\leq C\, \big(\inf_x J_{S^n(-t)}^t(x)\big)^{-1}\leq 
C\,J^u_{\lfloor t \rfloor}(\ep'_{n-\lfloor t \rfloor}\cdots \ep'_n)\,,\qquad 0\leq t\leq n-1\,.
\end{equation}
When differentiating $b_0^n(-t)$, the derivatives of the flow $g^t_{S^n(-t)}$ also appear.
Since $\cL^n(-t)$ is close to the weak unstable manifold, 
the derivatives become large as $t>>1$:
$$
|\partial^\alpha_x g^{t}_{S^{n}(-t)}(x)|\leq C(\alpha)\,\e^{t\lambda_+}\,,\quad\text{where}\ 
\lambda_+\defeq\lambda_{\max}(1+\delta'/2)\,,\quad 0\leq t\leq n-1,\quad x\in \pi\cL^n(-t)\,.
$$
Hence, for
any $t\leq n-1$ and index $0\leq m\leq 2N$ the
$m$-derivatives of the principal symbol can be bounded as follows:
\begin{equation}
\begin{split}
\forall t\leq n-1,\qquad |d^m b^n_0(-t,x)|&\leq C\,\big(J_{S^n(-t)}^{t}(x)\big)^{1/2}\;
|dg^{t}_{S^n(-t)}(x)|^m\,\norm{b^n_0(0)}_{C^m}\\
&\leq C\,J^u_{\lfloor t \rfloor}(\ep'_{n-\lfloor t \rfloor}\cdots \ep'_n)^{-1/2}\;\e^{tm\lambda_+}\,\norm{b^n_0(0)}_{C^m}\\
&\leq C\,J^u_{n-\lfloor t \rfloor}(\ep'_0\cdots\ep'_{n-\lfloor t \rfloor})^{1/2}\;\e^{tm\lambda_+}\,.
\end{split}
\end{equation}
In the last line we used the estimates of Lemma~\ref{l:transfer} for $\norm{b^n(0)}_{C^m}$.
From now on we will abbreviate $J^u_{n-\lfloor t \rfloor}(\ep'_0\cdots\ep'_{n-\lfloor t \rfloor})$ 
by $J^u_{n-\lfloor t \rfloor}(\bep')$.
By iteration, we similarly estimate the derivatives of the higher-order symbols
($k< N,\; m\leq 2(N-k)$):
\begin{equation}\label{e:M3bound}
\forall t\leq n-1,\qquad |d^m b^n_k(-t,x)|
\leq C\,J^u_{n-\lfloor t \rfloor}(\bep')^{1/2}\;\e^{t(m+2k)\lambda_+}\,.
\end{equation}
We see that the higher-order symbols may grow faster (with $t$) than the principal one.
As a result, when $t$ becomes too large, the expansion \eqref{e:v^n} does not make sense 
any more, since the remainder in \eqref{e:hatR_N} becomes larger than the main term. 
From \eqref{e:Duham}, this remainder is bounded by
$$
\norm{\hat R_N(-t)}\leq \frac{\hbar^N}2\,\int_0^t\norm{\lap b^n_{N-1}(-s)}\,ds
\leq C\,\hbar^N\;\e^{t\,2N\,\lambda_+}\,J^u_{n-\lfloor t \rfloor}(\bep')^{1/2}\,.
$$
This remainder remains smaller than the previous terms if 
$t\leq n_E(\hbar)/2$. Since we assume 
$n\leq n_E(\hbar)$, the WKB expansion still makes
sense if we take $t=n/2$.
To ease the notations in the following sections, we call
$w^{n/2}\defeq v^{n}(-n/2)$ the WKB state approximating $U^{-n/2}v^n(0)$, its phase function
$S^{n/2}=S^n(-n/2)$ and its symbol
$c^{n/2}_\hbar(x)\defeq b^n_\hbar(-n/2,x)$, all these data depending on $\bep'$.
The above discussion shows that
\begin{equation}\label{e:u-n/2}
\norm{U^{-n/2}\,v^n(0;\bep')-w^{n/2}(\bep')}=
\norm{\hat R_N(-n/2)}\leq C\,\hbar^{N\delta'/2}\;J^u_{n/2}(\bep')^{1/2}\,.
\end{equation}
We will select an integer $N$ large enough ($N\delta'>>1$), so that the 
above remainder is smaller than 
the estimate $J^u_n(\bep')^{1/2}$ we have on $\norm{v^n(\bep')}$.

\subsubsection{Evolution under $U^{n/2}$}\label{s:decompo} 
We now study the forward evolution $U^{n/2}\,v^{n}(0;\bep)$. From now on we omit to indicate
the dependence in the parameter $t=0$.
Using the smooth partition \eqref{e:partition}, we decompose $U^{n/2}$ as:
$$
U^{n/2} = \sum_{\alpha_i, 1\leq i\leq n/2} P_{\alpha_{n/2}}^2\,U\,P_{\alpha_{n/2-1}}^2\,U\cdots 
P_{\alpha_1}^2\,U\defeq \sum_{\bal} Q_{\bal}\,.
$$
The operators $(Q_{\bal})$ are very similar
with the $(P_{\bal})$ of Eq.~\eqref{e:P_bep}: the cutoffs $P_k$ are replaced by their squares $P_k^2$.
As a result, the iterative WKB method presented in the previous sections can be 
adapted
to obtain approximate expressions for each state
$Q_{\bal}\,v^n(\bep)$, similarly as in \eqref{e:WKB-n}:
$$
Q_{\bal}\,v^n(\bep)=v^{\frac32 n}(\bep\bal)+\cO_{L^2}(\sqrt{J^u_n(\bep)}\,\hbar^N)\,,\qquad 
v^{\frac32 n}(x;\bep\bal)=b^{\frac32 n}_\hbar(x;\bep\bal)\,
\e^{\frac{i}{\hbar}S^{\frac32 n}(x;\bep\bal)}\,.
$$ 
Here $\bep\bal$ is the sequence of length $3n/2$ with elements 
$\ep_0\cdots\ep_n\alpha_1\cdots\alpha_{n/2}$.
That state is localized on the Lagrangian manifold $\cL^{\frac32 n}(\bep\bal)$. 
The symbols $b^{\frac32 n}_k(\bep\bal)$
and their derivatives satisfy the bounds of Lemma~\ref{l:transfer}.
The state $U^{n/2} v^n(\bep)$ is therefore given by a sum of contributions
\begin{equation}\label{e:decompo2}
U^{n/2} v^n(\bep)=\sum_{\bal} v^{\frac32 n}(\bep\bal)+
\cO_{L^2}\big(\hbar^{N-N_K}\big)\,.
\end{equation}
Here $N_K$ is a constant depending on the cardinal $K$ of the partition $\cP^{(0)}$, and we
assumed $n\leq n_E(\hbar)$. The integer $N$ will be taken large enough,
such that $\hbar^{N-N_K}$ is smaller than the remainder appearing in \eqref{e:u-n/2}.

\subsubsection{Grouping terms into connected Lagrangian leaves}\label{s:disjoint}
\begin{figure}
\begin{center}
\includegraphics[width=10cm]{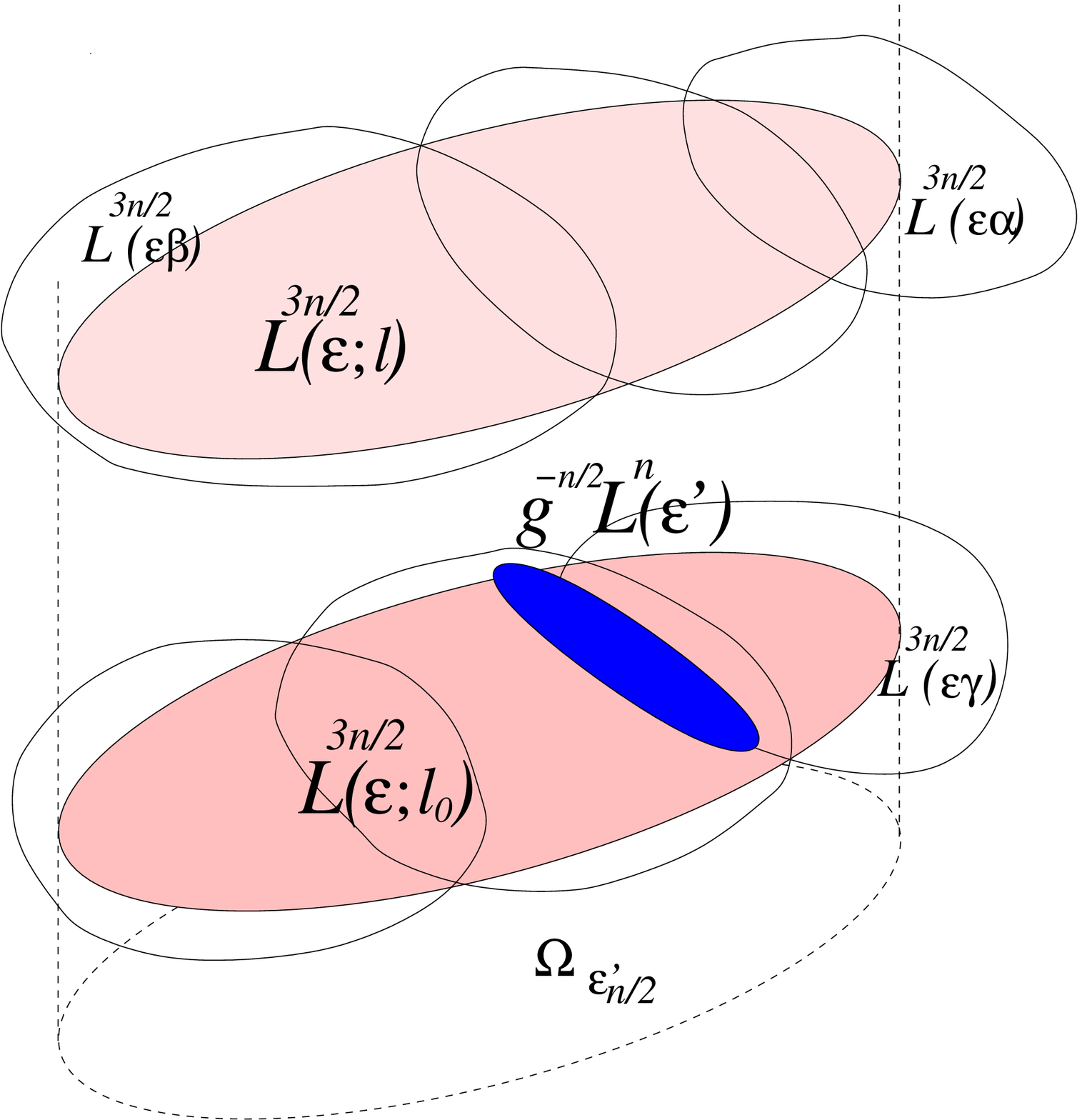}
\caption{\label{f:decompo} Decomposition of $\big(g^{n/2} \cL^{n}(\bep)\big)\cap T^*\Omega_{\ep'_{n/2}}$
into connected leaves (here we show two of them, in light pink). The leaf $\ell$ contains
the components $\cL^{\frac32 n}(\bep\bal)$, $\cL^{\frac32 n}(\bep\blds{\beta})$ 
while the leaf $\ell_0$ contains
$\cL^{\frac32 n}(\bep\blds{\gamma})$. We also show the elongated leaf $g^{-n/2}\cL^n(\bep')$
supporting the state $w^{n/2}(\bep')$ (dark blue). This state might 
interfere with $v^{\frac32 n}(\bep,\ell_0)$,
but not with $v^{\frac32 n}(\bep,\ell)$ or any other leaf above $\Omega_{\ep'_{n/2}}$.
}
\end{center}
\end{figure}
To compute the overlap \eqref{e:overlap}, we do not need the full sum \eqref{e:decompo2}, but only the
components $\bal$ such that the support of $v^{\frac32 n}(\bep\bal)$ intersects the support of 
$w^{n/2}(\bep')$, which is inside $\Omega_{\ep'_{n/2}}$. 
Thus, we can restrict ourselves to the set of sequences
$$
A\defeq \set{\bal\; : \; 
\pi\cL^{\frac32 n}(\bep\bal)\cap\Omega_{\ep'_{n/2}}\neq\emptyset}\subset\set{1,\ldots,K}^{n/2}\,.
$$
For $n>>1$, the Lagrangian $\bigcup_{\bal\in A}\cL^{\frac32 n}(\bep\bal)$, which 
is a strict subset of $g^{n/2} \cL^{n}(\bep)$,
is the disjoint union of a large number of {\em connected leaves}, 
which we denote by $\cL^{\frac32 n}(\bep,\ell)$, $\ell\in[1,L]$ (see Figure~\ref{f:decompo}).
Each leaf $\cL^{\frac32 n}(\bep,\ell)$ corresponds to geodesics of length $n/2$
from $\Omega_{\ep_n}$ to $\Omega_{\ep'_{n/2}}$ in a definite homotopy class.
As a consequence, if $\rho$, $\rho'$ belong to two different leaves $\ell\neq\ell'$, 
there must be a time $0< t< \frac{n}2$ such that the backwards images $g^{-t}\rho$, $g^{-t}\rho'$ 
are at a distance larger than $D>0$ ($D$ is related to the 
injectivity radius). The total number of leaves above $\Omega_{\ep'_{n/2}}$ can grow at most
like the full volume of $g^{n/2}\cL(\bep)$, so that
$$
L\leq C\,\e^{n(d-1)\lambda_+/2}\leq C\,\hbar^{-(d-1)/2}\,.
$$
Each leaf $\cL^{\frac32 n}(\bep,\ell)$ is the union of a certain number of 
components $\cL^{\frac32 n}(\bep\bal)$, and we group the corresponding sequences 
$\bal$ into the subset $A_\ell\subset \set{1,\ldots,K}^{n/2}$: 
$$
\cL^{\frac32 n}(\bep,\ell)=\bigcup_{\bal\in A_\ell} \cL^{\frac32 n}(\bep\bal)\,.
$$
We obviously have $A=\bigsqcup_{\ell}A_\ell$.
All components  $\cL^{\frac32 n}(\bep\bal)$ with $\bal\in A_\ell$ are generated by the
same phase function 
$S^{\frac32 n}(\bep\bal)\defeq S^{\frac32 n}(\bep,\ell)$, so that the state
\begin{equation}\label{e:decompo3}
v^{\frac32 n}(x;\bep,\ell)\defeq \sum_{\bal\in A_\ell} v^{\frac32 n}(x;\bep\bal) = 
b^{\frac32 n}_{\hbar}(x;\bep,\ell)\;\e^{\frac{i}{\hbar}S^{\frac32 n}(x;\bep,\ell)}
\end{equation}
is a Lagrangian state supported on $\cL^{\frac32 n}(\bep,\ell)$, with symbol
$$
b^{\frac32 n}_{\hbar}(x;\bep,\ell)=\sum_{\bal\in A_\ell} b^{\frac32 n}_{\hbar}(x;\bep\bal)\,.
$$
By inspection one can check that, at each 
point $\rho\in\cL^{\frac32 n}(\bep,\ell)$, the above sum over $\bal\in A_\ell$ has 
the effect to insert
partitions of unity $\sum_k P^2_k=1$ at each preimage $g^{-j}(\rho)$, $j=0,\ldots,\frac{n}2-1$. 
As a result,
the principal symbol will satisfy the same type of upper bound as in \eqref{e:prod-Jac}:
$$
|b^{\frac32 n}_{0}(x;\bep,\ell)|\leq |b^n(g_{S}^{-n/2}(x))|\;
J_{S}^{-\frac12 n}(x)^{1/2}
\leq C\,J_{S}^{-\frac32 n}(x)^{1/2}\,,\qquad \text{with}\ \ S=S^{\frac32 n}(\bep,\ell)\,.
$$
The same argument holds for the higher-order terms and their derivatives.
Besides, because the action of $g^{-3n/2}$ on $\cL^{\frac32 n}(\bep,\ell)$ is contracting, for any
$x\in \Omega_{\ep'_{n/2}}$ the Jacobian
$J_{S}^{-\frac32 n}(x)$ is of the order of
$J^u_{\frac32 n}(\bep\bal)$, where $\bal$ can be any sequence in $A_\ell$ (all these Jacobians
are of the same order). Defining
$$
J^u_{\frac32 n}(\bep,\ell)=\max_{\bal\in A_\ell}J^u_{\frac32 n}(\bep\bal)\geq 
\frac1{C}\min_{\bal\in A_\ell}J^u_{\frac32 n}(\bep\bal)\,,
$$ 
the full symbol
$b^{\frac32 n}_{\hbar}(x;\bep,\ell)$ satisfies similar bounds as in Lemma~\ref{l:transfer}:
\begin{equation}\label{e:est3/2}
|d^m b^{\frac32 n}_{k}(x;\bep,\ell)|\leq 
C\,n^{m+3k}\,J^u_{\frac32 n}(\bep,\ell)^{1/2}\,,\qquad k\leq N-1,\ m\leq 2(N-k)\,.
\end{equation}

\subsection{Overlaps between the Lagrangian states}\label{s:overlaps}
Putting together (\ref{e:u-n/2}, \ref{e:decompo3}, \ref{e:decompo2}), the overlap \eqref{e:overlap} 
is approximated by the following sum:
\begin{align}\label{e:decompo4}
\left\la U^{-n/2}v^{n}(\bep'), U^{n/2}\,v^{n}(\bep)\right\ra&=
\sum_{\ell=1}^L \la w^{n/2}(\bep'),\,v^{\frac32 n}(\bep,\ell)\ra+\cO(\hbar^{N\delta'/2})\,,\qquad\text{where}\\
\la w^{n/2}(\bep'),\,v^{\frac32 n}(\bep,\ell)\ra&=
\int \e^{\frac{i}{\hbar}\big(S^{\frac32 n}(x;\bep,\ell)-S^{n/2}(x;\bep')\big)}\, 
\bar c^{n/2}_\hbar(x;\bep')\,b^{\frac32 n}_\hbar(x;\bep,\ell)\,.\label{e:integral}
\end{align}
Each term is the overlap between the WKB state $w^{n/2}(\bep')$ supported on 
$g^{-n/2}\cL^{n}(\bep')$,
and the WKB state $v^{\frac32 n}(\bep,\ell)$ supported on $\cL^{\frac32 n}(\bep,\ell)$, both Lagrangians 
sitting above $\Omega_{\ep'_{n/2}}$ (see Figure~\ref{f:decompo}). 
The sup-norms of these two 
states, governed by the principal symbols $c^{n/2}_0(\bep')$, 
$b^{\frac32 n}_0(\bep,\ell)$, are bounded by
\begin{equation}
\norm{w^{n/2}(\bep')}_{L^\infty}\leq C\,J^u_{n/2}(\bep')^{1/2},\qquad
\norm{v^{\frac32 n}(\bep,\ell)}_{L^\infty}\leq C\,J^u_{\frac32 n}(\bep,\ell)^{1/2}\,.
\end{equation}
Here $C>0$ is independent of all parameters, including the diameter $\dia$ of
the partition.
The integral \eqref{e:decompo4} takes place on the support of $c^{n/2}_\hbar(x;\bep')$,
that is (see \eqref{e:volume}), on a set of volume $\cO(J^u_{n/2}(\ep'_{n/2}\cdots \ep'_{n}))$. 
It follows that each overlap \eqref{e:integral} is bounded by 
\begin{equation}\label{e:rough}
| \la w^{n/2}(\bep'),\,v^{\frac32 n}(\bep,\ell)\ra | \leq
C\,J^u_{n/2}(\bep')^{1/2}\; J^u_{\frac32 n}(\bep,\ell)^{1/2}\;
J^u_{n/2}(\ep'_{n/2}\cdots\ep'_{n})\,. 
\end{equation}
We show below that the above
estimate can be improved for almost all leaves $\ell$, when one takes into account
the phases in the integrals \eqref{e:integral}. 
Actually, for times $n\leq n_E(\hbar)$,
there is at most a single
term $\ell_0$ in the sum \eqref{e:decompo4} for which the above bound is sharp; for all other terms $\ell$,
the phase oscillates fast enough to make the integral negligible. Geometrically, this phase oscillation
means that the Lagrangians  $\cL^{\frac32 n}(\bep,\ell)$,  
$g^{-n/2}\cL^{n}(\bep')\subset \cL^{n/2}(\bep')$ are ``far enough'' from
each other (see Fig.~\ref{f:decompo}). The ``distance'' between two Lagrangians above $\Omega_{\ep'_{n/2}}$
is actually measured by the {\em height}
$$
H\big(\cL^{\frac32 n}(\bep,\ell), \cL^{n/2}(\bep')\big)\defeq 
\inf_{x\in \Omega_{\ep'_{n/2}}} |dS^{\frac32 n}(x;\bep,\ell)-dS^{n/2}(x;\bep')|\,.
$$
The overlap between ``distant'' leaves can be estimated through a
nonstationary phase argument:
\begin{lem}
Assume that, for some $\delta''<\delta'/2$, for some $\hbar>0$ and some time 
$n\leq n_E(\hbar)$, the height
$$
H\big(\cL^{\frac32 n}(\bep,\ell), \cL^{n/2}(\bep')\big)\geq \hbar^{\frac{1-\del''}2}\,.
$$
Then, provided $\hbar$ is small enough, the overlap 
\begin{equation}\label{e:sharp}
|\la w^{n/2}(\bep'),\,v^{\frac32 n}(\bep,\ell)\ra|\leq C\,
\hbar^{N\delta''}\,\sqrt{J^u_{n/2}(\bep')J^u_{\frac32n}(\bep,\ell)}\,.
\end{equation}
The constant $C>0$ is uniform with respect to $\bep'$, $\bep$ and the implicit
parameters $z,z',\eta_1,\eta'_1$.
\end{lem}
\begin{proof}
Let us call $s(x)=S^{\frac32 n}(x;\bep,\ell)-S^{n/2}(x;\bep')$ the phase function appearing in the
integral \eqref{e:integral}. Notice that the assumption on the height means that 
$|ds(x)|\geq \hbar^{\frac{1-\del''}2}$
for all $x$. 
We then expand the product $\bar c^{n/2}_\hbar\,b^{\frac32 n}_\hbar$ and keep only the
first $N$ terms:
$$
\bar c^{n/2}_\hbar(x;\bep')\,b^{\frac32 n}_\hbar(x;\bep,\ell)=
a_\hbar(x)+Rem_N(x)\,,\qquad a_\hbar(x)=\sum_{k=0}^{N-1}\hbar^k\,a_k(x)\,.
$$
From the estimates (\ref{e:M3bound},\ref{e:est3/2}), we control the sup-norm of the remainder:
$$
\norm{Rem_N}_{L^\infty}\leq C\,\hbar^{N\delta'/2}\,\sqrt{J^u_{n/2}(\bep')J^u_{\frac32n}(\bep,\ell)}\,.
$$ 
Through the Leibniz rule we control the derivatives of $a_k$:
$$
\norm{a_k}_{C^m}\leq C\,n^{m+3k}\,\sqrt{J^u_{n/2}(\bep')J^u_{\frac32n}(\bep,\ell)} \; 
\e^{\frac{n}{2}(m+2k)\lambda_+}\,,\qquad k\leq N-1,\ \ m\leq 2(N-k)\,.
$$
For each $k< N$ and $m\leq 2(N-k)$, we have at our disposal the following nonstationary phase estimate 
\cite[Section~7.7]{Horm}:
\begin{align*}
\Big|\int
a_k(x)\,\exp\Big(\frac{i}{\hbar}s(x)\Big)\,dx\Big|&\leq C\,\hbar^{m}\sum_{m'\leq m}
\sup_{x}\Big(\frac{|d^{m'} a_k(x)|}{|ds(x)|^{2m-m'}}\Big)\\
&\leq C\,\hbar^{m\delta''-k(1-\delta'/2)}\,\sqrt{J^u_{n/2}(\bep') J^u_{\frac32n}(\bep,\ell)}\,.
\end{align*}
Here we used the assumption on $|ds(x)|$ and the fact that $\delta''<\delta'/2$. 
By taking $m=N-k$ for each $k$ and summing the estimate over $k$, we get:
$$
\Big|\int a_\hbar(x)\,\exp\Big(\frac{i}{\hbar}s(x)\Big)\,dx \Big|\leq C\,\hbar^{N\delta''}\,
\sqrt{J^u_{n/2}(\bep')J^u_{\frac32n}(\bep,\ell)}\,.
$$
Since $\delta'/2>\delta''$, the remainder $Rem_N$ yields a smaller contribution, which ends the proof.
\end{proof}

We now show that there
is at most one Lagrangian leaf $\cL^{\frac32 n}(\bep,\ell_o)$ which can be very close to
$\cL^{n/2}(\bep')$:
\begin{lem}
Take as above $\delta''<\delta'/2$, assume the diameter $\dia$ is much smaller than the injectivity
radius, and for $\hbar$ small enough
take $n\leq \frac{(1-\delta')|\log\hbar|}{\lambda_{\max}}$.

If there is some $\ell_o\in\set{1,\ldots,L}$ such that
the height $H\big(\cL^{\frac32 n}(\bep,\ell_o), \cL^{n/2}(\bep')\big)\leq \hbar^{\frac{1-\del''}2}$,
then for any $\ell\neq \ell_o$ we must have 
$H\big(\cL^{\frac32 n}(\bep,\ell), \cL^{n/2}(\bep')\big) > \hbar^{\frac{1-\del''}2}$.
\end{lem}
\begin{proof}
Assume {\em ab absurdo} the existence of 
$\rho_o\in \cL^{\frac32 n}(\bep,\ell_o)$,  $\rho\in \cL^{\frac32 n}(\bep,\ell)$ and
$\rho'_1,\rho'_2\in\cL^{n/2}(\bep')$,
such that the Riemannian distances
$d(\rho_o , \rho'_1)\leq \hbar^{\frac{1-\del''}2}$ and 
$d(\rho,\rho'_2)\leq \hbar^{\frac{1-\del''}2}$.
When applying the backwards flow for times $0\leq t\leq \frac{n}2$, these points
depart at most like
\begin{align*}
d(g^{-t}\rho_o,d^{-t}\rho'_1)&\leq C\,\e^{t\lambda_+}\,\hbar^{\frac{1-\del''}2}
\leq C\,\hbar^{\delta'/4-\delta''/2}\,,\\
d(g^{-t}\rho,d^{-t}\rho'_2)&\leq C\,\e^{t\lambda_+}\,\hbar^{\frac{1-\del''}2}
\leq C\,\hbar^{\delta'/4-\delta''/2}\,.
\end{align*}
Besides, on this time interval the
points $g^{-t}\rho'_1$, $g^{-t}\rho'_2$ remain in the small Lagrangian piece $g^{-t}\cL^{n/2}(\bep')$
of diameter $\leq \dia$, so that $d(g^{-t}\rho_o,g^{-t}\rho)\leq \dia$. 
Since $\dia$ has been chosen small,
this contradicts the property that the points $g^{-t}\rho_o$, $g^{-t}\rho$ must depart
at a distance $\geq D$ (see the discussion at the beginning of \S\ref{s:disjoint}).
\end{proof}

If there exists a leaf $\ell_o$ such that 
$H(\cL^{\frac32 n}(\bep,\ell_o),\cL^{n/2}(\bep'))\leq \hbar^{\frac{1-\del''}2}$, there is
a point $\rho_o\in \cL^{\frac32 n}(\bep,\ell_o)$ such that $g^{-j}\rho_o$ stays at small
distance from $\cL^{n/2-j}(\bep')$ for all $j=0,\ldots,n/2-1$, and therefore satisfies
$\pi g^{-j}\rho_o\in \Omega_{\ep'_{n/2-j}}$. This shows that the set $A_{\ell_o}$ contains
the sequence $(\ep'_{1}\cdots\ep'_{n/2})\defeq \tilde\bep'$. 
The overlap corresponding to this leaf is bounded as in \eqref{e:rough}, and after replacing 
$J^u_{\frac32n}(\bep,\ell_o)$ by $J^u_{\frac32n}(\bep\tilde\bep')$ we obtain
\begin{equation}\label{e:single}
|\la w^{n/2}(\bep'),\,v^{\frac32n}(\bep;\ell_o)\ra|
\leq C\,J^u_n(\bep')\,J^u_n(\bep)^{1/2}\,.
\end{equation}
According to the above two Lemmas, all the remaining leaves are ``far from'' $\cL^{n/2}(\bep')$, and their
contributions to \eqref{e:decompo4} sum up to 
$$
\sum_{\ell\neq\ell_o} \la w^{n/2}(\bep'),v^{\frac32 n}(\bep;\ell)\ra =\cO(\hbar^{N\delta''-(d-1)/2})\,.
$$
We take $N$ large enough (say, $N\del''>>1$), 
such that this is negligible compared with \eqref{e:single}. We finally get,
whether such an $\ell_o$ exists or not:
$$
|\la U^{-n/2}v^n(z',\eta_1',\bep'),\,U^{n/2}v^{n}(z,\eta_1,\bep)\ra|\leq C\,J^u_n(\bep')\,J^u_n(\bep)^{1/2}\,.
$$
To finish the proof of Lemma~\ref{l:main}, there remains to integrate over the parameters 
$\eta_1,\eta'_1$ in \eqref{e:decompo1}. Since $\chi^{(n)}$ (resp. $\chi^{(4n)}$) is supported 
on an interval of length $\hbar^{1-\del}\e^{n\del}$ (resp. $\hbar^{1-\del}\e^{4n\del}$),
the overlap of Lemma~\ref{l:main} finally satisfies the following bound:
$$
|\la U^{-n/2}P_{\bep'}\,\delta^{(4n)}_{j'}(z'),\, U^{n/2}\, P_{\bep}\, \delta^{(n)}_j(z) \ra|\leq
C\,\hbar^{-(d+1)}\,\e^{5\delta n}\,\hbar^{2-2\delta}\,J^u_n(\bep')\,J^u_n(\bep)^{1/2}\,.
$$
This is the estimate of Lemma~\ref{l:main}, with $c=2+5/\lambda_{\max}$.
Proposition~\ref{p:main} and Theorem~\ref{t:main} follow. $\hfill\square$

\section{Subadditivity}\label{s:subadditivity}
The aim of this section is to prove Proposition \ref{p:subadd}.
It is convenient here to use some notions of symbolic dynamics. Starting
from our partition of unity $(P_k)_{k=1,\ldots, K}$, we introduce a symbolic space
$\Sigma=\{1,\ldots, K\}^{\IN}$. The shift $\sigma$
acts on $\Sigma$ by shifting a sequence $\bep=\ep_0\ep_1\ldots$ to the left and deleting the first symbol. 
For $\bep=(\ep_0 \ldots \ep_n)$,
we denote $[\bep]\subset \Sigma$ the subset ($n$-cylinder) formed of sequences starting with the symbols
$\ep_0\ldots\ep_n$ (throughout this section the integer $n$ will generally differ from $n_E(\hbar)$).

To any normalized
eigenfunction $\psi_\hbar$ we can associate a probability measure $\mu^\Sigma_\hbar$ on $\Sigma$
by letting, for any $n$-cylinder $[\bep]$,
$$
\mu^\Sigma_\hbar([\bep])\defeq
\norm{P_{\ep_n }P_{\ep_{n-1}}(1)\ldots P_{\ep_0}(n)\,\psi_\hbar}^2=
\norm{P_{\ep_n }(-n)P_{\ep_{n-1}}(-(n-1))\ldots P_{\ep_0}\,\psi_\hbar}^2\,.
$$ 
If we denote $\barbep=(\eps_n\eps_{n-1} \cdots \ep_0)$, this quantity is equal to
$\norm{\tP^*_{\barbep}\, \psi_\hbar}^2=\norm{P^*_{\barbep}\, \psi_\hbar}^2$ (see \eqref{e:weights}).
To ensure that this defines a probability measure on $\Sigma$, one needs to check the
following compatibility condition
\bequ\label{e:compatibility}
\mu^\Sigma_\hbar([\ep_0\ldots \ep_n])=\sum_{\ep_{n+1}=1}^K\mu^\Sigma_\hbar([\ep_0\ldots \ep_n \ep_{n+1}])
\end{equation}
for all $n$ and all $\ep_0\ldots \ep_n$. This identity is obvious from \eqref{e:partition}.

\subsection{Invariance until the Ehrenfest time}
By the Egorov theorem, if $\mu$ is the weak-$*$ limit of the Wigner measures 
$W_{\psi_\hbar}$ on $T^* M$, then
for every $n$ and any fixed $n$-cylinder $[\bep]\subset \Sigma$ we have
$\mu^\Sigma_\hbar([\bep])\hto0 \mu(\{\barbep\})$,
where $\{\barbep\}$ was defined in \S\ref{s:subadd} as the function
$P^2_{\ep_n}\,(P^2_{\ep_{n-1}}\circ g^1)\ldots (P^2_{\ep_{0}}\circ g^{n})$ on $T^* M$.
This means that the measures $\mu^\Sigma_\hbar$ converge to a measure
$\mu^\Sigma_0$ defined by $\mu^\Sigma_0([\bep])\defi\mu(\{\barbep\})$.

Since the $\psi_\hbar$ are eigenfunctions, $\mu$ is localized on $\cE$ and is $(g^t)$-invariant
(Prop.~\ref{e:semiclass-measure}), so that
$\mu^\Sigma_0$ is $\sigma$-invariant. For $\hbar>0$ the measures 
$\mu^\Sigma_\hbar$ are not exactly $\sigma$-invariant; yet,
we show below that $\mu^\Sigma_\hbar$ is {\em almost invariant} under the shift, until the Ehrenfest
time.

For small $\gamma, \nu>0$ we introduce 
the time
$T_{\nu,\gamma,\hbar}\defi \frac{(1-\gamma)|\log\hbar|}{2(1+\nu)\lambda_{\max}}$.
\begin{prop} \label{p:invariance} 
For any given $n_o\in\N$, for any small enough $\hbar$ and any $n\in\N$ such that 
$n+n_o\leq 2\,T_{\nu, \gamma, \hbar}$,
for any cylinder $[\bep]=[\ep_{0}\ep_{1}\ldots\ep_{n_o}]$ of length $n_o$,
one has
\begin{equation*}
\sum_{\ep_{i},-n\leq i\leq -1}
\mu^\Sigma_\hbar([\ep_{-n}\ldots\ep_{-1} \ep_0\ep_1\ldots\ep_{n_o}])
=\mu_\hbar^\Sigma([\ep_{0}\ep_{1}\ldots \ep_{n_o}])+\cO(\hbar^{\gamma/2})\,.
\end{equation*}
The implied constant is uniform with respect to $n_o$ and $n$ in the allowed interval.
In other words, the measure $\mu^\Sigma_\hbar$ is almost $\sigma$-invariant:
$$
\sigma^n_\sharp\,\mu^\Sigma_\hbar([\bep])\defeq
\mu^\Sigma_\hbar(\sigma^{-n}[\bep])=\mu^\Sigma_\hbar([\bep])+\cO(\hbar^{\gamma/2})\,.
$$
\end{prop}
\begin{proof}
For simplicity we prove the result for $n_o=0$; the argument can easily be 
adapted to any $n_o>0$.

We use an estimate on the norm of commutators, proved in Lemma~\ref{l:commutator}.
If $A$ is an operator on $L^2(M)$, remember that we denote $A(t)=U^{-t}A U^t$.
According to Lemma \ref{l:commutator},
for any smooth observables $a, b$ supported inside $\cE^\nu=\cE(1/2-\nu,1/2+\nu)$, one has
\begin{equation}\label{e:ref-egorov}
\norm{[\Op_\hbar(a)(t), \Op_\hbar(b)(-t)]}_{L^2(M)}=\cO(\hbar^\gamma)\,,
\end{equation}
or equivalently 
$$
\norm{[\Op_\hbar(a)(2t), \Op_\hbar(b)]}_{L^2(M)}=\cO(\hbar^\gamma),
$$
for any time $|t|\leq T_{\nu,\gamma,\hbar}$. 
This result will be applied to the observables $a=P_{\ep_{0}}\,\cut$, 
$b=P_{\ep_{j}}\,\cut$, where $\cut$ is compactly supported in $\cE^\nu$ and identically $1$
near $\cE$. 
According to Remark \ref{r:cutoffs}, inserting the cutoff $\cut$ after each $P_{\ep_j}$
only modifies $\mu_\hbar^\Sigma([\bep])$ by an amount $\cO(\hbar^\infty)$. In the
following, we will omit to indicate these insertions and the 
$\cO(\hbar^\infty)$ errors.

To prove Proposition \ref{p:invariance}, we first write
\begin{align*}
\sum_{\ep_{i},-n\leq i\leq -1}\mu_\hbar^\Sigma( & [\ep_{-n}\ep_{-(n-1)}\ldots\ep_0])=
\sum_{\ep_{i},-n\leq i\leq -1}\norm{P_{\ep_0}P_{\ep_{-1}}(1)\ldots P_{\ep_{-n}}(n)\,\psi_\hbar}^2
\\&=\sum \la P_{\ep_{-1}}(1)P_{\ep_0}^2 P_{\ep_{-1}}(1)
\tP^*_{[\ep_{-2}\ldots \ep_{-n}]}(2)\,\psi_\hbar, \,  \tP^*_{[\ep_{-2}\ldots \ep_{-n}]}(2)\, \psi_\hbar\ra
\\&=\sum \la P_{\ep_0}^2 P_{\ep_{-1}}(1)^2
\tP^*_{[\ep_{-2}\ldots \ep_{-n}]}(2)\,\psi_\hbar, \tP^*_{[\ep_{-2}\ldots \ep_{-n}]}(2)\,\psi_\hbar\ra
\\&\qquad +\cO(\hbar^\gamma)
\Big[ \sum_{\ep_{i},-n\leq i\leq -2}  \norm{ \tP^*_{[\ep_{-2}\ldots \ep_{-n}]}(2)\,\psi_\hbar}^2\Big]
\\&=\sum_{\ep_{i},-n\leq i\leq -2} \la P_{\ep_0}^2 \tP^*_{[\ep_{-2}\ldots \ep_{-n}]}(2)\,\psi_\hbar, 
\tP^*_{[\ep_{-2}\ldots\ep_{-n}]}(2)\,\psi_\hbar\ra +\cO(\hbar^\gamma)\,.
\end{align*}
We have used the identities $\sum_{\ep_{-1}} P_{\ep_{-1}}(1)^2=I$ and 
$ \sum_{\ep_{-n},\ldots,\ep_{-2}}    \norm{ \tP^*_{[\ep_{-2}\ldots\ep_{-n}]}\,\psi_\hbar}^2=1$.
We repeat the procedure:
\begin{align*}
\sum_{\ep_{i},-n\leq i\leq -2} \la P_{\ep_0}^2
\tP^*_{[\ep_{-2}\ldots \ep_{-n}]} & (2)\,\psi_\hbar, \tP^*_{[\ep_{-2}\ldots \ep_{-n}]}(2)\,\psi_\hbar\ra
\\&=\sum \la P_{\ep_{-2}}(2)P_{\ep_0}^2 P_{\ep_{-2}}(2)
\tP^*_{[\ep_{-3}\ldots\ep_{-n}]}(3)\,\psi_\hbar, \tP^*_{[\ep_{-3}\ldots\ep_{-n}]}(3)\,\psi_\hbar\ra
\\&=\sum \langle P_{\ep_0}^2  P_{\ep_{-2}}(2)^2
\tP^*_{[\ep_{-3}\ldots \ep_{-n}]}(3)\,\psi_\hbar, \tP^*_{[\ep_{-3}\ldots\ep_{-n}]}(3)\,\psi_\hbar\ra
\\&\qquad +\cO(\hbar^\gamma)
\Big[\sum_{\ep_{i},-n\leq i\leq -3}  \norm{ \tP^*_{[\ep_{-3}\ldots \ep_{-n}]}(3)\,\psi_\hbar}^2\Big]
\\&=\sum_{\ep_{i},-n\leq i\leq -3} \la P_{\ep_0}^2 \tP^*_{[\ep_{-3}\ldots\ep_{-n}]}(3)\,\psi_\hbar, 
\tP^*_{[\ep_{-3}\ldots\ep_{-n}]}(3)\,\psi_\hbar\ra + \cO(\hbar^\gamma)\,.
\end{align*}
Iterating this procedure $n$ times we obtain
$$ 
\sum_{\ep_{i},-n\leq i\leq -1} \mu_\hbar^\Sigma([\ep_{-n}\ep_{-(n-1)}\ldots\ep_0])=
\la P_{\ep_0}^2 \psi_\hbar, \psi_\hbar\ra + n\,\cO(\hbar^\gamma)\,,
 $$
which proves the Proposition for $n_0=0$, since $n=\cO(|\log\hbar|)$. The proof for
any fixed $n_0>0$ is identical.
\end{proof}

\subsection{Proof of Proposition \ref{p:subadd}}

For $\psi_\hbar$ an eigenstate of the Laplacian, the entropy $h_n(\psi_\hbar)$ 
introduced in \eqref{e:entropy} can be expressed in terms of the measure $\mu_\hbar^\Sigma$:
\bequ\begin{split}\label{e:phi-muSigma}
h_n(\psi_\hbar)&=-\sum_{|\bep|=n}\norm{\tP_{\bep}^*\,\psi_\hbar}^2\,\log \norm{\tP_{\bep}^*\,\psi_\hbar}^2
=-\sum_{|\bep|=n} \mu^\Sigma_\hbar([\barbep])\,\log\mu^\Sigma_\hbar([\barbep])\\
&=-\sum_{|\bep|=n}\mu^\Sigma_\hbar([\bep])\log\mu^\Sigma_\hbar([\bep])\defi h_n(\mu_\hbar^\Sigma)\,.
\end{split}\end{equation}
In ergodic theory, the last term is called the entropy of the measure 
$\mu_\hbar^\Sigma$ with respect to the partition of $\Sigma$ into
$n$-cylinders.
Before using the results of the previous section, we choose the parameters $\nu,\gamma$ appearing
in Proposition~\ref{p:invariance} such that $\nu=\gamma=\delta'/2$, where $\delta'$ is the small
parameter in Proposition~\ref{p:subadd}. This ensures that the time 
$2\,T_{\nu,\gamma,\hbar}\geq n_E(\hbar)$ (see \eqref{e:Ehrenf}).

We then have, for any $n_o$ and $n$ such that $n+n_o\leq T_{\nu, \gamma, \hbar}$,
\begin{equation}\label{e:subadd-h}
h_{n_o+n}(\mu_\hbar^\Sigma)
\leq  h_{n-1}(\mu_\hbar^\Sigma )+h_{n_o}(\sigma^{n}_\sharp\,\mu_\hbar^\Sigma)
=  h_{n-1}(\mu_\hbar^\Sigma) +h_{n_o}(\mu_\hbar^\Sigma )+\cO_{n_o}(\hbar^{\delta'/4})\,.
\end{equation}
The notation $\cO_{n_o}$ means that the last term is bounded by $C_{n_o} \hbar^{\delta'/4}$,
with a constant $C_{n_o}$ depending on $n_o$.
The first inequality is a general property of the entropy, due to the concavity of the logarithm.
The second equality comes from the almost invariance of $\mu_\hbar^\Sigma$ (Proposition \ref{p:invariance}) 
and the continuity of the function $x\mapsto -x\log x$. 
The pressure for $\psi_\hbar$ (see \eqref{e:Jpressures}) also involves sums of the type
$$
\sum_{\bep=\ep_0\ldots\ep_{n_o+n}} \mu_\hbar^{\Sigma}([\bep])\, \log J^u_{n_o+n}(\bep)\defeq
\mu_\hbar^{\Sigma}(\log J^u_{n_o+n}) \,.
$$
Using the factorization \eqref{e:multi} of the Jacobian, this sum can be split into
\bequ\label{e:subadd-J}
\begin{split}
\mu_\hbar^{\Sigma}(\log J^u_{n_o+n})&=\mu_\hbar^{\Sigma}(\log J^u_{n-1})
+\sigma^{n-1}_\sharp\,\mu_\hbar^{\Sigma}(\log J^u_1)
+\sigma^n_\sharp\,\mu_\hbar^{\Sigma}(\log J^u_{n_o})\\
&= \mu_\hbar^{\Sigma}(\log J^u_{n-1})
+\mu_\hbar^{\Sigma}(\log J^u_1)
+\mu_\hbar^{\Sigma}(\log J^u_{n_o})+\cO_{n_o}(\hbar^{\delta'/4})\,.   
\end{split}
\end{equation}
We used once more the quasi-invariance of $\mu_\hbar^{\Sigma}$ to get the second equality.
Combining the inequalities (\ref{e:subadd-h},\ref{e:subadd-J}) with \eqref{e:phi-muSigma}, 
we obtain the Proposition~\ref{p:subadd} with the constant
$$
R= 3\,\max_{\rho\in \cE^{\dia}}|\log J^u_1(\rho)|\,.
$$
$\hfill\square$

\section{Some results of pseudodifferential calculus}\label{s:psiDO}
\subsection{Pseudodifferential calculus on a manifold}\label{s:PDO}
In this section we present the standard Weyl quantization of observables defined on the cotangent
of the compact $d$-dimensional manifold $M$ (see for instance \cite{EvZw06}). 
The manifold can be equipped with an atlas $\set{f_\ell,V_\ell}$, such that the $V_\ell$ form an open cover
of $M$, and for each $\ell$, $f_\ell$ is a diffeomorphism from $V_\ell$ to a bounded open set $W_\ell\subset\IR^d$.
Each $f_\ell$ induces a pullback $f_\ell^*:C^\infty(W_\ell)\to C^\infty(V_\ell)$.
We denote by $\tilde f_\ell$ the induced canonical map between
$T^*V_\ell$ and $T^*W_\ell$: 
$$
(x,\xi)\in T^*V_\ell\mapsto\tilde f_\ell(x,\xi)=(f_\ell(x),(D f_\ell(x)^{-1})^T\xi)\in T^*W_\ell\,,
$$ 
($A^T$ is the transposed of $A$) and by $\tilde f_\ell^*: C^\infty(T^*W_\ell)\to C^\infty(T^*V_\ell)$ 
the corresponding pull-back. 
One then chooses a smooth partition of unity on $M$ adapted to the charts $\set{V_\ell}$, namely
a set of functions $\phi_\ell\in C^\infty_c(V_\ell)$ such that $\sum_\ell\phi_\ell=1$ on $M$.

Any observable $a\in C^\infty(T^*M)$ can now be split into $a=\sum_j a_\ell$, with $a_\ell=\phi_\ell\,a$,
each term being pushed to $\tilde a_\ell=(\tilde f_\ell^{-1})^*a_\ell\in C^\infty(T^*W_\ell)$.
If $a$ belongs to a nice class of functions (possibly depending on $\hbar$), for instance 
the space of symbols
\begin{equation}\label{e:S^mk}
a\in S^{m,k}=S^k(\la\xi\ra^m)\defeq
\set{a=a_\hbar\in C^\infty(T^*M),\ |\partial_x^\alpha\partial_\xi^\beta a |\leq 
C_{\alpha,\beta} \hbar^{-k}\,\la\xi\ra^{m-|\beta|}},
\end{equation}
then Weyl-quantization associates to each $\tilde a_\ell$ a pseudodifferential operator on $\cS(\IR^d)$:
\bequ\label{e:Weyl}
\forall u\in \cS(\IR^d)\,,\qquad
\Op^w_\hbar(\tilde a_\ell)\,u(x)= \frac{1}{(2\pi \hbar)^d}\int e^{\frac{i}{\hbar}\langle x-y,\xi\rangle}
\tilde a_\ell\left(\frac{x+y}2,\xi;\hbar\right)\,u(y)\, dy\,d\xi\,.
\eequ
To pull this pseudodifferential operator back on $C^\infty(V_\ell)$, one takes
a smooth cutoff $\psi_\ell\in C^\infty_c(V_\ell)$ such that $\psi_\ell(x)=1$
close to $\supp\phi_\ell$. The quantization of $a\in S^{m,k}$ is finally defined as follows:
\begin{equation}\label{e:Op}
\forall u\in C^\infty(M),\quad
\Op_\hbar(a)\,u=\sum_\ell \psi_\ell\times f_\ell^*\circ
\Op_\hbar^w(\tilde a_\ell)\circ(f_\ell^{-1})^*(\psi_\ell\times u)\,.
\end{equation}
The space of pseudodifferential operators image of $S^{m,k}$ through this quantization 
is denoted by $\Psi^{m,k}(M)$. 
The quantization obviously depends on the cutoffs $\phi_\ell$, $\psi_\ell$.
However, this dependence only appears at second order in $\hbar$, and the principal symbol map
$\sigma:\Psi^{m,k}(M)\to S^{m,k}/S^{m,k-1}$ is intrinsically defined.
All microlocal properties of pseudodifferential operators on $\IR^d$ are carried over to
$\Psi^{m,k}(M)$. The Laplacian $-\hbar^2\lap$ belongs to $\Psi^{2,0}(M)$, with principal symbol
$\sigma(-\hbar^2\lap)=|\xi|^2_x$. 

We actually need to consider symbols more general than \eqref{e:S^mk}.
Following \cite{DS99}, for any $0\leq \eps <1/2$ we introduce the symbol class 
\bequ\label{e:symbol-eps}
S_\eps^{m,k}\defeq\set{a\in C^\infty(T^*M),\ |\partial_x^\alpha \partial_{\xi}^\beta a|\leq 
C_{\alpha,\beta}\, \hbar^{-k-\eps|\alpha+\beta|}\,\la\xi\ra^{m-|\beta|}}\,.
\eequ
The induced functions $\tilde a_\ell$ will then belong to the corresponding class on $T^*W_\ell$, for which
we can use the results of \cite{DS99}. For instance, the quantization of any $a\in S_\eps^{0,0}$ leads
to a bounded operator on $L^2(M)$ (the norm being bounded uniformly in $\hbar$). 

\subsection{Egorov theorem up to logarithmic times}\label{s:egorov}
We need analogous estimates to Bouzouina-Robert's \cite{BouzRob} concerning the quantum-classical
equivalence for long times. Our setting is more general,
since we are interested in observables on $T^*M$ for an arbitrary manifold $M$. 
On the other hand, we will only be interested in the first order term in the Egorov
theorem, whereas \cite{BouzRob} described the complete asymptotic expansion in power
of $\hbar$.

The evolution is given by the propagator $U^t$ on $L^2(M)$, which quantizes the flow $g^t$ on $T^*M$.
We will consider smooth observables $a\in C^\infty_c(T^*M)$ supported in a thin neighbourhood of 
the energy layer $\cE$, say inside the energy strip
$\cE^\nu=\cE([1/2-\nu,1/2+\nu])$ for some small 
$\nu>0$. This strip is invariant through the flow, so the evolved
observable $a_t=a\circ g^t$ will remain supported inside $\cE^\nu$. 
If $\lambda_{\max}$ is the maximal expansion rate of the flow on $\cE$ (see the definition
in Theorem~\ref{thethm}), then by homogeneity 
the maximal expansion rate
inside $\cE^\nu$ is $\sqrt{1+2\nu}\lambda_{\max}$. If we let
$\lambda_\nu\defeq(1+\nu)\lambda_{\max}$, the successive derivatives of the
flow on $\cE^\nu$ are controlled as follows:
\begin{equation}
\forall t\in\IR,\quad\forall\rho\in \cE^\nu,\quad \norm{\partial_\rho^\alpha g^t(\rho)}
\leq C_\alpha\,\e^{\lambda_\nu|\alpha\,t|}\,.
\end{equation}
Obviously, the derivatives of the evolved observable also satisfy
\begin{equation}\label{e:growth}
\forall t\in\IR,\quad\forall\rho\in \cE^\nu,\quad 
\norm{\partial^\alpha a_t(\rho)}\leq C_{a,\alpha}\,
\e^{\lambda_\nu|\alpha\, t|}\,.
\end{equation}
For times of the order of $|\log\hbar|$,
each derivative is bounded by some power of $\hbar^{-1}$. More precisely, 
for any $\gamma\in (0,1]$ and any $\hbar\in(0,1/2)$,
we call $T_{\nu,\gamma,\hbar}$ the following time:
\begin{equation}
T_{\nu,\gamma,\hbar}=\frac{(1-\gamma)|\log\hbar|}{2\lambda_\nu}=
\frac{(1-\gamma)|\log\hbar|}{2(1+\nu)\lambda_{\max}}\,.
\end{equation}
Starting from a smooth observable $a=a_0$, the bounds \eqref{e:growth} show that 
the family of function $\set{a_t=a\circ g^t\ :\ |t|\leq T_{\nu,\gamma,\hbar}}$ remains in the symbol class 
$S_{\eps}^{-\infty,0}$, with $\eps=\frac{1-\gamma}2$. 
Furthermore, any quasi-norm is uniformly bounded within the family. 
To prove a Egorov estimate, we start as usual from the identity
\begin{align}\label{e:Duhamel}
U^{-t}\,\Op_\hbar(a)\,U^t-\Op_\hbar(a\circ g^t)&=\int_0^t ds\,U^{-s}\,(\Diff a_{t-s})\,U^s,\\
\text{with} \quad \Diff a_t&\defeq\frac{i}{\hbar}[-\hbar^2\lap,\Op_\hbar(a_t)]-\Op_\hbar(\set{H,\,a_t})\,.
\end{align}
Since $-\hbar^2\lap$ belongs to $\Psi^{2,0}\subset \Psi_{\eps}^{2,0}$ and
$\Op_\hbar(a_t)\in \Psi_{\eps}^{-\infty,0}$ for times $|t|\leq T_{\nu,\gamma,\hbar}$,
the semiclassical calculus of \cite[Prop. 7.7]{DS99} (performed locally on each chart $V_j$) shows that
$\Diff a_t \in \Psi_{\eps}^{-\infty,-\alpha}$, with $\alpha=1-\eps=\frac{1+\gamma}2$.
From the Calderon-Vaillancourt theorem on $\Psi_{\eps}^{-\infty,-\alpha}$ \cite[Thm. 7.11]{DS99}, we extract
a constant $C_a>0$ such that, 
for any small enough $\hbar>0$ and any time $|t|\leq T_{\nu,\gamma,\hbar}$,
$$ 
\norm{\Diff a_t}\leq C_{a}\,\hbar^{\alpha}=C_{a}\,\hbar^{\frac{1+\gamma}2}\,.
$$
We can finally combine the above estimate in \eqref{e:Duhamel} and use the 
unitarity of $U^t$ (Duhamel's principle)
to obtain the following Egorov estimate.
\begin{prop}\label{p:Egorov}
Fix $\nu,\gamma\in(0,1]$.
Let $a$ be a smooth, $\hbar$-independent observable supported in $\cE^\nu$. Then, there
is a constant $C_a$ such that, for any time $|t|\leq T_{\nu,\gamma,\hbar}$, one has
\bequ
\norm{U^{-t}\,\Op_\hbar(a)\,U^t-\Op_\hbar(a\circ g^t)}\leq C_{a}\,|t|\,\hbar^{\frac{1+\gamma}2}\,. 
\eequ
\end{prop}
Let us now consider 
two observables $a,\ b\in C_c^\infty(\cE^\nu)$, evolve one in the future, 
the other in the past. The
calculus in $S_{\eps}^{-\infty,0}$ (with again $\eps=\frac{1-\gamma}2$) 
shows that, for any time $|t|\leq T_{\nu,\gamma,\hbar}$,
one has
$$
[\Op_\hbar(a\circ g^t),\,\Op_\hbar(b\circ g^{-t})]\in S_{\eps}^{-\infty,-\gamma}\,.
$$
Together with the above Egorov estimate and the Calderon-Vaillancourt theorem on 
$\Psi_{\eps}^{-\infty,-\gamma}$, this shows the following  
\begin{lem}\label{l:commutator}
Fix $\nu,\gamma\in (0,1]$. 
Let $a,b\in C_c^\infty(\cE^\nu)$ be independent of $\hbar$. 
Then there is a constant $C>0$ such that,
for small $\hbar$ and any time $|t|\leq T_{\nu,\gamma,\hbar}$,
$$
\norm{[U^{-t}\,\Op_\hbar(a)\,U^t,\,U^{t}\,\Op_\hbar(b)\,U^{-t}]}\leq C\,\hbar^\gamma\,.
$$
\end{lem}

\subsection{Cutoff in a thin energy strip}\label{s:cutoff}

As explained in \S\ref{s:energy}, we need an energy cutoff $\chi^{(0)}$ 
localizing in the energy strip of width $\sim\hbar^{\eps}$ around $\cE$, 
with $\eps\in[0,1)$ arbitrary close to $1$.
As a result, the $m$-th derivatives of $\chi$ transversally to $\cE$ 
will grow like $\hbar^{-m\eps}$.
The symbol classes \eqref{e:symbol-eps} introduced in the previous sections
do not include such functions if $\eps>1/2$. Yet, because the fluctuations
occur close to $\cE$ and only transversally, it is possible
to work with a ``second-microlocal'' pseudodifferential calculus which includes such fast-varying,
anisotropic symbols. 
We summarize here the treatment of this problem performed in \cite[Section 4]{SZ99}.

\subsubsection{Local behavior of the anisotropic symbols}

For any $\ep\in [0,1)$, we introduce a class of symbols $S^{m,k}_{\cE,\eps}$, made of
functions $a=a_\hbar$ satisfying the following properties:
\begin{itemize}
\item for any family of smooth vectors fields $V_1,\ldots, V_{l_1}$ tangent
to $\cE$, and of smooth vector fields $W_{1},\ldots, W_{l_2}$, one
has in each energy strip $\cE^{\nu}=\cE([1/2-\nu,1/2+\nu])$:
$$
\sup_{\rho\in\cE^\nu}|V_1\ldots V_{l_1}\, W_1\ldots W_{l_2}\,a(\rho)|=\cO(h^{-k-\eps\, l_2})\,.
$$
\item away from $\cE$, we have
$|\partial_x^\alpha \partial_\xi^\beta a (\rho)|=\cO(h^{-k} \la \xi \ra ^{m-|\beta|})$.
\end{itemize}
Notice that $S^{m,k}\subset S^{m,k}_{\cE,\eps'}\subset S^{m,k}_{\cE,\eps}$ if $1>\eps>\eps'\geq 0$.

To quantize this class of symbols, we 
cover a certain neighbourhood $\cE^{\nu}$ of $\cE$ by a family of bounded open sets 
$\cV_j$, such that
for each $j$, $\cV_j$ is mapped by a canonical diffeomorphism $\kappa_j$
to a bounded open set $\cW_j\subset T^*\R^{d}$, with $(0,0)\in\cW_j$.
We will denote by
$(x,\xi)$ the local coordinates on $\cV_j\subset T^*M$, and $(y,\eta)$ the image coordinates on $\cW_j$.
The canonical map $\kappa_j$ is chosen such that $H\circ\kappa_j^{-1}=\eta_1+1/2$. 
In particular, the image of $\cE\cap\cV_j$ is a 
piece of the hyperplane $\set{\eta_1=0}$. 

We consider a smooth cutoff function $\phi$ supported inside $\cE^{\nu}$, with
$\phi\equiv 1$ in $\cE^{\nu/2}$, and a smooth partition of unity 
$(\varphi_j)$ such that $1=\sum_j\varphi_j$ on $\cup_j\cV_j$, 
and $\supp \varphi_j\Subset \cV_j$.
For any symbol $a\in S^{m,k}_{\cE,\eps}$, the function 
$a(1-\phi)$ is supported outside $\cE^{\nu/2}$, and it belongs to the standard class 
$S^{m,k}$ of \eqref{e:S^mk}.
On the other hand, for each index $j$ the function
$$
a_j\defeq (a\,\phi\,\varphi_j)\circ\kappa_j^{-1}
$$ 
is compactly supported inside $\cW_j\subset T^*\IR^d$.
That function can be Weyl-quantized as in \eqref{e:Weyl}. 
Although $a_j(y,\eta)$ can oscillate at a rate $\hbar^{-\eps}$ along the
coordinate $\eta_1$ near $\set{\eta_1=0}$, 
for $a,b\in S^{m,k}_{\cE,\eps}$ the product $\Op^w_\hbar(a_j)\Op^w_\hbar(b_j)$
is still of the form $\Op^w_\hbar(c_j)$, where the function 
$c_j(y,\eta)$ is given by the Moyal product $a_j\sharp b_j$ and satisfies an asymptotic
expansion in powers of $\hbar^{1-\eps}$ and $\hbar$.

Mimicking the proof of the Calderon-Vaillancourt theorem in \cite[Thm.~7.11]{DS99}, we use the 
isometry (in $L^2(\IR^d)$) between $\Op_\hbar^w(A)$ and $\Op_1^w(A\circ T_\hbar)$, 
where the rescaling\\ 
$T_\hbar(y,\eta)=(y_1\hbar^{\frac{1-\eps}2},y'\hbar^{1/2};\eta_1\hbar^{\frac{1+\eps}2},\eta'\hbar^{1/2})$
ensures that the derivatives of $a_j\circ T_\hbar$ 
are uniformly bounded in $\hbar$. As a consequence we get the following
\begin{prop}
There exist $N_d$ and $C>0$ such that the following 
bound holds. 
For any symbol $a\in S^{m,k}_{\cE,\eps}$ and any $j$, the operator $\Op_\hbar^w(a_j)$
acts continuously on $L^2 (\IR^d)$, and its norm is bounded as follows:
$$
\norm{\Op^w_\hbar(a_j)}\leq \norm{a_j}_{L^\infty} + C\,
\sum_{1\leq |\alpha|+|\beta|\leq N_d} \hbar^{\frac12(|\alpha'|+|\beta'|+(1-\eps)\alpha_1+(1+\eps)\beta_1)}\;
\norm{\partial_y^\alpha\partial_\eta^\beta a_j}_{L^\infty}\,.
$$
\end{prop}

\subsubsection{Global quantization of the anisotropic symbols}

We now glue together the various pieces of $a\in S^{m,k}_{\cE,\eps}$ to define its global
quantization.
First of all, since $a(1-\phi)$ belongs to the standard class $S^{m,k}$ of \eqref{e:S^mk},
we can quantize it as in \S\ref{s:PDO}.

Then, for each index $j$ we select a Fourier integral operator
$U_{\kappa_j}: L^2(\pi(\cV_j))\to L^2(\pi(\cW_j))$, elliptic near
$\supp\varphi_j\times \kappa_j(\supp\varphi_j)\subset\cV_j\times \cW_j$, and associated with
the diffeomorphism $\kappa_j$ (an explicit expression is given in \S\ref{s:U_kappa}). 
Since $a_j$ describes the symbol $a$ in the coordinates $(y,\eta)$, 
it makes sense to pull $\Op^w_\hbar(a_j)$ back to the original coordinates $(x,\xi)$ using
$U_{\kappa_j}$.
The quantization of the global symbol $a\in S^{m,k}_{\cE,\eps}$ is then defined as follows:
\bequ\label{e:Op-Sigma}
\Op_{\cE,\hbar}(a)\defi\Op_\hbar(a(1-\phi))+\sum_j U_{\kappa_j}^{*}\,\Op^w_\hbar(a_j)\,U_{\kappa_j}\,.
\eequ
The Fourier integral operators $(U_{\kappa_j})$ can and will be chosen such that 
$\Op_{\cE,\hbar}(1)=Id+\cO_{L^2\to L^2}(\hbar^\infty)$.
The operators $\Op_{\cE,\hbar}(a)$ make up a space $\Psi^{m,k}_{\cE,\eps}$ of 
pseudodifferential operators on $M$.
The quantization $\Op_{\cE,\hbar}$ depends on the choice of the cutoffs $\phi$, $\varphi_j$, the
diffeomorphisms $\kappa_j$ and the associated FIOs $(U_{\kappa_j})$. 
It is equal to the quantization
$\Op_\hbar$ for symbols $a$ supported outside the energy strip $\cE^\nu$; 
otherwise, it differs from $\Op_\hbar$ by higher-order terms.

The space $\Psi^{-\infty,k}_{\cE,\eps}$ is invariant under conjugation
by FIOs which preserve the energy layer $\cE$. We will apply that property to the propagator 
$U=\e^{i\hbar\lap/2}$, which quantizes the flow $g^1$. One actually has 
a Egorov property
$$
U^{-1}\Op_{\cE,\hbar}(a)\,U=\Op_{\cE,\hbar}(b)\,,\quad\text{with}\quad
b-a\circ g\in S^{-\infty,k-1+\eps}_{\cE,\eps}\,.
$$
One is naturally lead to the definition of an $\hbar$-dependent {\em essential support} 
of a symbol $a_\hbar\in S^{m,k}_{\cE,\eps}$ (we will only consider the {\em finite part} of the
essential support, the infinite part at $|\xi|=\infty$ being irrelevant for our purposes).
A family of sets $(V_\hbar\subset T^*M)_{\hbar\to 0}$ does not intersect 
${\rm ess-supp}a_\hbar$ iff there exists $\chi_\hbar\in S^{-\infty,0}_{\cE,\eps}$, with $\chi_\hbar\geq 1$
on $V_\hbar$, such that $\chi_\hbar\,a_\hbar\in S^{-\infty,-\infty}_{\cE,\eps}$.
The essential support of $a_\hbar$ is also the {\em wavefront set} of its quantization,
$WF_\hbar(\Op_{\cE,\hbar}(a_\hbar))$.

The above Egorov property can be iterated to all orders, showing that
the wavefront set of an operator $A\in\Psi^{-\infty,k}_{\cE,\eps}$ is transported classically:
\begin{equation}\label{e:ess-supp}
WF_\hbar\big( U^{-1}\,A\,U \big)= g^{-1}(WF_\hbar(A))\,.
\end{equation}

\subsection{Properties of the energy cutoffs}\label{s:cutoff-props}

Take some small $\del>0$ and  $C_\del>0$ as in \S\ref{s:energy}, and define
$\eps=1-\delta$. One 
can easily check that
the cutoffs $\chi^{(n)}$ defined in \eqref{e:chi_n}, with $n\leq C_\del|\log\hbar|$, all belong 
to the symbol class $S^{-\infty,0}_{\cE,\eps}$.
From the above results, their quantizations $\Op(\chi^{(n)})=\Op_{\cE,\eps}(\chi^{(n)})$
are continuous operators on $L^2(M)$, of norms 
\bequ\label{e:norm}
\norm{\Op(\chi^{(n)})}=1+\cO(\hbar^{\delta/2})\,,
\eequ
with an implied constant independent of $n$.
We want to check that these cutoffs have little influence 
on an eigenstate $\psi_\hbar$ satisfying 
\eqref{e:eigenstate}. For this, we invoke the ellipticity of 
$(-\hbar^2\lap-1)\in\Psi^{2,0}\subset \Psi^{2,0}_{\cE,\eps}$ away from $\cE$. Using
\cite[Prop.~4.1]{SZ99}, one can adapt the standard division lemma to show the
following
\begin{prop}\label{p:division}
For $\hbar>0$ small enough and any $n\in\IN$, $0\leq n\leq C|\log\hbar|$, 
there exists $A^{(n)}_\hbar \in \Psi_{\cE,\eps}^{-2,\eps}$ 
and $R^{(n)}_\hbar\in \Psi^{-\infty,-\infty}_{\cE,\eps}$ such that
$$
\Op_{\cE,\hbar}(1-\chi^{(n)})= A^{(n)}_\hbar (-\hbar^2\Lap-1)+R^{(n)}_\hbar\,.
$$
As a result, for any eigenstate $\psi_\hbar= -\hbar^2\lap\psi_\hbar$, one has
$$
\norm{\psi_\hbar-\Op_{\cE,\hbar}(\chi^{(n)})\psi_\hbar}=\cO(\hbar^\infty)\,\norm{\psi_\hbar}\,.
$$
The implied constant is uniform with respect to $n$.
\end{prop}
This result contains in particular the estimate \eqref{e:local}.

\bigskip

We end this section by proving some properties of the cutoffs $\chi^{(n)}$. 
The general idea is that an eigenstate $\psi_\hbar$ is localized in
an energy strip of width $\hbar$, so that inserting cutoffs $\chi^{(n)}$ in expressions
of the type $\Op(a)\psi_\hbar$ has a negligible effect.

\begin{lem}\label{l:highlytechnical}
The following estimates are uniform for $\hbar>0$ small enough and $0\leq n\leq C_\del |\log\hbar|$:
\begin{align*}
&\qquad 
\norm{(1- \Op(\chi^{(n+1)}))\,U \Op(\chi^{(n)})}=\cO(\hbar^\infty)\,,\\
\forall k=0,\ldots,K,
&\qquad\norm{(1- \Op(\chi^{(n+1)}))\,U\,P_k\Op(\chi^{(n)})}=\cO(\hbar^\infty)\,.
\end{align*}
Here $P_k$ is any element of the partition of unity \eqref{e:partition}.
\end{lem}
\begin{proof}
For the symbols $\chi^{(n)}$ the essential support (which has been defined
above in a rather indirect way) coincides with the support.
The first statement of the Lemma uses the classical transport of the wavefront set \eqref{e:ess-supp},
applied to $\Op(\chi^{(n)})$. Since $\chi^{(n)}$
is invariant through the geodesic flow, $U\,\Op(\chi^{(n)})\,U^{-1}$ has the same wavefront set as
$\Op(\chi^{(n)})$. From the definition \eqref{e:chi_n}, the support
of $(1-\chi^{(n+1)})$ is at a distance $\geq C\,\hbar^{\eps}$ from the support of $\chi^{(n)}$.
The calculus on $S^{0,0}_{\cE,\eps}$ then implies that the product $(1- \Op(\chi^{(n+1)}))\Op(\chi^{(n)})$ is
in $\Psi^{-\infty,-\infty}_{\cE,\eps}$.

The second statement is a consequence of the first: the calculus on 
$\Psi^{0,0}_{\cE,\eps}$, which contains
the cutoffs $\Op(\chi^{(n)})$ and the multiplication operators 
$P_k$, shows that $\Op(\chi^{(n)})$ and $P_k\,\Op(\chi^{(n)})$ have the same wavefront set.
\end{proof}
We draw from this Lemma two properties which we use in the text (see \eqref{e:P_bep} for the definition
of $P_{\bep}$).
\begin{cor}\label{c:highlytechnical} 
For any sequence $\bep$ of length $n\leq C_\del|\log\hbar|$, one has
$$
\norm{(1-\Op(\chi^{(n)}))\,P_{\bep}\,\Op(\chi^{(0)})}=\cO(\hbar^\infty)\,.
$$
For any two sequence $\bep,\bep'$ of length $n\leq C_\del|\log\hbar|/4$, one has
$$
\norm{(1-\Op(\chi^{(4n)}))\,P_{\bep'}^*\,U^n\,P_{\bep}\,\Op(\chi^{n}))}=\cO(\hbar^\infty)\,.
$$
\end{cor}

\section{The entropic uncertainty principle: an application of complex interpolation}\label{s:proofWEUP}

In this section we prove the weighted entropic uncertainty principle, namely theorem~\ref{t:WEUP},
by adapting the original proof of \cite{MaaUff}.
 
We consider a complex Hilbert space $(\cH, \la.,.\ra)$, and denote 
the associated norm by $\norm{\psi}=\sqrt{\la\psi,\psi\ra}$. 
The same notation $\norm{\cdot}$ will also be used for the operator norm on $\cL(\cH)$.

Let $(\alpha_k)_{k=1,\ldots,\cN}$ be a family of positive numbers.
We consider the weighted $l_{p}$--norms on $\cH^{\cN}\ni\Psi=(\Psi_{1},\ldots,\Psi_{\cN})$:
 \begin{equation}\label{e:p-norms}
\norm{\Psi}^{(\alpha)}_p\defeq\left(\sum_{k=1}^{\cN} \alpha_j^{{p-2}}\norm{\Psi_k}^p \right)^{1/p}, \quad
1\leq p<\infty\,,
\text{ and}\quad
\norm{\Psi}^{(\alpha)}_\infty\defeq\max_k(\alpha_k\, \norm{\Psi_k})\,.
\end{equation}
For $p=2$, this norm does not depend on $(\alpha_k)$ and coincides with the Hilbert
norm deriving from the scalar product
\[
\la\Psi,\Phi\ra_{\cH^{\cN}}=\sum_{k}\la\Psi_{k},\Phi_{k}\ra_{\cH}.
\]
If $\Psi\in \cH^{\cN}$ has Hilbert norm unity, we define its entropy as
$$
h(\Psi)=-\sum_{k=1}^{\cN} \norm{\Psi_k}^2\log\norm{\Psi_k}^2\,,
$$
and its pressure with respect to the weights $(\alpha_k)$ is defined by
\begin{equation}\label{eq:pressures}
p_\alpha(\Psi)=-\sum_{k=1}^{\cN} 
\norm{\Psi_k}^2\log\norm{\Psi_k}^2-\sum_{k=1}^N \norm{\Psi_k}^2\log \alpha_k^2\,.
\end{equation}
This is the derivative of $\norm{\Psi}_p^{(\alpha)}$ with respect to $p$,
evaluated at $p=2$.

Similarly, let $(\beta_j)_{j=1,\ldots,\cM}$ be a family of weights.
They induce the following $l_{p}^{(\beta)}$--norms on $\cH^{\cM}\ni\Phi=(\Phi_{1},\ldots,\Phi_{\cM})$:
\begin{equation}
\norm{\Phi}^{(\beta)}_p\defeq\left(\sum_{j=1}^{\cM} \beta_j^{{p-2}}\norm{\Phi_j}^p \right)^{1/p}, \quad
1\leq p<\infty\,,
\text{ and}\quad
\norm{\Phi}^{(\beta)}_\infty\defeq\max_j(\beta_j\, \norm{\Phi_j})\,.
\end{equation}
We can define the entropy of a normalized vector $\Phi\in \cH^{\cM}$, and its pressure
$p_\beta(\Phi)$ with respect to the weights $(\beta_j)_{j=1,\ldots,\cM}$.
The standard $l_p-l_q$ duality \cite[Thm.IV.8.1]{DunSchw58} reads as follows in the present context: 
\begin{prop}\label{p:duality}
For any $1<p,q<\infty$ such that $\frac1p+\frac1q=1$, then 
\begin{equation}
\sup_{\left\Vert \Psi\right\Vert ^{(\alpha)}_{p}=1}|\la\Lambda,\Psi\ra|=
\left\Vert \Lambda\right\Vert^{(\alpha)}_{q}.\label{eq:duality}
\end{equation}
\end{prop}

\subsection{Complex interpolation}
A bounded operator $T:\cH^{\cN}\to\cH^{\cM}$ can be represented by
a $\cM\times \cN$ matrix $\left(T_{j\, k}\right)$ of bounded operators
on $\cH$. For $1\leq p,q\leq\infty$ we denote by 
$\left\Vert T\right\Vert^{(\alpha, \beta)}_{p,q}$ the norm of $T$ 
from $l_{p}^{(\alpha)}(\cH^{\cN})$
to $l_{q}^{(\beta)}(\cH^{\cM})$. 
We assume that $\left\Vert T\right\Vert_{2,2}=1$, which implies in particular
that $\norm{T_{jk}}\leq 1$ for all $k,j$.

\begin{ex}\label{mainex} 
Suppose we have two partitions of unity $(\pi_k)_{k=1}^{\cN}$ and $(\tau_j)_{j=1}^{\cM}$ on $\cH$,
that is, two families of operators such that 
\begin{equation}\label{eq:partition}
\sum_{k=1}^{\cN}\pi_{k}\pi_{k}^{*}=Id,\qquad\sum_{j=1}^{\cM}\tau_{j}\tau_{j}^{*}=Id.
\end{equation}
The main example we have in mind is the case where $\cU$ is a unitary operator on $\cH$ 
and $T_{j\, k}\defeq\tau_{j}^{*}\cU\pi_{k}$.
\end{ex}
   
\bigskip

Let $O$ be a bounded operator on $\cH$, and let $\eps \geq 0$. We will be interested
in the action of $T$ on the cone
$$
\cC(O, \eps)=\{\Psi\in \cH^{\cN},\  \norm{O\Psi_k-\Psi_k}\leq \eps\Vert\Psi\Vert_2
\mbox{ for all }k=1,\ldots,\cN\}\subset \cH^{\cN}\,.
$$
Notice that the cone $\cC(O, \eps)$ coincides with $\cH^{\cN}$ in the special case $O=Id$, $\eps=0$,
which is already an interesting case.

We introduce the positive number
$$
c_O(T)=\max_{j, k}\alpha_k\beta_j\norm{T_{jk}O}_{\cL(\cH)}\,,
$$
and also $A=\max_k \alpha_k$, $B=\max_j \beta_j$. The following theorem extends the
result of \cite{MaaUff}.
\begin{thm}\label{t:WEUPC}
For all $\Psi\in\cC(O,\eps)$ such that $\norm{\Psi}_2=1$ and $\norm{T\Psi}_2=1$, we have
$$
p_\beta(T\Psi)+p_\alpha(\Psi)\geq - 2\log \big(c_O(T)+\cN A B\eps\big)\,.
$$ 
\end{thm}

The proof of this theorem follows the standard proof of the 
Riesz-Thorin theorem \cite[sec.VI.10]{DunSchw58}.
In particular, one uses the following convexity property of complex analytic
functions. 

\begin{lem}[3-circle theorem]\label{l:3-circle}
Let $f(z)$ be analytic and bounded in the strip $\left\{ 0<x<1\right\} $,
and continuous on the closed strip. Then, the function $\log M(x)=\log\sup_{y\in\IR}\left|f(x+iy)\right|$
is convex in the interval $0\leq x\leq1$.
\end{lem}

We will define an appropriate analytic function in the unit strip. 
Let $\Psi \in \cC(O, \eps)$ with $\norm{\Psi}_2=1$. Fix $t\in [0, 1]$,
close to $0$, and let
$$
\tilde\Psi=\frac{\Psi}{\norm{\Psi}^{(\alpha)}_{\frac2{1+t}}}\,.
$$
From the definition of the norm and H\"older's inequality, we have
$$
\norm{\Psi}^{(\alpha)}_{\frac2{1+t}}\geq A^{-t}\,.
$$
Consider any state $\Phi\in\cH^{\cM}$ such that
$\norm{\Phi}^{(\beta)} _{\frac2{1+t}}\leq 1$.
For each $z=x+iy$ in the strip $\set{0\leq x\leq 1}$, we define
$$
a(z)=\frac{1+z}{1+t},
$$
and the states
\begin{align*}
\tilde\Psi(z)&=\left(\tilde\Psi(z)_{k}=\tilde\Psi_{k}
\norm{\tilde\Psi_{k}}^{a(z)-1}\alpha_k^{a(z)-1}\right)_{k=1\ldots\cN},\\
\Phi(z)&=\left(\Phi(z)_{j}=\Phi_{j}\norm{\Phi_{j}}^{a(z)-1}\beta_j^{a(z)-1}\right)_{j=1\ldots\cM}\,.
\end{align*}
By construction, we have
$$
\forall z=x+iy,\qquad\norm{ \tilde\Psi(z)}^{(\alpha)}_{\frac2{1+x}}=1\quad\textrm{and}\quad
\norm{\Phi(z)}^{(\beta)}_{\frac2{1+x}}\leq 1\,.
$$
In particular, for any $y\in\IR$ we have
\begin{equation}\label{e:b0}
\norm{\tilde\Psi(iy)}_{2} =1\;\textrm{and}\;\norm{\Phi(iy)}_{2}\leq1\Longrightarrow
\left|\la T\tilde\Psi(iy),\Phi(iy)\ra\right|\leq \norm{T}_{2,2}\,.
\end{equation}
Similarly, for any $y\in\IR$,
$$
\norm{\Phi(1+iy)}^{(\beta)} _{1}\leq1
\Longrightarrow\big|\la T\tilde\Psi(1+iy),\Phi(1+iy)\ra\big|
\leq \norm{T\tilde\Psi(1+iy)}^{(\beta)}_\infty.
$$
We decompose the right hand side by inserting the operator $O$:
\begin{multline*}
\norm{T\tilde\Psi(1+iy)}^{(\beta)}_\infty=\max_j \beta_j\norm{\sum_k T_{jk} \tilde\Psi(1+iy)_k}\\
\leq \max_j \beta_j \norm{\sum_k T_{jk} O\tilde\Psi(1+iy)_k}+
\max_j \beta_j\norm{\sum_k T_{jk} (Id-O) \tilde\Psi(1+iy)_k}\,.
\end{multline*}
The first term on the right hand side is bounded above by 
$c_O(T)\norm{\tilde\Psi(1+iy)}^{(\alpha)}_{1}=c_O(T)$. 
For the second term, we remark that
$$
\norm{\tilde\Psi(1+iy)_k}=|\alpha_k|^{\frac{1-t}{1+t}}\norm{\tilde\Psi_{k}}^{\frac2{1+t}}
=\frac{|\alpha_k|^{\frac{1-t}{1+t}}\norm{\Psi_{k}}^{\frac2{1+t}}}
{\big(\norm{\Psi}^{(\alpha)}_{\frac2{1+t}}\big)^{\frac2{1+t}}}.
$$
On the one hand, $\norm{ \Psi_k} \leq \norm{ \Psi}_2\leq 1$ and 
$|\alpha_k|^{\frac{1-t}{1+t}}\leq A^{\frac{1-t}{1+t}}$. 
On the other hand we have already stated that 
$\norm{\Psi}^{(\alpha)}_{\frac2{1+t}}\geq A^{-t}$. Putting these bounds together and using
the fact that $\Psi\in \cC(O,\eps)$, we get 
$$
\forall k=1,\ldots,\cN,\qquad \norm{(Id-O)\tilde\Psi(1+iy)_k}\leq A\,\eps\,.
$$
Summing over $k$ and using $\norm{T_{jk}}\leq 1$, we find
$$
\max_j\beta_j \norm{\sum_{k=1}^{\cN} T_{jk} (Id-O) \tilde\Psi(1+iy)_k}\leq \cN\,A\,B\,\eps\,.
$$
We have proved that for all $y\in\IR$,
\begin{equation}
\label{e:b1}\big|\la T\tilde\Psi(1+iy),\Phi(1+iy)\ra\big|\leq c_O(T)+\cN\,A\,B\,\eps\,.
\end{equation}
The function $z\mapsto \la T\tilde\Psi(z),\Phi(z)\ra$ is
bounded and analytic in the strip $\left\{ 0\leq x\leq1\right\} $: this is the function
to which we apply the 3-circle theorem (Lemma~\ref{l:3-circle}).
Taking in to account (\ref{e:b0},\ref{e:b1}), we obtain for any $x\in[0,1],\ y\in \IR$,
\begin{align*}
\log\big|\la T\tilde\Psi(x+iy),\Phi(x+iy)\ra\big|&\leq(1-x)\log \norm{T}_{2,2}+x\log(c_O(T)+\cN A B\eps)\\
&\leq x\log\big(c_O(T)+\cN A B\eps\big).
\end{align*}
The last inequality is due to our assumption $\norm{T}_{2,2}=1$.
In particular, taking $x+iy=t$, and exponentiating, we get
$$
\big|\la T\tilde\Psi,\Phi\ra\big|\leq\big(c_O(T)+\cN A B\eps\big)^t \,.
$$
Taking the supremum over $\set{\Phi\in\cH^{\cM},\ \norm{\Phi}_{\frac{2}{1+t}}^{(\beta)}\leq 1}$ 
and using the $l^{(\beta)}_{\frac2{1+t}}-l^{(\beta)}_{\frac2{1-t}}$ duality (Prop.~\ref{p:duality}), we obtain
$$
\norm{T\tilde\Psi}^{(\beta)}_{\frac2{1-t}}\leq \big(c_O(T)+\cN AB\eps\big)^t\,,
$$
and by homogeneity
\begin{equation}\label{e:ineq2}
\norm{T\Psi}^{(\beta)}_{\frac2{1-t}}\leq \big(c_O(T)+\cN A B \eps\big)^t\,\norm{\Psi}^{(\alpha)}_{\frac2{1+t}}\,.
\end{equation}
We may now take the limit $t\to0$
in this inequality. 
Using the assumption $\norm{\Psi}_2=1$, we notice that
$$
\log\norm{\Psi}^{(\alpha)}_{\frac{2}{1+t}}\sim\frac{1+t}{2}\log\left(\sum_k \norm{\Psi_k}^2
\exp\{-t\log\norm{\Psi_k}^2 - t\log\alpha_k^2\}\right)\sim\frac{t}{2}\,p_\alpha(\Psi).
$$
Similarly, $\log\norm{T\Psi}^{(\beta)} _{\frac{2}{1-t}}\sim-\frac{t}{2}\, p_\beta(T\Psi)$.
Therefore, in the limit $t\to0$, \eqref{e:ineq2} implies Theorem~\ref{t:WEUPC}.

$\hfill\square$

\subsection{Specialization to particular operators $T$ and states $\Psi$}
We now come back to the case of Example~\ref{mainex}.
\begin{lem}
Let $\cU:\cH\to\cH$ be a bounded operator. Using the two partitions of Example~\ref{mainex}, we
construct the operator $T:\cH^\cN\to \cH^\cM$ through its components
$T_{j k}=\tau_{j}^{*}\cU\pi_{k}$.
Then the two following norms are equal:
$$
\left\Vert T\right\Vert _{2,2}=\left\Vert \cU\right\Vert _{\cL(\cH)}.
$$
\end{lem}
\begin{proof}
The operator $T$ may be described as follows. Consider a line and
column vectors of operators on $\cH$:
$$
L\defi\left(\pi_{1},\ldots,\pi_{N}\right),\quad\textrm{respectively}\quad C=\left(\begin{array}{c}
\tau_{1}^{*}\\
\vdots\\
\tau_{M}^{*}\end{array}\right).
$$
We can write $T=C\cU L$. We insert this formula in the identity
$$
\left\Vert T\right\Vert _{2,2}^{2}=\left\Vert T^{*}T\right\Vert _{\cL(\cH^{\cN})}=
\left\Vert L^{*}\cU^{*} C^{*} C \cU L\right\Vert _{\cL(\cH^{\cN})}
$$
Using the resolution of identity of the $\tau_{j}$, we notice that $C^{*}C=Id_{\cH}$,
so that the above norm reads
$$
\left\Vert L^{*}\cU^{*}\cU L\right\Vert _{\cL(\cH^{\cN})}.
$$
Then, using the resolution of identity of the $\pi_{k}$, we get
$$
\left\Vert (\cU L)^{*}(\cU L)\right\Vert _{\cL(\cH^{\cN})}=
\left\Vert (\cU L)(\cU L)^{*}\right\Vert _{\cL(\cH)}=
\left\Vert (\cU L)L^{*}\cU^{*}\right\Vert _{\cL(\cH)}=\left\Vert \cU \cU^{*}\right\Vert _{\cL(\cH)}\,.
$$
\end{proof}
Therefore, if $\cU$ is contracting (resp. $\left\Vert \cU\right\Vert _{\cL(\cH)}=1$)
one has $ \left\Vert T\right\Vert _{2,2}\leq1$ (resp. $\left\Vert T\right\Vert _{2,2}=1$). 

We also specialize the vector $\Psi\in\cH^{\cN}$ by taking $\Psi_{k}=\pi_{k}^{*}\psi$
for some normalized $\psi\in\cH$. From the resolution of identity
on the $\pi_{k}$, we check that $\left\Vert \Psi\right\Vert _{2}=\left\Vert \psi\right\Vert $,
and also $(T\Psi)_{j}=\tau_{j}^{*}U\psi$. Thus, if $\left\Vert \cU\psi\right\Vert =1$,
the second resolution of identity induces $\left\Vert T\Psi\right\Vert _{2}=\left\Vert \cU\psi\right\Vert =1$.
With this choice for $T$ and $\Psi$, Theorem \ref{t:WEUPC}
reads as follows:
\begin{thm}
We consider the setting of Example~\ref{mainex}.
Let $\cU$ be an isometry on $\cH$.\\ 
Define $c_O^{(\alpha, \beta)}(U)\defi\sup_{j,k}\alpha_k\beta_j \norm{\tau_j^*\,\cU\,\pi_{k}\, O}_{\cL(\cH)}$.
 
Then, for any normalized $\psi\in\cH$ satisfying
$$
\forall k=1,\ldots,\cN,\qquad \norm {(Id-O)\pi^*_k\psi}\leq \eps\,,
$$
and defining the pressures as in \eqref{eq:pressures}, we have
$$
p_\beta\big((\tau_{j}^{*}\cU\psi)_{j=1\ldots \cM}\big) + p_\alpha\big((\pi_{k}^{*}\psi)_{k=1\ldots \cN}\big)
\geq - 2 \log \big(c_O^{(\alpha, \beta)}(U)+\cN\,A\,B\,\eps\big)\,.
$$
\end{thm}
This theorem implies Theorem \ref{t:WEUP}, if we take the same partition $\pi=\tau$ (in particular $\cN=\cM$), and if we 
remark that the pressures $p_\alpha\big((\pi_{k}^{*}\psi)_{k=1\ldots \cN}\big)$ and 
$p_\beta\big((\pi_{j}^{*}\cU\psi)_{j=1\ldots \cN}\big) $ are the same as the quantities 
$p_{\pi, \alpha}(\psi)$, $p_{\pi, \beta}(U\psi)$ appearing in the theorem.



\begin{thebibliography}{999999}

\bibitem{An}
N.~Anantharaman, {\em Entropy and the localization of eigenfunctions}, to appear in Ann. of Math.

\bibitem{AN06}
N.~Anantharaman, S.~Nonnenmacher, {\em Entropy of semiclassical measures of the Walsh-quantized baker's map}, 
Ann. H. Poincar\'e {\bf 8}, 37--74 (2007)

\bibitem{berry77}
M.V.~Berry,  {\it Regular and irregular
semiclassical wave functions}, J.Phys. {\bf A 10}, 2083--2091 (1977)

\bibitem{bohigas} O.~Bohigas, {\it Random matrix theory and chaotic dynamics}, in
 M.J.~Giannoni, A.~Voros and J.~Zinn-Justin eds.,
{\it Chaos et physique quantique}, 
(\'Ecole d'\'et\'e des Houches, Session LII, 1989), North Holland, 1991



\bibitem{BouzRob}
A.~Bouzouina  and D.~Robert: {\em Uniform Semi-classical Estimates for the Propagation of
Quantum Observables}, Duke Math. J. {\bf 111}, 223--252 (2002)

\bibitem{CdV85} 
Y.~Colin~de~Verdi\`ere, 
{\it Ergodicit\'e et fonctions propres du laplacien}, 
Commun. Math. Phys. {\bf 102}, 497--502 (1985)

\bibitem{CdVP94} Y.~Colin~de~Verdi\`ere, B.~Parisse, {\em \'Equilibre
instable en r\'egime semi-classique I : concentration microlocale}, 
Comm. Partial Differential Equations {\bf 19} no. 9-10, 1535--1563 (1994)

\bibitem{DS99}
M.~Dimassi, J.~Sj\"ostrand, {\em
Spectral asymptotics in the semi-classical limit},
London Mathematical Society Lecture Note Series, 268.
Cambridge University Press, Cambridge, 1999.

\bibitem{DunSchw58}
N.~Dunford and J.T.~Schwartz, {\it Linear Operators, Part I}, Interscience, New York, 1958.

\bibitem{EvZw06} L.C.~Evans and M.~Zworski, {\em Lectures on semiclassical analysis} (version 0.2),
available at {\tt http://math.berkeley.edu/$\sim$zworski}

\bibitem{FNdB03}
F.~Faure, S.~Nonnenmacher and S.~De~Bi\`evre,
{\it Scarred eigenstates for quantum cat maps of minimal periods},
Commun. Math. Phys. {\bf 239}, 449--492 (2003).

\bibitem{FN04}
F.~Faure and S.~Nonnenmacher, {\it On the maximal scarring for quantum cat map eigenstates},
Commun. Math. Phys. {\bf 245}, 201--214 (2004)

\bibitem{Horm}
L.~H\"ormander, {\it The analysis of linear partial differential operators I}, Springer 
(Heidelberg), 1983.

\bibitem{KatHas95}
A.~Katok and B.~Hasselblatt, {\it Introduction to the modern theory of
dynamical systems}, Encyclopedia of Mathematics and its applications vol.54,
Cambridge University Press, 1995.

\bibitem{Kelmer05}
D.~Kelmer, {\it Arithmetic quantum unique ergodicity for symplectic linear
maps of the multidimensional torus}, to appear in Ann. of Math., \texttt{math-ph/0510079}

\bibitem{Kelmer06}
D.~Kelmer, {\it Scarring on invariant manifolds for perturbed quantized hyperbolic toral automorphisms},
preprint, \texttt{math-ph/0607033}

\bibitem{Kl74} W.P.A.~Klingenberg, {\it Riemannian manifolds with geodesic
flows of Anosov type}, Ann. of Math. (2) {\bf 99}, 1--13 (1974)

\bibitem{Kraus87}
K.~Kraus, {\it Complementary observables and uncertainty relations},
Phys. Rev. {\bf D 35}, 3070--3075 (1987)  

\bibitem{LY85} F.~Ledrappier, L.-S.~Young, {\it 
 The metric entropy of diffeomorphisms. I. Characterization of measures satisfying Pesin's entropy formula},
Ann. of Math. (2) {\bf 122} (1985), no. 3, 509--539.

\bibitem{Linden06}
E.~Lindenstrauss, {\it Invariant measures and arithmetic quantum unique ergodicity}, 
Annals of Math. {\bf 163}, 165-219 (2006)

\bibitem{MaaUff}
H.~Maassen and J.B.M.~Uffink, {\it Generalized entropic uncertainty relations},
Phys. Rev. Lett. {\bf 60}, 1103--1106 (1988)
 
\bibitem{RudSar94}
Z.~Rudnick and P.~Sarnak, {\em The behaviour of eigenstates of arithmetic
hyperbolic manifolds}, Commun. Math. Phys. {\bf 161}, 195--213 (1994)

\bibitem{Shni74} 
A.~Schnirelman, {\it Ergodic properties of eigenfunctions}, 
Usp. Math. Nauk. {\bf 29}, 181--182 (1974)

\bibitem{SZ99} J.~Sj\"ostrand and M.~Zworski, {\em Asymptotic distribution of resonances
for convex obstacles}, Acta Math. {\bf 183}, 191--253 (1999)
 
\bibitem{voros77}
A.~Voros, {\it Semiclassical ergodicity of quantum eigenstates in the
  Wigner representation}, Lect. Notes Phys. {\bf 93}, 326-333 (1979) 
in: {\em Stochastic Behavior in Classical and Quantum Hamiltonian
  Systems}, G.~Casati, J.~Ford, eds., Proceedings of the Volta
Memorial Conference, Como, Italy, 1977, Springer, Berlin

\bibitem{Wol01}
S.A.~Wolpert, {\it The modulus of continuity for $\Gamma_0(m)/\IH$ semi-classical
limits}, Commun. Math. Phys. {\bf 216}, 313--323 (2001)

\bibitem{Zel87}
S.~Zelditch, 
{\it Uniform distribution of the eigenfunctions on compact hyperbolic surfaces},
Duke Math. J. {\bf 55}, 919--941 (1987)







\end{thebibliography}
\end{document}